\documentclass[smallextended]{svjour3a}

\usepackage{graphicx,amsmath,amssymb,hyperref,nicefrac,xcolor,bm,enumitem,xfrac}
\usepackage[paperwidth=6.11in,paperheight=9.26in,text={4.7in,7.75in},centering,top=.75in]{geometry}

\usepackage[normalem]{ulem}
\usepackage{esint}
\usepackage{fancyhdr}
\usepackage{hyperref}
\hypersetup{
 colorlinks,
 linkcolor={blue!90!black},
 citecolor={red!80!black},
 urlcolor={blue!50!black}
}

\numberwithin{equation}{section}
\numberwithin{theorem}{section}

\usepackage{mathpazo}
\usepackage[T1]{fontenc}

\newcommand{\ignore}[1]{}
\newcommand{\opn}{\operatorname}

\newcommand{\re}{\operatorname{Re}}
\newcommand{\im}{\operatorname{Im}}

\newcommand{\mbb}[1]{\mathbb{#1}}
\newcommand{\mb}[1]{\mathbf{#1}}
\newcommand{\mc}[1]{\mathcal{#1}}
\newcommand{\pa}{\partial}
\newcommand{\der}[2]{\frac{\partial #1}{\partial #2}}

\newcommand{\jt}{\textstyle}

\newcommand{\e}[1]{{(#1)}}

\newcommand{\bu}{\mathbf{u}}
\newcommand{\fs}{0}
\DeclareMathOperator{\diver}{div}

\newcommand{\la}{\langle}
\newcommand{\ra}{\rangle}

\fancypagestyle{plain}{%
\fancyhf{}
\fancyfoot[C]{\fontsize{10pt}{10pt}\selectfont\thepage}

}

\begin{document}

\title{Numerical Algorithms for Water Waves with Background Flow over
  Obstacles and Topography}

\titlerunning{Water waves over obstacles}

\author{David M. Ambrose \and Roberto Camassa \and Jeremy L. Marzuola \and
  Richard M. McLaughlin \and Quentin Robinson \and Jon Wilkening}

\institute{D.M. Ambrose \at
  Department of Mathematics, Drexel University, Philadelphia, PA 19104 USA\\
  \email{dma68@drexel.edu} \and
  R. Camassa, J.L. Marzuola, R. McLaughlin \at
  Department of Mathematics, University of North Carolina at Chapel Hill\\
  \email{camassa@email.unc.edu, marzuola@math.unc.edu, rmm@email.unc.edu} \and
  Q. Robinson \at
  Department of Mathematics and Physics, North Carolina Central University\\
  \email{qrobinson5@nccu.edu} \and
  J. Wilkening \at
  Department of Mathematics, University of California, Berkeley, CA 94720-3840\\
  \email{wilkening@berkeley.edu}
}

\date{\today}

\maketitle

\begin{abstract}
We present two accurate and efficient algorithms for solving the
incompressible, irrotational Euler equations with a free surface in
two dimensions with background flow over a periodic,
multiply-connected fluid domain that includes stationary obstacles and
variable bottom topography. One approach is formulated in terms of the
surface velocity potential while the other evolves the vortex sheet
strength. Both methods employ layer potentials in the form of
periodized Cauchy integrals to compute the normal velocity of the free
surface, are compatible with arbitrary parameterizations of the free
surface and boundaries, and allow for circulation around each
obstacle, which leads to multiple-valued velocity potentials but
single-valued stream functions.  We prove that the resulting
second-kind Fredholm integral equations are invertible, possibly after
a physically motivated finite-rank correction. In an angle-arclength
setting, we show how to avoid curve reconstruction errors that are
incompatible with spatial periodicity. We use the proposed methods to
study gravity-capillary waves generated by flow around several
elliptical obstacles above a flat or variable bottom boundary. In each
case, the free surface eventually self-intersects in a splash
singularity or collides with a boundary.  We also show how to evaluate
the velocity and pressure with spectral accuracy throughout the fluid,
including near the free surface and solid boundaries. To assess the
accuracy of the time evolution, we monitor energy conservation and the
decay of Fourier modes and compare the numerical results of the two
methods to each other. We implement several solvers for the
discretized linear systems and compare their performance. The fastest
approach employs a graphics processing unit (GPU) to construct the
matrices and carry out iterations of the generalized minimal residual
method (GMRES).
\end{abstract}


\thispagestyle{plain}
\fancyfoot[C]{\fontsize{10pt}{10pt}\selectfont\thepage}

\section{Introduction}

Many interesting phenomena in fluid mechanics occur as a result of the
interaction of a fluid with solid or flexible structures.  Most
numerical algorithms to study such problems require discretizing the
bulk fluid \cite{alben:flap:05,froehle:persson,zahr:16} or are
tailored to the case of slender bodies \cite{tornberg:04}, flexible
filaments \cite{alben:flags:09,alben:flutter:20} or unbounded domains
\cite{hoogedoorn:10}.  In the present paper, we propose a robust
boundary integral framework for the fast and efficient numerical
solution of the incompressible, irrotational Euler equations in
multiply-connected domains that have numerous fixed obstacles,
variable bottom topography, a background current, and a free
surface. We present two methods within a common boundary integral
framework, one in which the surface velocity potential is evolved
along with the position of the free surface and another where the
vortex sheet strength is evolved. Treating the methods together in a
unified framework consolidates the work in analyzing the schemes,
reveals unexpected connections between the integral equations that
arise in the two approaches, and provides strong validation through
comparison of the results of the two codes.

Studies of fluid flow over topography of various forms is a rather
classical problem, and any attempt to give a broad overview of the
history of the problem would inevitably fall short within a limited
space. We give here a brief discussion, including many articles that
point to further relevant citations to important works on the topics.
The linear response to a background current for water waves driven by
gravity and surface tension was studied long ago and is present in now
classical texts such as
\cite{lamb1932hydrodynamics,whitham2011linear}.  In the case of
cylindrical obstacles, Havelock
\cite{havelock1927method,havelock1929vertical} carries out an analysis
using the method of successive images.  Further nonlinear studies of
the gravity wave case are undertaken in works such as
\cite{dagan1972two,miloh1993nonlinear,peregrine1976interaction,scullen1995nonlinear,tuck1965effect}.
Capillary effects are considered in
\cite{forbes1983free,grandison2006truncation,milewski1999time}.
Algorithms using point sources for cylindrical obstacles are
introduced and studied in
\cite{moreira2001interactions,moreira2012nonlinear}.  Analytic
solutions in infinite water columns exterior to a cylinder are given
in \cite{crowdy2006analytical}.  Flows in shallow water with variable
bottom topography are studied in various contexts as forced
Korteweg-de Vries equations in
\cite{camassa1991stability,el2006unsteady,el2009transcritical,grimshaw1986resonant,hirata2015numerical}
and again recently in \cite{robinson_thesis}.  An algorithm for
computing the Dirichlet-Neumann operator (DNO) in three-dimensions
over topography has recently been proposed by Andrade and Nachbin
\cite{andrade:nachbin}.


Computational boundary integral tools are developed and implemented in
\cite{baker1982generalized}, for instance, and have been made quite
robust in the works
\cite{ambrose2010computation,baker:nachbin:98,ceniceros2001efficient,hou2007computing,HLS1,HLS2,water2}
and many others. Analysis of these types of models and schemes is
carried out in \cite{akers2013traveling,ambrose2003well}. In two
dimensions, complex analysis tools have proved useful for summing over
periodic images and regularizing singular integrals; early examples of
these techniques date back to Van de Vooren \cite{vande:vooren:80},
Baker \emph{et al.} \cite{BMO} and Pullin \cite{pullin:82}.

More
recently, the conformal mapping framework of Dyachenko \emph{et al.}
\cite{dyachenko1996analytical} has emerged as one of the simplest and
most efficient approaches to modeling irrotational water waves over
fluids of infinite depth
\cite{choi1999exact,li2004numerical,milewski:10,zakharov2002new}. The
conformal framework extends to finite depth with flat
\cite{turner:bridges} or variable bottom topography
\cite{viotti2014conformal} and can also handle quasi-periodic boundary
conditions \cite{quasi:trav,quasi:ivp}.  However, at large amplitude,
these me\-th\-ods suffer from an anti-resolution problem in which the
gridpoints spread out near wave crests, especially for overturning
waves, which is precisely where more gridpoints are needed to resolve
the flow. There are also major technical challenges to formulating and
implementing conformal mapping methods in multiply-connected domains
with obstacles, and of course they do not have a natural extension to
3D. By contrast, boundary integral methods are compatible with
adaptive mesh refinement \cite{water2}, can handle multiply-connected
domains (as demonstrated in the present work), and can be extended to
3D via the theory of layer potentials; see Appendix~\ref{sec:3d}.

In multiply-connected domains, the integral equations of potential
theory sometimes possess nontrivial kernels
\cite{folland1995introduction}. This turns out to be the case in
  the present work for the
velocity potential formulation but not for the vortex sheet
formulation. We propose a physically motivated finite-rank correction
in the velocity potential approach to eliminate the kernel and compute
the constant values of the stream function on each of the obstacles
relative to the bottom boundary, which is taken as the zero contour of
the stream function. These stream function values are needed anyway
(in both the velocity potential and vortex sheet formulations) to
compute the energy. This stream-function technique does not generalize
to 3D, but the challenge of a multiple-valued velocity potential also
vanishes in 3D, alleviating the need to introduce a stream
function to avoid having to compute line integrals through the fluid
along branch cuts of the velocity potential in the energy formula.
Our study of the solvability of the integral equations that arise is
rigorous, generalizing the approach in \cite{folland1995introduction}
to the spatially periodic setting and adapting it to different sets of
boundary conditions than are treated in
\cite{folland1995introduction}.

In our numerical simulations, we find that gravity-capillary waves
interacting with rigid obstacles near the free surface often evolve to
a splash singularity event in which the curve self-intersects. In
rigorous studies of such singularities
\cite{castro2013finite,castro2012splash}, the system is prepared in a
state where the curve intersects itself. Time is then reversed
slightly to obtain an initial condition that will evolve forward to
the prepared splash singularity state. Here we start with a flat wave
profile and the free surface dynamics is driven by the interaction of
the background flow with the obstacles and bottom boundary.
The same qualitative results occur for different choices of parameters
governing the circulation around the obstacles, though in one
  case the free surface collides with an obstacle rather than
  self-intersecting.  Thus, if we widen the class of splash
  singularities to include boundary collisions, they seem to
be a robust eventual outcome, at least for sufficiently large
background flow.

Of course, the circulation around obstacles in real fluids would be
affected by viscosity and the shedding of a wake, which can be modeled
as a vortex sheet. For bodies with sharp edges the circulation can be
assigned within a potential flow formulation using the so-called
`Kutta condition' at the edge by choosing the circulation to eliminate
a pressure singularity there. Note that for time dependent problems,
this condition would have to be applied dynamically in time, which
adds additional steps in the solution method.  We will not pursue this
here, and leave this generalization to future work.

We find that the angle-arclength parameterization of Hou, Lowengrub
and Shelley (HLS) \cite{HLS1,HLS2} is particularly convenient for
overturning waves. Nevertheless, we formulate our boundary integral
methods for arbitrary parameterizations. This allows one to switch to
a graph-based parameterization of the free surface, if appropriate,
and can be combined with any convenient parameterization of the bottom
boundary and obstacles --- it is not necessary to parameterize these
boundaries uniformly with respect to arclength even if a uniform
parameterization is chosen for the free surface. One could
  also build upon this framework to employ adaptive mesh refinement in
  angle-arclength variables along the lines of what was done in
  \cite{water2} in a graph-based setting. We use explicit 8th
order Runge-Kutta timestepping in the examples presented in
Section~\ref{sec:numerics}, though it would be easy to implement a
semi-implicit Runge-Kutta scheme \cite{carpenter} or exponential
time-differencing scheme \cite{cox:matthews,quasi:ivp} using the HLS
small-scale decomposition. The \raisebox{-0.5pt}{$\sfrac{3}{2}$}%
\hspace*{1pt}-\hspace*{1pt}order CFL condition of this problem
\cite{ambrose2003well,HLS1,HLS2} is a borderline case where explicit
time-stepping is competitive with semi-implicit methods if the surface
tension is not too large.

One challenge in using the HLS angle-arclength parameterization in a
Runge-Kutta framework is that internal Runge-Kutta stages are only
accurate to $O(\Delta t^2)$. When the tangent angle function and
arclength are evolved as ODEs, this can lead to discontinuities in the
curve reconstruction that excite high spatial wave numbers that do not
cancel properly over a full timestep to yield a higher order
method. Hou, Lowengrub and Shelley avoid this issue by using an
implicit-explicit multistep method \cite{ascherRuuthWetton}. In the
present paper, we propose a more flexible solution in which only the
zero-mean projection of the tangent angle is evolved via an ODE. The
arclength and the mean value of the tangent angle are determined
algebraically from periodicity constraints. This leads to properly
reconstructed curves even in interior Runge-Kutta stages, improving
the performance of the timestepping algorithm.

To aid in visualization, we derive formulas for the velocity and
pressure in the fluid that remain spectrally accurate up to the
boundary. For this we adapt a technique of Helsing and Ojala
\cite{helsing} for evaluating layer potentials in 2D near boundaries.
Details are given in Appendix~\ref{sec:helsing}.

This paper is organized as follows.  First, in Section
\ref{sec:surface}, we establish notation for parameterizing the free
surface and solid boundaries and show how to modify the HLS
angle-arclength representation to avoid falling off the constraint
manifold of angle functions and arclength elements that are compatible
with spatial periodicity. In Section~\ref{sec:cauchy} we describe the
velocity potential formulation and introduce multi-valued complex
velocity potentials to represent background flow and circulation
around obstacles. In Section~\ref{sec:laypotalg} we describe the
vortex sheet formulation and derive the evolution equation for the
vortex sheet strength on the free surface. Connections are made with
the velocity potential method. In Section~\ref{sec:summary:upE},
  we summarize the methods, show how to implement different choices of
  the tangential velocity, derive formulas for the energy that
  remain valid for multi-valued velocity potentials, and show how to
compute the velocity and pressure in the interior of the fluid from
the surface variables that are evolved by the time-stepping scheme.
In Section~\ref{sec:solvability}, we analyze the solvability of the
velocity potential and vortex sheet methods and prove that the
resulting integral equations are invertible after a finite-rank
modification of the integral operator for the velocity potential
method. We also show that the systems of integral equations for the
two methods are adjoints of each other after modifying one of them to
evaluate each layer potential by approaching the boundary from the
``wrong'' side.

In Section~\ref{sec:numerics}, we present numerical results for
four scenarios of free surface flow over elliptical obstacles
with a flat or variable bottom boundary.  In each case, the mesh
  is refined several times in the course of evolving the solution. We
  stop at the point that the solution is still resolved with spectral
  accuracy but cannot be evolved further on the finest mesh due to a
  self-intersection event or collision with the boundary that appears
  imminent. The results are validated by monitoring energy
  conservation, decay of spatial Fourier modes, and quantitative
  comparison of the results of the velocity potential and vortex sheet
  methods. We then discuss the performance of the algorithms using
  Gaussian elimination or the generalized minimal residual method
  (GMRES) in the integral equation solvers. Our fastest implementation
  employs a graphics processing unit (GPU) to accelerate the
  computation of the integral equation matrices and perform GMRES
  iterations.  Concluding remarks are given in
Section~\ref{sec:conclude}, followed by seven appendices containing
further technical details. In particular, Appendix~\ref{sec:3d}
discusses progress and challenges in extending the algorithms to
multiply-connected domains in 3D.

\section{Boundary Parameterization and Motion of the Free Surface}
\label{sec:surface}

We consider a two-dimensional fluid whose velocity and pressure
satisfy the incompressible, irrotational Euler equations.  The fluid
is of finite vertical extent, and is bounded above by a free surface
and below by a solid boundary.  The location of the free surface is
given by the parameterized curve
\begin{equation*}
  (\xi(\alpha,t),\eta(\alpha,t)),
\end{equation*}
with $\alpha$ the parameter along the curve and with $t$ the time.  We
denote this free surface by $\Gamma,$ or to be very precise, we may
call it $\Gamma(t).$ We will also write $\xi_\fs$, $\eta_\fs$ and
$\Gamma_\fs$ when enumerating the free surface as one of the domain
boundaries.  We consider the horizontally periodic case in which
\begin{equation}\label{eq:per0}
  \xi (\alpha+2\pi,t)=\xi (\alpha,t)+2\pi,\quad
  \eta (\alpha+2\pi,t)=\eta(\alpha,t),\quad  (\alpha\in\mbb R,\; t\ge0).
\end{equation}
The bottom boundary, $\Gamma_1$, is time-independent. Its
location is given by the parameterized curve $(\xi_1 (\alpha),\eta_1
  (\alpha)),$ which is horizontally periodic with the same
period,
\begin{equation}\label{eq:per1}
  \xi_1(\alpha+2\pi)=\xi_1(\alpha)+2\pi, \qquad
  \eta_1(\alpha+2\pi)=\eta_1(\alpha),
\qquad (\alpha\in\mbb R).
\end{equation}

One may also consider one or more obstacles in the flow, such as a
cylinder.  As we are considering periodic boundary conditions, in fact
there is a periodic array of obstacles.  We denote the location of
such objects by the parameterized curves
\begin{equation}
  (\xi_j(\alpha),\eta_j(\alpha)), \qquad\quad (2\le j\le N),
\end{equation}
where $N$ is the number of solid boundaries. Like the bottom boundary,
these curves are time-independent. We denote these curves by
$\Gamma_j$, $2\le j\le N$.  We have simple periodicity of the location
of the obstacles,
\begin{equation}\label{eq:per2N}
\xi_j(\alpha+2\pi)=\xi_j(\alpha),\qquad
\eta_j(\alpha+2\pi)=\eta_j (\alpha),\qquad (2\le j\le N, \;
  \alpha\in\mathbb{R}).
\end{equation}
While the periodic images of the free surface and bottom boundary are
swept out by extending $\alpha$ beyond $[0,2\pi)$, the periodic images
of the obstacles can only be obtained by discrete horizontal
translations by $2\pi\mathbb{Z}$.  We take the parameterization of the
solid boundaries to be such that the fluid lies to the left, i.e., the
normal vector $(-\eta_{j,\alpha}, \xi_{j,\alpha})$ points into the
fluid region for $1\le j\le N$,
where an $\alpha$-subscript denotes differentiation.
Thus, the bottom boundary is
parameterized left to right and the obstacles are parameterized
clockwise.  The free surface is also parameterized left to right, so
the fluid lies to the right and the normal vector points away from the
fluid.  This is relevant for the Plemelj formulas later.

Since each of these boundaries is described by a parameterized curve,
there is no restriction that any of them must be a graph; that is, the
height of the free surface and the height of the bottom need not be
graphs with respect to the horizontal.  Similarly, the shapes of the
obstacles need not be graphs over the circle.  We denote the length of
one period of the free surface by $L(t)$ or $L_0(t),$ the length of
one period of the bottom boundary by $L_1,$ and the circumference of
the $j$th obstacle by $L_j.$ We will often benefit from a complexified
description of the location of the various surfaces, so we introduce
the following notations:
\begin{equation}\label{eq:zeta:cx}
\begin{aligned}
  \zeta(\alpha,t) =
  \zeta_\fs(\alpha,t) &= \xi(\alpha,t)+i \eta(\alpha,t), \\
  \zeta_j(\alpha) &= \xi_j (\alpha)+i\eta_j (\alpha),
  \qquad\quad (1\le j\le N).
\end{aligned}
\end{equation}

\begin{figure}[t]
  \begin{center}
    \includegraphics[scale=.45]{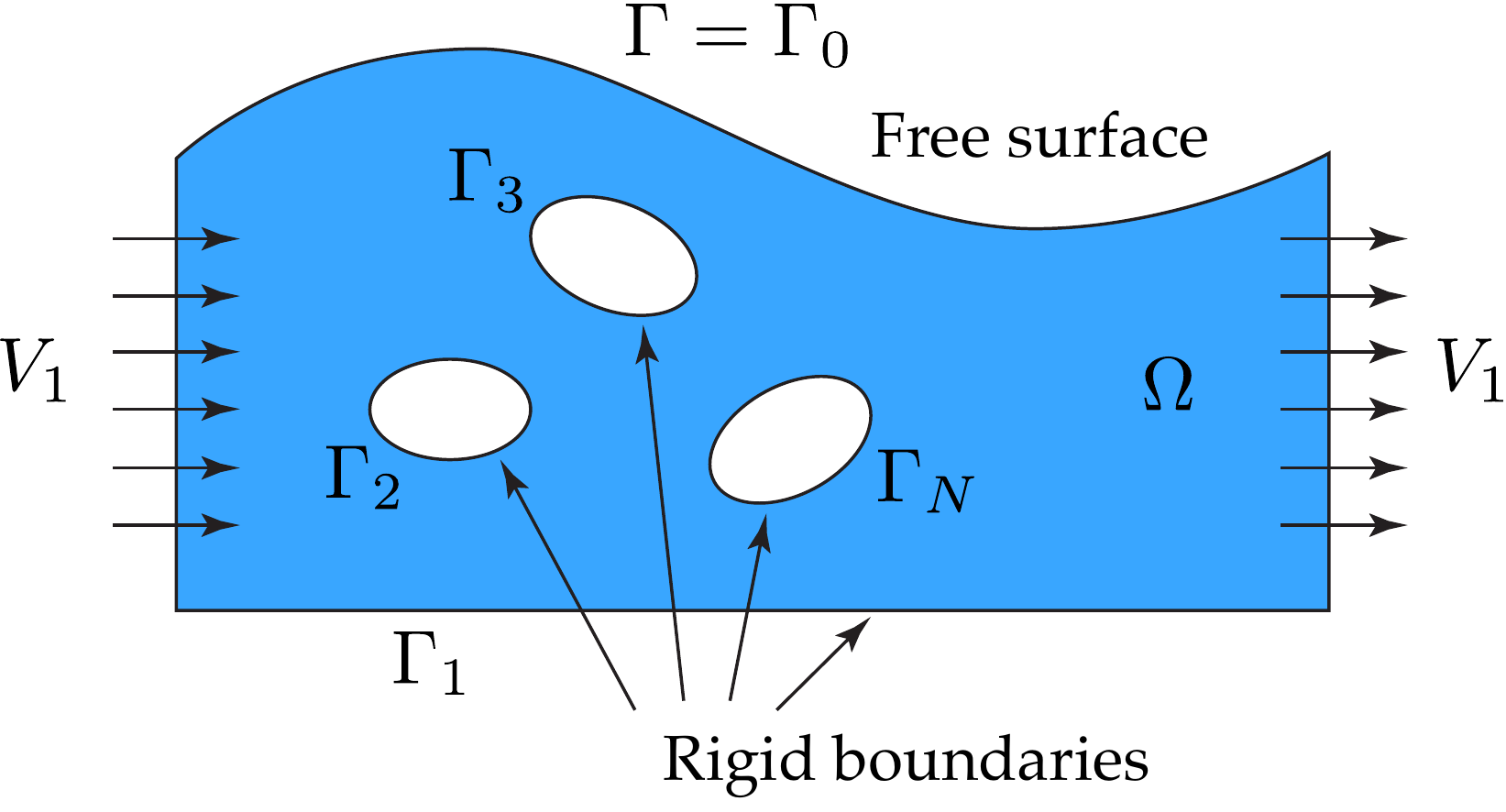}
  \end{center}
  \caption{ \label{fig:setup} The fluid region is bounded above by a
    free surface, $\Gamma(t)$, below by a solid boundary, $\Gamma_1$,
    and internally by obstacles $\Gamma_2$,\dots,$\Gamma_N$. The
    domain is spatially periodic with all components having the same
    period, normalized to be $2\pi$.  We allow for a background flow
    in which the velocity potential increases by $2\pi V_1$ when $x$
    increases by $2\pi$ along a path passing above each of the
    obstacles.}
\end{figure}

\subsection{Graph-based and angle-arclength parameterizations
  of the free surface}

At a point $(\xi(\alpha,t), \eta(\alpha,t))$ we have unit tangent and
normal vectors. Suppressing the dependence on $(\alpha,t)$ in the
notation, they are
\begin{equation}\label{eq:tn:hat}
  \mathbf{\hat{t}}=\frac{(\xi_{\alpha},\eta_{\alpha})}{
    |(\xi_{\alpha},\eta_{\alpha})|},
  \qquad
  \mathbf{\hat{n}}=\frac{(-\eta_{\alpha},\xi_{\alpha})}{
    |(\xi_{\alpha},\eta_{\alpha})|}.
\end{equation}
We describe the motion of the free surface using the generic evolution
equation
\begin{equation}\label{genericEvolution}
  (\xi,\eta)_{t}=U\mathbf{\hat{n}}+V\mathbf{\hat{t}}.
\end{equation}
Here $U$ is the normal velocity and $V$ is the tangential velocity of
the parameterization.  One part of the Hou, Lowengrub and Shelley
(HLS) \cite{HLS1,HLS2} framework is the idea that $V$ need not be
chosen according to physical principles, but instead may be chosen to
enforce a favorable parameterization on the free surface.  The normal
velocity, however, must match that of the fluid.

In Sections~\ref{sec:cauchy} and~\ref{sec:laypotalg} below, we present
two methods of computing the normal velocity $U=\pa\phi/\pa n$ of the
fluid on the free surface, where $\phi(x,y,t)$ is the velocity
potential. A simple approach for cases when the free surface is not
expected to overturn or develop steep slopes is to set
$\xi(\alpha)=\alpha$ and evolve $\eta(x,t)$ in time.  Setting
$\xi_t=0$ in \eqref{genericEvolution} and using \eqref{eq:tn:hat}
gives $V=\eta_\alpha U$ and
\begin{equation}\label{eq:graph:eta:t}
  \eta_t = \sqrt{1+\eta_\alpha^2}\,U, \qquad U = \der\phi n.
\end{equation}
This is the standard graph-based formulation
\cite{craig:sulem:93,zakharov68} of the water wave equations, where
the Dirichlet-Neumann operator mapping the velocity potential on the
free surface to the normal velocity now involves solving the Laplace
equation on a multiply-connected domain.  Mesh refinement can be
introduced by choosing a different function $\xi(\alpha)$ such that
$\xi_\alpha(\alpha)$ is smaller in regions requiring additional
resolution. This is done in \cite{water2} for the case without
obstacles to resolve small-scale features at the crests of
large-amplitude standing water waves.

Hou, Lowengrub and Shelley \cite{HLS1,HLS2} proposed a flexible
alternative to the graph-based representation that allows for
overturning waves and simplifies the treatment of surface tension.
Rather than evolving the Cartesian coordinates $\xi(\alpha,t)$ and
$\eta(\alpha,t)$ directly, the tangent angle $\theta(\alpha,t)$
of the free surface relative to the horizontal is evolved in time.
In the complex representation \eqref{eq:zeta:cx}, we have
\begin{equation}\label{eq:za:zt}
  \zeta_\alpha = s_\alpha e^{i\theta}, \qquad \zeta_t = (V+iU)e^{i\theta},
\end{equation}
where $s_{\alpha}(\alpha,t)$ is the arclength element, defined by
$s_{\alpha}=|\zeta_\alpha|=\sqrt{\xi_{\alpha}^{2}+\eta_{\alpha}^{2}}.$
Equating $\zeta_{\alpha t}=\zeta_{t\alpha}$ in \eqref{eq:za:zt}, one
finds that
\begin{equation}\label{eq:theta:t}
  \theta_t = \frac{U_\alpha + V\theta_\alpha}{s_\alpha}, \qquad
  s_{\alpha t}=V_{\alpha}-\theta_{\alpha}U.
\end{equation}
One can require a uniform parameterization in which
$s_\alpha(\alpha,t)=L(t)/2\pi$ is independent of $\alpha$,
where $L(t)$ is the length of a period of the interface.
This gives
\begin{equation}\label{definitionOfV}
  V_{\alpha}=\theta_{\alpha}U-\frac{1}{2\pi}
  \int_{0}^{2\pi}\theta_{\alpha}U\ d\alpha.
\end{equation}
By taking the tangential velocity, $V,$ to be a solution of
\eqref{definitionOfV}, we ensure that the normalized arclength
parameterization is maintained at all times.  When solving
\eqref{definitionOfV} for $V,$ a constant of integration must be
chosen.  Three suitable choices are (a) the mean of $V$ can be taken
to be zero; (b) $V(0,t)=0;$ or (c) $\xi(0,t)=0$.  We usually prefer
(c) as it conveniently anchors the coordinate system
  for describing the surface.

\subsection{Staying on the constraint manifold}
\label{sec:constman}

In solving the evolution of the surface profile in the HLS framework,
we must ensure that a periodic profile arises at each stage of the
iteration. As we have described the HLS method so far, the curve
$\zeta(\alpha)$ is represented by $\theta(\alpha)$ and
$s_\alpha=L/2\pi$ together with two integration constants, which we
take to be $\xi(0)=0$ and $\la\eta\ra = \frac1{2\pi}\int_0^{2\pi}
\eta(\alpha)\xi_{\alpha}(\alpha)\,d\alpha = 0$. The latter quantity is the
average height of the free surface, which, by incompressibility,
remains constant in time and can be set to 0 by a suitable vertical
adjustment of the initial conditions and solid boundaries if
necessary.  The problem is that not every function $\theta$ and number
$s_\alpha$ are the tangent angle and arclength element of a periodic
curve (in the sense of (\ref{eq:per0})). We refer to those that are as
being on the constraint manifold.

A drawback of the HLS formulation is that numerical error can cause
the solution to deviate from this constraint manifold, e.g., in
internal Runge-Kutta stages or when evolving the interface over many
time steps.  Internal Runge-Kutta stages typically contain $O(h^2)$
errors that cancel out over the full step if the solution is smooth
enough; thus, it is critical that the curve reconstruction not
introduce $O(h^2)$ grid oscillations.

Our idea is to evolve only $P\theta$ in time and select $P_0\theta$
and $s_\alpha$ as part of the reconstruction of $\zeta(\alpha)$ to
enforce $\zeta(2\pi)=\zeta(0)+2\pi$. Here $P_0$ is the orthogonal
projection in $L^2(0,2\pi;d\alpha)$ onto the constant functions while
$P$ projects onto functions with zero mean,
\begin{equation}
  P = I-P_0, \qquad P_0f = \frac1{2\pi}\int_0^{2\pi}f(\alpha)\,d\alpha.
\end{equation}
Note that $P_0\eta$ is the mean of $\eta$ with respect to $\alpha$ on
$[0,2\pi]$, which differs from the mean in physical space, $\la\eta\ra
= P_0[\eta\xi_\alpha]$.  Given $P\theta$, we define
\begin{equation}\label{eq:CS:th}
  \begin{aligned}
    C &= P_0\big[\cos P\theta], \\
    S &= P_0\big[\sin P\theta],
  \end{aligned} \qquad
  \begin{aligned}
    P_0\theta &= \opn{arg}(C-iS), \\
    \theta &= P\theta + P_0\theta
  \end{aligned} \qquad
  s_\alpha = (C^2+S^2)^{-1/2}.
\end{equation}
We then define $\zeta_\alpha = s_\alpha e^{i\theta}$ and note
that
\begin{equation}\label{eq:z2p:z0}
  \zeta(2\pi)-\zeta(0)
  = (s_\alpha e^{iP_0\theta})\int_0^{2\pi}e^{i(P\theta)(\alpha)}d\alpha
  = \frac{C-iS}{C^2+S^2}\big[2\pi(C+iS)\big] = 2\pi.
\end{equation}
Thus, any antiderivative $\zeta(\alpha)$ of
$\zeta_\alpha=\xi_\alpha+i\eta_\alpha$ will lie on the constraint
manifold. We also note for future reference that
\begin{equation}\label{eq:P:cos:sin}
  P\big[\cos\theta\big] = \cos\theta - s_\alpha^{-1}, \qquad
  P\big[\sin\theta\big] = \sin\theta.
\end{equation}
Next, we compute the zero-mean antiderivatives
\begin{equation*}
  \xi^\text{aux} = \int [\xi_\alpha-1]\,d\alpha, \qquad
  \eta^\text{aux} =
  \int \eta_\alpha\,d\alpha
\end{equation*}
via the FFT.  Both integrands have zero mean due to (\ref{eq:z2p:z0}),
so $\xi^\text{aux}$ and $\eta^\text{aux}$ are $2\pi$-periodic. The
conditions $\xi(0)=0$ and $\la\eta\ra=0$ are achieved by adding
integration constants
\begin{equation}\label{eq:xi:eta:recon}
  \xi(\alpha)=\alpha+\xi^\text{aux}(\alpha)-\xi^\text{aux}(0), \qquad
  \eta(\alpha)=\eta^\text{aux}(\alpha) - P_0[\eta^\text{aux}\xi_\alpha].
\end{equation}
The $\alpha$ term in $\xi(\alpha)$ accounts for the 1 in the integrand
in the formula for $\xi^\text{aux}$.  This completes the
reconstruction of $\zeta(\alpha) = \xi(\alpha)+i\eta(\alpha)$ from
$P\theta$.

We compute the normal velocity, $U$, of the fluid on the reconstructed
curve $\zeta(\alpha)$ as described in Sections~\ref{sec:cauchy}
or~\ref{sec:laypotalg} below.  The evolution of $P\theta$ is obtained
by applying $P$ to the first equation of (\ref{eq:theta:t}),
\begin{equation}\label{eq:Pth:t}
  (P\theta)_t = P\left(\frac{ U_\alpha + V\theta_\alpha }{s_\alpha} \right).
\end{equation}
In Appendix~\ref{appendix:HLS}, we show that $\theta$ and $s_\alpha$
from (\ref{eq:CS:th}) satisfy (\ref{eq:theta:t}) even though
$P_0\theta$ and $s_\alpha$ are computed algebraically rather than by
solving ODEs.  We also show that
  \eqref{eq:theta:t} implies that the curve kinematics are
  correct, i.e., $(\xi_t,\eta_t)=U\mb n + V\mb t$.  As far as the
authors know, this approach of evolving $P\theta$ via (\ref{eq:Pth:t})
and computing $P_0\theta$ and $s_\alpha$ algebraically (rather than
  evolving them) is an original formulation (although a different
  algebraic formula for just $s_{\alpha}$ has been used previously
  \cite{akers2013traveling}).

We reiterate that the advantage of computing $P_{0}\theta$ and
$s_{\alpha}$ from $P\theta$ is that the reconstructed curve is always
on the constraint manifold. This avoids loss of accuracy in internal
Runge-Kutta stages or after many steps due to grid oscillations that
arise when computing periodic antiderivatives from functions with
non-zero mean.

\section{Cauchy Integrals and the Velocity Potential Formulation}
\label{sec:cauchy}

As explained above, we let the fluid region, $\Omega,$ be
$2\pi$-periodic in $x$ with free surface $\Gamma=\Gamma_\fs$, bottom
boundary $\Gamma_1$, and cylinder boundaries $\Gamma_2$, \dots
$\Gamma_N$.  The cylinder boundaries need not be circular, but are
assumed to be smooth.  We view $\Omega$ as a subset of the complex
plane. Let us decompose the complex velocity potential
$\Phi(z)=\phi(z)+i\psi(z)$ as
\begin{equation}\label{eq:Phi:decomp}
  \Phi(z) = \tilde\Phi(z) + \Phi_\text{mv}(z),
\end{equation}
where $\tilde\Phi(z) = \Phi_\fs(z) + \cdots + \Phi_N(z)$ and
\begin{equation}\label{eq:Phi:mv}
  \Phi_\text{mv}(z) =
  V_1 z + \sum_{j=2}^N a_j\Phi_\text{cyl}(z-z_j), \qquad
  \Phi_\text{cyl}(z) = -i\log\big( 1 - e^{iz} \big).
\end{equation}
Here $z_j$ is a point inside the $j$th cylinder; $V_1$ and the $a_j$
are real parameters corresponding to the background flow
    strength and circulation around each cylinder, divided by $2\pi$,
  which allow $\phi(z)$ (but not $\psi(z)$) to be
multi-valued on $\Omega$; and $\tilde\Phi(z)$ is
represented by Cauchy integrals:
\begin{align}\label{eq:w123}
    \Phi_\fs(z) &= \frac{1}{2\pi i}\int_0^{2\pi}
    \frac{1}{2}\cot\frac{\zeta_\fs(\alpha) - z}{2}\omega_\fs(\alpha)
    \,\zeta_\fs'(\alpha)\,d\alpha, & & \text{(free surface)}, \\
    \notag
    \Phi_j(z) &= \frac{1}{2\pi i}\int_0^{2\pi}
    \frac{1}{2}\cot\frac{\zeta_j(\alpha) - z}{2}
    i\omega_j(\alpha)\,\zeta_j'(\alpha)\,d\alpha,
    & & \left(\begin{gathered}\text{solid boundaries} \\
        1\le j\le N\end{gathered}\right).
\end{align}
Here $\omega_j(\alpha)$ is a real-valued function for $0\le j\le N$,
and we use primes interchangeably with $\alpha$ subscripts
  to denote derivatives of $\zeta_j(\alpha)$, $\omega_j(\alpha)$,
etc. We refer to these as Cauchy integrals as they correspond to a
principal value sum of the Cauchy kernel over periodic images via a
Mittag-Leffler formula \cite{ablowitzFokas}, namely
\begin{equation}
  PV\sum_k\frac1{\zeta+2\pi k-z}=\frac12\cot\frac{\zeta-z}2.
\end{equation}
We have temporarily dropped $t$ from $\zeta_\fs(\alpha,t)$ since time
may be considered frozen when computing the velocity potential. The
subscript $\fs$ is optional only for $\zeta_\fs$, $\xi_\fs$,
$\eta_\fs$, $\Gamma_\fs$ and for quantities such as $s_\alpha$ and
$\theta$ defined in terms of $\eta_\alpha$ and $\xi_\alpha$.  In
particular, $\Phi,\phi,\psi$ are not the same as
$\Phi_\fs,\phi_\fs,\psi_\fs$ in our notation.  The real and imaginary
parts of $\tilde\Phi$, $\Phi_\text{mv}$, $\Phi_j$ and
$\Phi_\text{cyl}$ will be denoted by $\tilde\phi$, $\tilde\psi$,
$\phi_\text{mv}$, etc.

\subsection{Properties of $\Phi_\text{cyl}(z)$ and time independence
  of $V_1$, $a_2$, \dots, $a_N$}
\label{sec:phi:cyl:prop}

We regard $\Phi_\text{cyl}(z)$ as a multi-valued analytic function
defined on a Riemann surface with branch points $z\in2\pi\mbb Z$. On
the $n$th sheet of the Riemann surface, $\Phi_\text{cyl}(z)$ is given
by
\begin{equation}\label{eq:phi:up}
  \Phi_{\text{upper},n}(z) = -i\opn{Log}\big(1-e^{iz}\big) + 2\pi n,
  \qquad (n\in\mbb Z),
\end{equation}
where $\opn{Log}(z)$ is the principal value of the logarithm.  The
functions \eqref{eq:phi:up} are analytic in the upper half-plane and
have vertical branch cuts extending from the branch points down to
$-i\infty$. Their imaginary parts are all the same, given by
$-\ln|1-e^{iz}|$, which is continuous across the branch cuts (except
  at the branch points) and harmonic on $\mbb C\setminus2\pi\mbb
Z$. The real part of $\Phi_{\text{upper},n}(z)$ jumps from
$2\pi(n+1/2)$ to $2\pi(n-1/2)$ when crossing a branch cut from left to
right.  We obtain $\Phi_\text{cyl}(z)$ by gluing
$\Phi_{\text{upper},n}(z)$ on the left of each branch cut to
$\Phi_{\text{upper},n+1}(z)$ on the right. Equivalently, we can define
horizontal branch cuts $I_k=\big(2\pi k,2\pi(k+1)\big)\subset\mbb R$
for $k\in\mbb Z$ and glue $\Phi_{\text{upper},n}(z)$ to
\begin{equation}
  \Phi_{\text{lower},m}(z) = -i\opn{Log}\big(1-e^{-iz}\big)+z+(2m-1)\pi,
  \qquad (m\in\mbb Z)
\end{equation}
along $I_{n-m}$. $\Phi_{\text{lower},m}(z)$ is analytic in the lower
half-plane and has vertical branch cuts extending from the points
$z\in2\pi\mbb Z$ up to $+i\infty$.  Both $\Phi_{\text{upper},n}(z)$
and $\Phi_{\text{lower},m}(z)$ are defined and agree with each other
on the strip $z=x+iy$ with $y\in\mbb R$ and $x\in I_{n-m}$, so they
are analytic continuations of each other to the opposite half-plane
through $I_{n-m}$. To show this, one may check that
\begin{equation*}
  \Phi_{\text{upper},0}(x)= \left(\frac{x-\pi}2-i\ln\sqrt{2-2\cos
    x}\right)=\Phi_{\text{lower},0}(x), \quad (0<x<2\pi).
\end{equation*}
By the identity theorem,
$\Phi_{\text{upper},0}(z)=\Phi_{\text{lower},0}(z)$ on the strip
$z=x+iy$ with $x\in I_0$ and $y\in\mbb R$.  The result follows using
the property that $\Phi_{\text{upper},n}(z)$ is $2\pi$-periodic while
$\Phi_{\text{lower},m}(z+2\pi k)= \Phi_{\text{lower},m}(z)+2\pi k$
for $k\in\mbb Z$.

Following a path from some point $z_*$ to $z_*+2\pi$ that passes above
all the cylinders will cause $\Phi_\text{mv}(z)$ to increase by $2\pi
V_1$.  The free surface is such a path. If the path passes below all
the cylinders, $\Phi_\text{mv}(z)$ will increase by
$2\pi(V_1+a_2+\cdots+a_N)$. More complicated paths from $z_*$ to
$z_*+2\pi n_1$ that loop $n_j\in\mbb Z$ times around the $j$th
cylinder in the counter-clockwise ($n_j>0$) or clockwise ($n_j<0$)
direction relative to a path passing above all the cylinders will
cause $\Phi_\text{mv}(z)$ to change by
$2\pi(n_1V_1+n_2a_2+\cdots+n_Na_N)$.

The derivative of $\Phi_\text{cyl}(z)$ is single-valued and has poles
at the points $z\in2\pi\mbb Z$. Explicitly,
\begin{equation}\label{eq:d:phi:cyl}
  \Phi'_\text{cyl}(z) = \frac12 - \frac i2\cot\frac z2.
\end{equation}
A more evident antiderivative of this function is
\begin{equation}\label{eq:phi:cyl:alt}
  \frac z2 - i\log\sin\frac z2,
\end{equation}
which has the same set of possible values as
$\big[\Phi_\text{cyl}(z)+\frac\pi2+i\ln 2\big]$ for a given $z$.
However, using the principal value of the logarithm in
\eqref{eq:phi:cyl:alt} leads to awkward branch cuts that are difficult
to explain how to glue together to obtain a multi-valued function
$\Phi_\text{cyl}(z)$ on a Riemann surface.

It follows from the Euler equations for
$\mathbf{u}=\nabla\phi$,
\begin{equation}
  \rho \nabla\left( \phi_t + \frac12\|\nabla\phi\|^2 + \frac p\rho + g y\right) =
  \rho[\bu_t + \bu\cdot\nabla\bu] + \nabla p + \rho g\mathbf{\hat{y}} = 0,
\end{equation}
that $V_1$ and the $a_j$ are independent of time.
This is because the change in $\phi_t$ around a path encircling a
cylinder or connecting $(0,y_*)$ to $(2\pi,y_*)$ is the negative of the
change in $\frac12\|\nabla\phi\|^2 + \frac p\rho + g y$, which is
single-valued and periodic. For the cylinders, this also follows
from the Kelvin circulation theorem.

\subsection{Integral equations for the densities $\omega_j$}
\label{sec:int:eq:densities}

Evaluation of the Cauchy integrals in (\ref{eq:w123}) on the boundaries
via the Ple\-melj formulas \cite{singularIntegralEquations} gives the
results in Table~\ref{tbl:K:def}.  When $j=k\in\{0,\dots,N\}$ and
$\beta\rightarrow\alpha$, $K_{jj}(\alpha,\beta)\rightarrow\im\{
\zeta_j''(\alpha)/[2\zeta_j'(\alpha)]\}$, so the apparently singular
integrands are actually regular due to the imaginary part.
They are automatically regular when $j\ne k$ since $\zeta_j(\beta)$
and $\zeta_k(\alpha)$ are never equal. So far $G_{kj}$ only arises
with $j\ne k$; the regularizing term $(1/2)\cot[(\beta-\alpha)/2]$
will become relevant in (\ref{eq:d:phi:0}) below.

\begin{table}
  \begin{center}
  \fbox{
  \parbox{.8\linewidth}{
\begin{align*}
  \phi_\fs(\zeta_\fs(\alpha)^-) &= -\frac{1}{2}\omega_\fs(\alpha)
  + \frac{1}{2\pi}
  \int_0^{2\pi} K_{\fs\fs}(\alpha,\beta)\omega_\fs(\beta)\,d\beta, \\
  \phi_j(\zeta_\fs(\alpha)) &= \frac{1}{2\pi}\int_0^{2\pi}
  G_{\fs j}(\alpha,\beta)\omega_j(\beta)\,d\beta, \qquad (1\le j\le N) \\
  \psi_\fs(\zeta_k(\alpha)) &= \frac{1}{2\pi}\int_0^{2\pi}
  -G_{k\fs}(\alpha,\beta)\omega_\fs(\beta)\,d\beta, \qquad (1\le k\le N) \\
  \psi_j(\zeta_j(\alpha)^+) &= \frac{1}{2}\omega_j(\alpha) + \frac{1}{2\pi}
  \int_0^{2\pi} K_{jj}(\alpha,\beta)\omega_j(\beta)\,d\beta,
  \qquad (1\le j\le N) \\
  \psi_j(\zeta_k(\alpha)) &= \frac{1}{2\pi}\int_0^{2\pi}
  K_{kj}(\alpha,\beta)\omega_j(\beta)\,d\beta, \qquad (1\le j,k\le N,
    \;\; j\ne k) \\
  K_{kj}(\alpha,\beta) &= \im\left\{\frac{\zeta_j'(\beta)}{2}\cot\left(
    \frac{\zeta_j(\beta)-\zeta_k(\alpha)}{2}\right)\right\},
  \qquad (0\le j,k\le N) \\
  G_{kj}(\alpha,\beta) &= \re\left\{\frac{\zeta_j'(\beta)}{2}\cot\left(
    \frac{\zeta_j(\beta)-\zeta_k(\alpha)}{2}\right) -
    \delta_{kj}\frac12\cot\frac{\beta-\alpha}2\right\}.
  \end{align*}}}
  \end{center}
  \caption{\label{tbl:K:def} Evaluation of the Cauchy integrals on the
    boundaries. The Plemelj formulas \cite{singularIntegralEquations}
    are used to take one-sided limits from within $\Omega$, where the
    positive side of a parameterized curve lies to the left. In the
    last formula, $\delta_{kj}=1$ if $k=j$ and 0 otherwise. This term
    will be relevant in (\ref{eq:d:phi:0}) below. }
\end{table}

Next we consider the operator $\mbb B$ mapping the dipole densities
$\omega_j$ to the values of $\tilde\phi$ on $\Gamma_0^-$ and
$\tilde\psi$ on $\Gamma_k^+$ for $1\le k\le N$. Recall from
\eqref{eq:Phi:decomp} that a tilde denotes the contribution of the
Cauchy integrals to the velocity potential.  We regard the functions
$\omega_j$, $\tilde\phi\vert_{\Gamma_0^-}$ and
$\tilde\psi\vert_{\Gamma_k^+}$ as elements of the (real) Hilbert space
$L^2(0,2\pi)$. They are functions of $\alpha$, and we do not assume
the curves $\zeta_j(\alpha)$ are parameterized by arclength.
The operator $\mbb B$ has a block structure arising from the
  formulas in Table~\ref{tbl:K:def}. For example, when $N=2$, $\mbb
B$ has the form
\begin{equation}\label{eq:B:def}
  \mbb B\omega :=
  \begin{pmatrix} \tilde \phi\vert_{\Gamma_0^-} \\
    \tilde \psi\vert_{\Gamma_1^+} \\
    \tilde \psi\vert_{\Gamma_2^+} \end{pmatrix} =
  \left[
    \begin{pmatrix}
      -\frac{1}{2}\mbb I \\ & \frac{1}{2}\mbb I \\ & & \frac{1}{2}\mbb I
    \end{pmatrix} + 
    \begin{pmatrix}
      \mbb K_{00} & \mbb G_{01} & \mbb G_{02} \\
      -\mbb G_{10} & \mbb K_{11} & \mbb K_{12} \\
      -\mbb G_{20} & \mbb K_{21} & \mbb K_{22}
      \end{pmatrix}
    \right]
  \begin{pmatrix} \omega_\fs \\ \omega_1 \\ \omega_2 \end{pmatrix},
\end{equation}
where
\begin{equation}
  \mbb K_{kj}\omega_j = \frac1{2\pi}\int_0^{2\pi}
  K_{kj}(\cdot,\beta)\omega_j(\beta)\,d\beta, \qquad
  \mbb G_{kj}\omega_j = \frac1{2\pi}\int_0^{2\pi}
  G_{kj}(\cdot,\beta)\omega_j(\beta)\,d\beta.
\end{equation}
Here $k$ and $j$ are fixed; there is no implied summation over
repeated indices.  Up to rescaling of the rows by factors of $-2$ or
$2$, the operator $\mbb B$ is a compact perturbation of the identity,
so has a finite-dimensional kernel.  The structure of
  $\mbb B$ for $N>2$ is easily extended as in \eqref{eq:B:def}, with
  the entries on the diagonal continuing to be of the form
  $\frac{1}{2}\mbb I$ for each new obstacle.
The dimension of the kernel turns
out to be $N-1$, spanned by the functions $\omega=\mb 1_m$
given by
\begin{equation}\label{eq:ker:B}
  (\mb 1_m)_j(\alpha) = \left\{\begin{array}{cc} 1, & \quad j=m \\ 0, &
  \quad j\ne m \end{array}\right\}, \qquad (0\le j\le N, \quad
  2\le m\le N).
\end{equation}
Indeed, if $\omega=\mb 1_m$ with $m\ge2$, then each $\Phi_j(z)$ is
zero everywhere if $j\ne m$ and is zero outside the $m$th cylinder if
$j=m$, including along $\zeta_m(\alpha)^+$.  Summing over $j$ and
restricting the real part to $\Gamma_0^-$ or the imaginary part to
$\Gamma_k^+$, $1\le k\le N$, gives zero for each component of $\mbb
B\omega$ in \eqref{eq:B:def}. In Section~\ref{sec:solvability} we will
prove that all the vectors in $\ker\mbb B$ are linear combinations of
these, and that the range of $\mbb B$ is complemented by the same
functions $\mb 1_m$, $2\le m\le N$. The operator
\begin{equation}\label{eq:AB:cor}
  \mbb A\omega = \mbb B\omega - \sum_{m=2}^N \mb 1_m\la\mb 1_m,\omega\ra,
  \quad
  \la \mu,\omega\ra = \sum_{j=0}^N \frac1{2\pi}\int_0^{2\pi}
  \mu_j(\alpha)\omega_j(\alpha)\,d\alpha.
\end{equation}
is then an invertible rank $N-1$ correction of $\mbb B$.
We remark that (\ref{eq:AB:cor}) is tailored to the case where $V_1$,
$a_2$, \dots, $a_N$ in the representation (\ref{eq:Phi:decomp}) for
$\Phi$ are given and the constant values $\psi\vert_k$ are
unknown. The case when $\psi$ is completely specified on $\Gamma_k$
for $1\le k\le N$ is discussed in Appendix~\ref{sec:psi:specified}.

In the water wave problem, we need to evaluate the normal derivative
of $\phi$ on the free surface to obtain the normal velocity $U$.  In
the present algorithm, we evolve
$\tilde\varphi=\tilde\phi\vert_\Gamma$ in time, so its value is known
when computing $U$.  On the bottom boundary and cylinders, the stream
function $\psi$ should be constant (to prevent the fluid from
  penetrating the walls). Let $\psi\vert_k$ denote the constant value
of $\psi$ on the $k$th boundary. We are free to set $\psi\vert_1=0$ on
the bottom boundary but do not know the other $\psi\vert_k$ in
advance.  We claim that $\psi\vert_k=\la\mb 1_k,\omega\ra$ for
$2\le k\le N$.  From (\ref{eq:Phi:decomp}),
\begin{equation}\label{eq:psi:k}
  \psi(z) = \tilde\psi(z)+\psi_\text{mv}(z)=\psi\vert_k=\text{const}, \qquad\quad
  (z\in\Gamma_k^+\,,\; 1\le k\le N).
\end{equation}
This is achieved by solving
\begin{equation}\label{eq:Awb}
  \mbb A\omega=b, \qquad
  b_0(\alpha) = \tilde\varphi(\alpha), \quad
  b_k(\alpha) = -\psi_\text{mv}(\zeta_k(\alpha)), \quad (1\le k\le N),
\end{equation}
which gives
$\tilde\psi\vert_{\Gamma_k^+} = (\mbb B\omega)_k = b_k + \sum_{m\ge2} \delta_{km}
\la\mb 1_m,\omega\ra = -\psi_\text{mv}\vert_{\Gamma_k}
+ \la \mb 1_k,\omega\ra$
for $2\le k\le N$. Thus, $\psi(z)=\psi\vert_k=\la\mb1_k,\omega\ra$
  is constant for $z\in\Gamma_k^+$, as required. (For $k=1$,
  each $\delta_{km}$ is zero and $\psi\vert_k=0$.)

\subsection{Numerical discretization}
\label{sec:num:discr}

We adopt a collocation-based numerical method and replace the
integrals in Table~\ref{tbl:K:def} with trapezoidal rule sums. Let
$M_0$, \dots, $M_N$ denote the number of grid points chosen
  to discretize the free surface and solid boundaries, respectively. Let
$\alpha_{kl} = 2\pi l/M_k$ for $0\le l<M_k$, and define
$K_{kj,ml}=K_{kj}(\alpha_{km},\alpha_{jl})/M_j$ and
$G_{kj,ml}=G_{kj}(\alpha_{km},\alpha_{jl})/M_j$ so that
\begin{equation*}
  \mbb K_{kj}\omega_j(\alpha_{km}) \approx
  \sum_{l=0}^{M_j-1} K_{kj,ml}\omega_j(\alpha_{jl}), \qquad
  \mbb G_{kj}\omega_j(\alpha_{km}) \approx
  \sum_{l=0}^{M_j-1} G_{kj,ml}\omega_j(\alpha_{jl}).
\end{equation*}
When $N=2$, the system (\ref{eq:Awb}) becomes
\begin{equation}\label{eq:lin:sys}
  \left(
  \begin{array}{c|c|c}
    -\frac12I_0+K_{00} & G_{01} & G_{02} \\[2pt] \hline
    \raisebox{-2pt}{$-G_{10}$} &
    \raisebox{-2pt}{$\frac12I_1+K_{11}$} &
    \raisebox{-2pt}{$K_{12}$} \\[4pt] \hline
    \raisebox{-2pt}{$-G_{20}$} &
    \raisebox{-2pt}{$K_{21}$} &
    \raisebox{-2pt}{$\frac12I_2+K_{22}-E_2$}
  \end{array}\right)
  \begin{pmatrix} \omega_\fs \\ \omega_1 \\ \omega_2
  \end{pmatrix} =
  \begin{pmatrix} \tilde\varphi  \\
    -\psi_\text{mv}\vert_{\Gamma_1} \\
    -\psi_\text{mv}\vert_{\Gamma_2}
  \end{pmatrix},
\end{equation}
where $E_m=M_m^{-1}ee^T$ with $e=(1;1;\dots;1)\in\mbb R^{M_m}$
represents $\mb 1_m\la\mb 1_m,\cdot\ra$ and the right-hand side is
evaluated at the grid points.  For example, in \eqref{eq:lin:sys}
  with $N=2$,
$-\psi_\text{mv}\vert_{\Gamma_1}$ has components
$-V_1\eta_1(\alpha_{1m})-a_2\psi_{\text{cyl}}(\zeta_1(\alpha_{1m})-z_2)$
for $0\le m<M_1$.  The generalization to $N>2$ solid boundaries is
straightforward, with additional diagonal blocks of the form $\frac12I
+ K_{jj}-E_j$.

\subsection{Computation of the normal velocity}

Once the $\omega_j$ are known from solving \eqref{eq:lin:sys},
we can compute $U=\partial\phi/\partial
n$ on the free surface, which is what is needed to evolve both
$\tilde\varphi$ and $\zeta(\alpha,t)$ using the HLS machinery. From
\eqref{eq:d:phi:cyl}, we see that the multi-valued part of the
potential, $\phi_\text{mv}(z) = \re\{\Phi_\text{mv}(z)\}$, contributes
\begin{equation}\label{eq:d:phi:mv}
\begin{aligned}
  & s_{\alpha}\der{\phi_\text{mv}}{n} =
  \re\{(\phi_{\text{mv},x}-i\phi_{\text{mv},y})(n_1+in_2)s_{\alpha}\}
  =\re\{\Phi_\text{mv}'(\zeta(\alpha))i\zeta'(\alpha)\} \\
  & \quad\;\; = -\bigg(V_1+\frac12\sum_{j=2}^N a_j\bigg)\eta'(\alpha)
  + \sum_{j=2}^N a_j\re\left\{\frac12\cot\left(
    \frac{\zeta(\alpha)-z_j}2\right)\zeta'(\alpha)\right\},
\end{aligned}
\end{equation}
where $s_{\alpha} = |\zeta'(\alpha)|$ and
$\eta(\alpha)=\im\zeta(\alpha)$. This normal derivative (indeed the
  entire gradient) of $\phi_\text{mv}$ is single-valued.  We can
differentiate (\ref{eq:w123}) under the integral sign and integrate by
parts to obtain
\begin{equation}\label{eq:Phi:j:prime}
  \Phi_j'(z) = \frac1{2\pi}\int_0^{2\pi}
  \frac12\cot\frac{\zeta_j(\beta)-z}2\omega_j'(\beta)\,d\beta,
  \qquad (1\le j\le N).
\end{equation}
We can then evaluate
\begin{equation}\label{eq:d:phi:j}
  \begin{aligned}
  s_{\alpha}\der{\phi_j}{n} &=
  \re\{(\phi_{j,x}-i\phi_{j,y})(n_1+in_2)s_{\alpha}\}=
  \re\{\Phi_j'(\zeta(\alpha))i\zeta'(\alpha)\} \\
  &= \frac{1}{2\pi}\int_0^{2\pi} K_{j0}(\beta,\alpha)\omega_j'(\beta)
  \,d\beta, \qquad (1\le j\le N).
  \end{aligned}
\end{equation}
Note that the integration variable $\beta$ now appears in the first
slot of $K_{j0}$.  For $j=0$, after integrating (\ref{eq:w123}) by
parts, we obtain
\begin{equation}\label{eq:Phi:0:prime}
  \Phi_0'(z) = \frac{1}{2\pi i}\int_0^{2\pi}
  \frac{1}{2}\cot\frac{\zeta(\beta) - z}{2}
  \omega_0'(\beta)\,d\beta.
\end{equation}
Taking the limit as $z\rightarrow\zeta(\alpha)^-$
(or as $z\rightarrow\zeta(\alpha)^+$) gives
\begin{equation}\label{eq:plemelj:ver2}
  \begin{aligned}
    \Phi_0'(\zeta(\alpha)^\pm) &= \lim_{z\rightarrow \zeta(\alpha)^\pm}
    \frac1{2\pi i}\int_0^{2\pi} \frac{\zeta'(\beta)}2
    \cot\frac{\zeta(\beta)-z}2
      \left(\frac{\omega_0'(\beta)}{\zeta'(\beta)}\right)\,d\beta \\
      &= \pm\frac{\omega_0'(\alpha)}{2\zeta'(\alpha)} +
      \frac1{2\pi i}PV\!\!\int_0^{2\pi}
      \frac12\cot\frac{\zeta(\beta)-\zeta(\alpha)}2\,
      \omega_0'(\beta)\,d\beta
  \end{aligned}
\end{equation}
where PV indicates a principal value integral and we used the Plemelj
formula to take the limit.  Finally, we regularize the integral using
the Hilbert transform,
\begin{equation}
  \mbb Hf(\alpha) = \frac{1}{\pi} PV\int_0^{2\pi}
  \frac{1}{2}\cot\frac{\alpha-\beta}{2} f(\beta)\,d\beta
\end{equation}
to obtain
\begin{equation}\label{eq:hilb:regularize}
  \begin{aligned}
    & \zeta'(\alpha)\Phi_0'(\zeta(\alpha)^\pm) =
    \pm\frac{1}{2}\omega_0'(\alpha) +
    \frac{i}{2}\mbb H\omega_0'(\alpha) \\
    & \qquad\qquad + \frac{1}{2\pi i}\int_0^{2\pi}\left[
      \frac{\zeta'(\alpha)}{2}\cot\frac{\zeta(\beta) - \zeta(\alpha)}{2}
      - \frac{1}{2}\cot\frac{\beta-\alpha}{2}\right]
    \omega_0'(\beta)\,d\beta.
  \end{aligned}
\end{equation}
The term in brackets approaches $-\zeta''(\alpha)/[2\zeta'(\alpha)]$ as
$\beta\rightarrow\alpha$, so the integrand is not singular.  The
symbol of $\mbb H$ is $\hat H_k = -i\opn{sgn} k$, so it can be computed
accurately and efficiently using the FFT.  Using
$s_\alpha\partial\phi_0/\partial n = \re\{ i\zeta'(\alpha)
\Phi_0'(\zeta(\alpha)^-)\}$, we find
\begin{equation}\label{eq:d:phi:0}
  \begin{aligned}
  s_\alpha \der{\phi_0}{n} &= -\frac{1}{2}\mbb H\omega_0'(\alpha)
  - \frac{1}{2\pi}\int_0^{2\pi} G_{00}(\beta,\alpha)
  \omega_0'(\beta)\,d\beta ,
  \end{aligned}
\end{equation}
which we evaluate with spectral accuracy using the trapezoidal rule.
The desired normal velocity $U$ is the sum of (\ref{eq:d:phi:mv}),
(\ref{eq:d:phi:j}) for $1\le j\le N$, and (\ref{eq:d:phi:0}),
all divided by $s_\alpha$.

\subsection{Time evolution of the surface velocity potential}

The last step is to find the evolution equation for
$\tilde\varphi(\alpha,t) = \tilde\phi(\zeta(\alpha,t),t)$, where
$\tilde\phi$ is the component of the velocity potential represented by
Cauchy integrals. The chain rule gives
\begin{equation}\label{eq:convect:phi}
  \tilde\varphi_t = \nabla\tilde\phi\cdot \zeta_t + \tilde\phi_t, \qquad
  \zeta_t = U\mathbf{\hat{n}} + V\mathbf{\hat{s}}.
\end{equation}
We note that $\tilde\phi_t = \phi_t$, and
the unsteady Bernoulli equation gives
\begin{equation}
  \phi_t =
  -\frac{1}{2}|\nabla\phi|^2 - p/\rho - g \eta_0 + C(t),
\end{equation}
where $p$ is the pressure, $\rho$ is the fluid density, $g$ is the
acceleration of gravity, and $C(t)$ is an arbitrary function of time
but not space.  At the free surface, the Laplace-Young condition for
the pressure is $p=p_0-\rho\tau\kappa$, where
$\kappa=\theta_\alpha/s_\alpha$ is the curvature and $\rho\tau$ is the
surface tension. The constant $p_0$ may be set to zero without loss of
generality.  We therefore have
\begin{equation}\label{phitileq}
  \tilde\varphi_t = (\tilde\varphi_\alpha/s_\alpha)V +
               (\partial\tilde\phi/\partial n)U
               -\frac{1}{2}|\nabla\phi|^2 - g\,\eta(\alpha,t) +
               \tau\frac{\theta_\alpha}{s_\alpha} + C(t),
\end{equation}
where $\zeta=\xi+i\eta$, $\zeta_\alpha=s_\alpha e^{i\theta}$, and
$|\nabla\phi|^2 = (\varphi_\alpha/s_\alpha)^2 + (\partial\phi/\partial
  n)^2$. These equations are valid for arbitrary parameterizations
$\zeta_j(\alpha,t)$ and choices of tangential component of velocity
$V$ for the curve. In particular, they are valid in the HLS framework
described in Section~\ref{sec:surface}.  $C(t)$ can be taken to be 0
or chosen to project out the spatial mean of the right-hand side, for
example.

\section{Layer Potentials and the Vortex Sheet Strength Formulation}
\label{sec:laypotalg}

We now give an alternate formulation of the water wave problem in
which the vortex sheet strength is evolved in time rather than the
single-valued part of the velocity potential at the free surface.  We
also replace the constant stream function boundary conditions on the
rigid boundaries by the equivalent condition that the normal velocity
is zero there. Elimination of the stream function provides a pathway
for generalization to 3D. However, in the 2D algorithm presented here,
we continue to take advantage of the connection between layer
potentials and Cauchy integrals; see Appendix~\ref{sec:cauchy:laypot}.

\subsection{Vortex sheet strength and normal velocities on the boundaries}
\label{sec:vsnv}

In this section, the evolution equation at the free surface will be
written in terms of the vortex sheet strength
$\gamma_0(\alpha)=-\omega_0'(\alpha)$.  We also define
\begin{equation}\label{eq:gam:def}
  \gamma_0(\alpha)=-\omega_0'(\alpha), \qquad\quad
  \gamma_j(\alpha)=\omega_j'(\alpha), \qquad (1\le j\le N).
\end{equation}
Expressing (\ref{eq:Phi:j:prime}) and (\ref{eq:Phi:0:prime}) in terms
of the $\gamma_j$, we see that the $j$th boundary contributes a term
$\mb u_j = (u_j,v_j) = (\phi_{j,x},\phi_{j,y})$ to the fluid velocity
given by
\begin{equation}\label{eq:u:from:gam}
  \begin{aligned}
    (u_0 - iv_0)(z) &= \frac{1}{2\pi i}\int_0^{2\pi}
    \frac12\cot\frac{z-\zeta_0(\beta)}2\gamma_0(\beta)\,d\beta, \\
    (u_j - iv_j)(z) &= -\frac1{2\pi i}\int_0^{2\pi}
    \frac12\cot\frac{z-\zeta_j(\beta)}2 i\gamma_j(\beta)\,d\beta, \quad
    (1\le j\le N).
  \end{aligned}
\end{equation}
The calculation in (\ref{eq:plemelj:ver2}) and a similar one for
$\Phi_j'(\zeta_k(\alpha)^\pm)$ then gives
\begin{equation}\label{eq:u:on:bdries}
  \begin{alignedat}{2}
    (u_0 - iv_0)(\zeta_k(\alpha)^\pm) &=
    \mp\frac{\delta_{k0}}2\gamma_0(\alpha)\frac{\zeta'(\alpha)^*}{s_\alpha^2}
    + W_{k0}^*(\alpha), &\;\; & \;\; (0\le k\le N), \\
    (u_j - iv_j)(\zeta_k(\alpha)^\pm) &=
    \mp\frac{\delta_{kj}}2\gamma_j(\alpha)
    \frac{\big(i\zeta_j'(\alpha)\big)^*}{s_{j,\alpha}^2}
    + W_{kj}^*(\alpha), &\;\; & \left(\begin{gathered}0\le k\le N \\ 1\le j\le N
          \end{gathered}\right).
  \end{alignedat}
\end{equation}
Here $W_{kj}^*(\alpha)=W_{kj1}(\alpha)-iW_{kj2}(\alpha)$ are the
Birkhoff-Rott integrals obtained by substituting $z=\zeta_k(\alpha)$
in the right-hand side of (\ref{eq:u:from:gam}) and interpreting the
integral in the principal value sense if $k=j$; see (\ref{eq:BR:def})
in Appendix~\ref{appendix:gam:t}.  The resulting singular integrals (when
  $k=j$) can be regularized by the Hilbert transform, as we did in
(\ref{eq:hilb:regularize}).  The vector notation
$\mb{W}_{kj}(\alpha)=\big(W_{kj1}(\alpha),W_{kj2}(\alpha)\big)$ will
also be useful below.

Although there is no fluid outside the domain $\Omega$, we can still
evaluate the layer potentials and their gradients there. In
(\ref{eq:u:on:bdries}), the tangential component of $\mb u_0$ jumps by
$-\gamma_0(\alpha)/s_\alpha$ on crossing the free surface $\Gamma_0$
while the normal component of $\mb u_0$ and all components of the
other $\mb u_j$'s are continuous across $\Gamma_0$. By contrast, if
$1\le k\le N$, the normal component of $\mb u_k$ jumps by
$-\gamma_k(\alpha)/s_{k,\alpha}$ on crossing the solid boundary
$\Gamma_k$, whereas the tangential component of $\mb u_k$ and all
components of the other $\mb u_j$'s are continuous across $\Gamma_k$.
Here crossing means from the right $(-)$ side to the left $(+)$ side.

In this formulation, we need to compute $U=\pa\phi/\pa n$ to evolve
the free surface and set $\pa\phi/\pa n=0$ on all the other
boundaries. We have already derived formulas for $\pa\phi/\pa n$ on
the free surface in (\ref{eq:d:phi:mv}), (\ref{eq:d:phi:j}) and
(\ref{eq:d:phi:0}).  Nearly identical derivations in which $\Gamma_0$
is replaced by $\Gamma_k$ yield
\begin{align}
  \notag
  s_{k,\alpha}\der{\phi_\text{mv}}{n_k}
  &= -\bigg(V_1 + \frac12\sum_{j=2}^N a_j\bigg)\eta_k'(\alpha)
  + \sum_{j=2}^N a_j\re\left\{\jt
  \frac12\cot\left(\frac{\zeta_k(\alpha)-z_j}2\right)
  \zeta_k'(\alpha)\right\}, \\
  \label{eq:dphij:dnk}
  s_{k,\alpha}\der{\phi_0}{n_k} &= \frac{\delta_{k0}}2
  \mbb H\gamma_0(\alpha) +
  \frac1{2\pi}\int_0^{2\pi}
  G_{0k}(\beta,\alpha)\gamma_0(\beta)\,d\beta,
  \\ \notag
  s_{k,\alpha}\der{\phi_j}{n_k} &= -\frac{\delta_{kj}}2 \gamma_j(\alpha) +
  \frac1{2\pi}\int_0^{2\pi}
  K_{jk}(\beta,\alpha)\gamma_j(\beta)\,d\beta, \quad
  \left(\begin{aligned} 0\le k\le N \\ 1\le j\le N \end{aligned}\right),
\end{align}
where $0\le k\le N$ in the first two equations.
Here $K_{kj}(\alpha,\beta)$ and $G_{kj}(\alpha,\beta)$ are as in
Table~\ref{tbl:K:def} above.  Since $\gamma_0$ is evolved in time, it
is a known quantity in the layer potential calculations. Given
$\gamma_0$, we compute $\gamma_1,\dots,\gamma_N$ by solving the
coupled system obtained by setting $\pa\phi/\pa n=0$ on the solid
boundaries.  When $N=3$, the system looks like
\begin{equation}\label{eq:lin:sys2}
  \left(
  \begin{array}{c|c|c}
    -\frac12\mbb I+\mbb K_{11}^* & \mbb K_{21}^* & \mbb K_{31}^* \\[2pt]
    \hline
    \raisebox{-2pt}{$\mbb K_{12}^*$} &
    \raisebox{-2pt}{$-\frac12\mbb I+\mbb K_{22}^*$} &
    \raisebox{-2pt}{$\mbb K_{32}^*$} \\[4pt] \hline
    \raisebox{-2pt}{$\mbb K_{13}^*$} &
    \raisebox{-2pt}{$\mbb K_{23}^*$} &
    \raisebox{-2pt}{$-\frac12\mbb I+\mbb K_{33}^*$}
  \end{array}\right)\!\!
  \begin{pmatrix} \gamma_1 \\ \gamma_2 \\ \gamma_3
  \end{pmatrix} =
  \begin{pmatrix}
    -\mbb G_{01}^*\gamma_0 - s_{1,\alpha}\der{\phi_\text{mv}}{n_1} \\[3pt]
    -\mbb G_{02}^*\gamma_0 - s_{2,\alpha}\der{\phi_\text{mv}}{n_2} \\[3pt]
    -\mbb G_{03}^*\gamma_0 - s_{3,\alpha}\der{\phi_\text{mv}}{n_3}
  \end{pmatrix}\!,
\end{equation}
where
\begin{equation}
  \mbb K_{jk}^*\gamma_j = \frac1{2\pi}\int_0^{2\pi}
  K_{jk}(\beta,\cdot)\gamma_j(\beta)\,d\beta, \quad
  \mbb G_{jk}^*\gamma_j = \frac1{2\pi}\int_0^{2\pi}
  G_{jk}(\beta,\cdot)\gamma_j(\beta)\,d\beta.
\end{equation}
The system for $N>3$ has an identical structure.
The matrices representing $\mbb K_{jk}^*$ and $\mbb G_{jk}^*$ in the
collocation scheme have entries
\begin{equation}\label{eq:KG:tmat}
  \begin{aligned}
    (K_{jk}^\dagger)_{ml} &= (M_k/M_j)K_{jk,lm} =
    K_{jk}(\alpha_{jl},\alpha_{km})/M_j, \\
    (G_{jk}^\dagger)_{ml} &= (M_k/M_j)G_{jk,lm} =
    G_{jk}(\alpha_{jl},\alpha_{km})/M_j.
  \end{aligned}
\end{equation}
Here a dagger is used in place of a transpose symbol as a reminder to also
re-normalize the quadrature weights.  Once $\gamma_0,\dots,\gamma_N$
are known, the normal velocity $U$ is given by
\begin{equation}\label{eq:U:from:gam:j}
  U = \frac1{s_\alpha}\bigg[
    \frac12\mbb H\gamma_0 + \mbb G_{00}^*\gamma_0 +
    \sum_{j=1}^N \mbb K_{j0}^*\gamma_j
    + s_\alpha\der{\phi_\text{mv}}{n_0}\bigg].
\end{equation}
In Section~\ref{sec:solvability}, we will show that in the general case,
with $N$ arbitrary, the system (\ref{eq:lin:sys2}) is invertible. In
practice, the discretized version is well-conditioned once enough
grid points $M_j$ are used on each boundary $\Gamma_j$.

\subsection{The evolution equation for $\gamma_0$}
\label{sec:evol:gam}

In the case without solid boundaries (i.e., the case of a fluid of
  infinite depth and in the absence of the obstacles), we only have
$\gamma_0$ to consider.  The appendix of
\cite{ambroseMasmoudi1} details how to use the Bernoulli equation to
find the equation for $\gamma_0$ in this case.  This is a version of a
calculation contained in \cite{BMO}.  The argument of
\cite{ambroseMasmoudi1} and \cite{BMO} for finding the $\gamma_{0,t}$
equation considers two fluids with positive densities; after deriving
the equation, one of the densities can be set equal to zero.  We
present here an alternative calculation that only requires
consideration of a single fluid, accounts for the solid boundaries,
and leads to simpler formulas to implement numerically.  Connections
to the results of \cite{ambroseMasmoudi1} and \cite{BMO}
are worked out in Appendix~\ref{appendix:gam:t}.

The main observation that we use to derive an equation for $\gamma_{0,t}$
is that $\phi_t$ is a solution of the Laplace equation
in $\Omega$ with homogeneous Neumann conditions at the solid
boundaries and a Dirichlet condition (the Bernoulli equation) at the
free surface. Decomposing $\phi=\tilde\phi + \phi_\text{mv}$
as before, we have $\phi_t = \tilde\phi_t$ since $\phi_\text{mv}$
is time-independent. The Dirichlet condition at the free surface
is then
\begin{equation}\label{eq:bernoulli:gam:t}
  \tilde\phi_t = -\frac12|\nabla\phi|^2 - \frac p\rho - g\eta_0.
\end{equation}
Let
\begin{equation}\label{eq:W:tot}
  \mb{W}(\alpha) = \mb W_{00}(\alpha) + \mb W_{01}(\alpha) +
  \cdots+ \mb W_{0N}(\alpha) + \nabla\phi_\text{mv}(\zeta(\alpha))
\end{equation}
denote the contribution of the Birkhoff-Rott integrals from all the
layer potentials evaluated at the free surface, plus the velocity due
to the multi-valued part of the potential.  By (\ref{eq:u:on:bdries}),
\begin{equation}\label{eq:dphi:W:gam}
  \nabla\phi(\zeta(\alpha)) = \mb{W} +
  \frac{\gamma_0}{2s_\alpha}\mb{\hat t}, \qquad\quad
  \frac12|\nabla\phi|^2 = \frac12\mb{W}\cdot\mb{W} +
  \frac{\gamma_0}{2s_\alpha}\mb W\cdot\mb{\hat t} +
  \frac{\gamma_0^2}{8s_\alpha^2}.
\end{equation}
To evaluate the left-hand side of (\ref{eq:bernoulli:gam:t}), we
differentiate (\ref{eq:w123}) with $z$ fixed to obtain
\begin{equation}\label{eq:Phi:t}
  \begin{aligned}
    \Phi_{0,t}(z) &= \frac1{2\pi i}\int_0^{2\pi}
    \frac{\zeta'(\beta)}2\cot\frac{\zeta(\beta)-z}2
    \left(\omega_{0,t}(\beta) - \frac{\zeta_t(\beta)}{
        \zeta'(\beta)}\omega_0'(\beta)\right)\,d\beta, \\
    \Phi_{j,t}(z) &= \frac1{2\pi}\int_0^{2\pi} \frac{\zeta_j'(\beta)}2
    \cot\frac{\zeta_j(\beta)-z}2\omega_{j,t}(\beta)\,d\beta,
    \qquad (1\le j\le N),
  \end{aligned}
\end{equation}
where we used $\pa_t\big\{(\zeta'/2)\cot[(\zeta-z)/2]\big\} =
\pa_\beta\big\{ (\zeta_t/2)\cot[(\zeta-z)/2]\big\}$ and integrated by
parts.  Here a prime indicates $\pa_\beta$ (i.e.~$\pa_\alpha$) and a
subscript $t$ indicates $\pa_t$.  We continue to suppress $t$ in the
arguments of functions, keeping in mind that the solid boundaries do
not move.  Letting $z\rightarrow\zeta(\alpha)^-$ and using
(\ref{eq:gam:def}) as well as $\zeta_t = (V+iU)\zeta'/s_\alpha$, we
obtain
\begin{align}
  \Phi_{0,t}(\zeta(\alpha)^-) &=
  -\frac12\omega_{0,t}(\alpha) - \frac{V+iU}{2s_\alpha}\gamma_0(\alpha) \\
  \notag
  &\qquad\quad + \frac1{2\pi i}PV\!\!\int_0^{2\pi}
  \frac{\zeta'(\beta)}2\cot\frac{\zeta(\beta)-\zeta(\alpha)}2
  \omega_{0,t}(\beta)\,d\beta \\
  \notag
  &\qquad\quad + \frac1{2\pi i}PV\!\!\int_0^{2\pi}
  \frac{\zeta_t(\beta)}2\cot\frac{\zeta(\beta)-\zeta(\alpha)}2
  \gamma_0(\beta)\,d\beta, \\
  \notag
  \Phi_{j,t}(\zeta(\alpha)) &= \frac1{2\pi}\int_0^{2\pi}
  \frac{\zeta_j'(\beta)}2
  \cot\frac{\zeta_j(\beta)-\zeta(\alpha)}2\omega_{j,t}(\beta)\,d\beta,
  \qquad (1\le j\le N).
\end{align}
Next we take the real part, sum over $j\in\{0,\dots,N\}$ to get
$\tilde\phi_t$ at the free surface, and use the Bernoulli equation
(\ref{eq:bernoulli:gam:t}). The first $PV$ integral becomes regular
when the real part is taken, so we can differentiate under the
integral sign and integrate by parts in the next step.  Finally, we
differentiate with respect to $\alpha$, which converts
$-(1/2)\omega_{0,t}(\alpha)$ into $(1/2)\gamma_{0,t}(\alpha)$; integrate
by parts to convert the $\omega_{j,t}$ terms in the integrals into
$-\gamma_{j,t}$ terms; and use the boundary condition for the pressure
(the Laplace-Young condition, $p=p_0-\rho\tau\kappa$) to obtain
\begin{align}\label{eq:gam:t:final}
    &\left(\frac12\mbb I + \mbb K_{00}^*\right)\gamma_{0,t} -
  \sum_{j=1}^N \mbb G_{j0}^*\gamma_{j,t} \\
  \notag
    &\qquad\quad = \der{}{\alpha}\left( -\frac12\mb W\cdot\mb W +
      \frac{(V-\mb{W} \cdot\mathbf{\hat{t}})}{2s_{\alpha}}\gamma_0
      - \frac{\gamma_0^2}{8s_{\alpha}^{2}} + \tau\frac{\theta_\alpha}{s_\alpha}
      - g\eta_0 - \mbb F_{00}\gamma_0
      \right),
\end{align}
where
\begin{equation}\label{eq:F0:def}
  \mbb F_{00}\gamma_0(\alpha) = \frac1{2\pi}PV\!\!\int_0^{2\pi} \im\left\{
  \frac{\zeta_t(\beta)}2\cot\frac{\zeta(\beta)-\zeta(\alpha)}2\right\}
  \gamma_0(\beta)\,d\beta.
\end{equation}
The additional equations needed to solve for the $\gamma_{j,t}$ can be
obtained by differentiating (\ref{eq:lin:sys2}) with respect to time;
note that all the $K_{jk}$ terms correspond to rigid boundaries that
do not change in time. Equivalently, the $\gamma_{j,t}$ can be
interpreted as the layer potential densities needed to enforce
homogeneous Neumann boundary conditions on $\phi_t$ on the solid
boundaries,
\begin{equation}
  s_{k,\alpha}\der{\phi_t}{n_k} = \sum_{j=0}^N
  \re\{\Phi_{j,t}'(\zeta_k(\alpha)^+)i\zeta_k'(\alpha)\}=0, \qquad
      (1\le k\le N).
\end{equation}
Either calculation yields the same set of additional linear equations,
illustrated here in the $N=3$ case, with identical structure when
$N>3$:
\begin{equation}\label{eq:gam:t:aux}
  \left(
    \begin{array}{c|c|c|c}
      \mbb G_{01}^* & -\frac12\mbb I+\mbb K_{11}^* &
      \mbb K_{21}^* & \mbb K_{31}^* \\[2pt] \hline
      \raisebox{-2pt}{$\mbb G_{02}^*$} &
      \raisebox{-2pt}{$\mbb K_{12}^*$} &
      \raisebox{-2pt}{$-\frac12\mbb I+\mbb K_{22}^*$} &
      \raisebox{-2pt}{$\mbb K_{32}^*$} \\[4pt] \hline
      \raisebox{-2pt}{$\mbb G_{03}^*$} &
      \raisebox{-2pt}{$\mbb K_{13}^*$} &
      \raisebox{-2pt}{$\mbb K_{23}^*$} &
      \raisebox{-2pt}{$-\frac12\mbb I+\mbb K_{33}^*$}
  \end{array}\right)
  \begin{pmatrix} \gamma_{0,t} \\ \gamma_{1,t} \\ \gamma_{2,t} \\ \gamma_{3,t}
  \end{pmatrix} =
  \begin{pmatrix}
    -\pa_\alpha(\mbb F_{10}\gamma_0)  \\
    -\pa_\alpha(\mbb F_{20}\gamma_0)  \\
    -\pa_\alpha(\mbb F_{30}\gamma_0)
  \end{pmatrix},
\end{equation}
where
\begin{equation}\label{eq:Fk:def}
  \mbb F_{k0}\gamma_0(\alpha) = \frac1{2\pi}\int_0^{2\pi} \re\left\{
  \frac{\zeta_t(\beta)}2\cot\frac{\zeta(\beta)-\zeta_k(\alpha)}2\right\}
  \gamma_0(\beta)\,d\beta, \quad (1\le k\le N).
\end{equation}
The formulas (\ref{eq:F0:def}) and (\ref{eq:Fk:def}) can be
regularized (when $k=0$) and expressed in terms of $\mbb K$ and $\mbb
G$ operators by writing $\zeta_t=(V+iU)\zeta_\alpha/s_\alpha$. The
result is
\begin{equation}\label{eq:Fk:from:UV}
  \begin{aligned}
    \mbb F_{00}\gamma_0 &=
          -\frac12\mbb H\left(\frac{U\gamma_0}{s_\alpha}\right)
    + \mbb K_{00}\left(\frac{V\gamma_0}{s_\alpha}\right)
    + \mbb G_{00}\left(\frac{U\gamma_0}{s_\alpha}\right), \\
    \mbb F_{k0}\gamma_0 &=
    \mbb G_{k0}\left(\frac{V\gamma_0}{s_\alpha}\right)
    - \mbb K_{k0}\left(\frac{U\gamma_0}{s_\alpha}\right), \qquad (1\le k\le N).
  \end{aligned}
\end{equation}
In Appendix~\ref{appendix:gam:t}, we present an alternative derivation
of (\ref{eq:gam:t:final}) that involves solving (\ref{eq:dphi:W:gam})
for $\gamma_0$ and differentiating with respect to time.  The moving
boundary affects this derivative, which complicates the intermediate
formulas but has the advantage of making contact with results reported
elsewhere \cite{ambroseMasmoudi1,BMO} for the case with no obstacles
or bottom topography.

\section{Method Summary and the Computation of
  Velocity, Pressure and Energy}
\label{sec:summary:upE}

In this section, we show how to compute the fluid velocity and
  pressure accurately throughout the fluid, including near the free
  surface and boundaries, and how to compute the energy when the
  velocity potential is multiple-valued. But first we summarize the
  steps needed to evolve the water wave problem.  As it is more
  complicated, involving the computation of more integral kernels, we
  will focus on the vortex sheet strength formulation.  The
  implementation for the velocity potential approach is similar, with
  $\gamma_0$ replaced by $\tilde\varphi$ and evolved via
  \eqref{phitileq}. We have so far left the choice of $V$
unspecified. We consider two options here.  In both variants, the
bottom boundary $\zeta_1(\alpha)$ and obstacles $\zeta_2(\alpha)$,
\dots, $\zeta_N(\alpha)$ can be parameterized arbitrarily, though we
assume they are smooth and $2\pi$-periodic in the sense of
(\ref{eq:per1}) and (\ref{eq:per2N}) so that collocation via the
trapezoidal rule is spectrally accurate.

The simplest case is to assume $\xi(\alpha)=\alpha$ for all time.  At
the start of a timestep (and at intermediate stages of a Runge-Kutta
  method), $\eta(\alpha)$ and $\gamma_0(\alpha)$ are known (still
  suppressing $t$ in the arguments of functions), and we need to
compute $\eta_t$ and $\gamma_{0,t}$. We construct the curve
$\zeta(\alpha)=\alpha+i\eta(\alpha)$ and compute the matrices
$G_{jk}^\dagger$, $K_{jk}^\dagger$ in (\ref{eq:KG:tmat}).
Computing these matrices
is the most expensive step, but is trivial to parallelize in openMP
and straightforward to parallelize on a cluster using MPI or on a GPU
using Cuda.  We solve the linear system (\ref{eq:lin:sys2}) using
GMRES to obtain $\gamma_j(\alpha,t)$ for $1\le j\le N$ and compute the
normal velocity $U$ from (\ref{eq:U:from:gam:j}). From
\eqref{eq:graph:eta:t}, we know $V=\eta_\alpha U$ and $\eta_t =
\sqrt{1+\eta_\alpha^2}\,U$.  Once $U$ and $V$ are known, we compute
$\mbb F_{k0}\gamma_0$ via (\ref{eq:Fk:from:UV}) and solve
(\ref{eq:gam:t:final}) and (\ref{eq:gam:t:aux}) for $\gamma_{j,t}$,
$0\le j\le N$. This gives $\gamma_{0,t}$.

Alternatively, in the HLS framework, using the
improved algorithm of Section~\ref{sec:constman}, $P\theta(\alpha)$
and $\gamma_0(\alpha)$ are evolved in time.  At the start of each time
step (and at intermediate Runge-Kutta stages), the
arclength element $s_\alpha$ and curve $\zeta(\alpha)$ are
reconstructed from $P\theta(\alpha)$ using \eqref{eq:CS:th} and
\eqref{eq:xi:eta:recon}. We then compute the matrices $G_{jk}^\dagger$,
$K_{jk}^\dagger$ in (\ref{eq:KG:tmat}) in parallel using openMP, and,
optionally, MPI or Cuda. We solve the linear system
(\ref{eq:lin:sys2}) using GMRES to obtain $\gamma_j(\alpha)$ for $1\le
j\le N$ and compute the normal velocity $U$ from
(\ref{eq:U:from:gam:j}). We then solve
\begin{equation}\label{eq:V:from:Va}
  V = \pa_\alpha^{-1}\left(
    \theta_\alpha U - \frac1{2\pi}\int_0^{2\pi}
    \theta_\alpha U\,d\alpha\right), \qquad
  V(0) = U(0)\frac{\eta'(0)}{\xi'(0)},
\end{equation}
where the antiderivative is computed via the FFT and the condition on
$V(0)$ keeps $\xi(0)=0$ for all time. This formula can break down if
an overturned wave crosses $\alpha=0$, leading to $\xi'(0)=0$; in such
cases, one can instead choose the integration constant in
(\ref{eq:V:from:Va}) so that $\int_0^{2\pi} V\,d\alpha=0$ and evolve
$\xi(0)$ via the ODE
$\pa_t[\xi(0)]=\re\big\{(V(0)+iU(0))\zeta'(0)/s_\alpha\big\}$.  Once
$U$ and $V$ are known, we compute
$(P\theta)_t=P[(U_\alpha+V\theta_\alpha)/s_\alpha]$ in
  \eqref{eq:Pth:t} and obtain $\gamma_{0,t}$ by
solving (\ref{eq:gam:t:final}) and (\ref{eq:gam:t:aux}).
We also employ a 36th order filter
  \cite{HLS1,HLS2,hou2007computing} in which the $k$th Fourier modes of
  $P\theta$ and $\gamma_0$ are multiplied by
\begin{equation}\label{eq:filter}
  \exp\Big[ -36\big(|k|/k_\text{max}\big)^{36}\Big], \qquad
  |k|\le k_\text{max}=M_0/2.
\end{equation}
The filter is applied at the end of each Runge-Kutta timestep
  (but not in intermediate Runge-Kutta stages).

\begin{remark}
  The velocity potential and vortex sheet formulations thus give two
equivalent systems of evolution equations in the $(P\theta, \tilde
  \varphi)(\alpha,t)$ and $(P\theta, \gamma_0)(\alpha,t)$
representations, respectively.  As we will illustrate with specific
examples in Section \ref{sec:connections} below, it is important to recognize
that when $\Phi_\text{mv}$ is nonzero, $\tilde \varphi (\alpha,
  0)\equiv 0$ is not equivalent to $\gamma_0 (\alpha,0)\equiv 0$.
Indeed, to obtain equivalent initial data in the two systems,
one must compute $\gamma_0(\alpha)=-\omega_0'(\alpha)$ as in
\eqref{eq:gam:def}, where the $\omega_j$ terms are computed as in
\eqref{eq:Awb}.
\end{remark}

\subsection{Numerical evaluation of the fluid velocity and pressure}
\label{sec:fvp}

Though they are secondary variables in the velocity potential and
vortex sheet formulations, one often wishes to compute the fluid
velocity and pressure throughout the fluid. We do this as a
post-processing step, after $\zeta(\alpha,t)$ and
$\tilde\varphi(\alpha,t)$ or $\gamma_0(\alpha,t)$ have been computed
at a given time. To make contour plots such as in
Figures~\ref{fig:prob1}--\ref{fig:prob4p} in
Section~\ref{sec:numerics} below, we generate a triangular
mesh in the fluid region using the distmesh package \cite{distmesh}
and compute
\begin{equation}\label{uv:phi:phi:t}
  u(x,y,t)-iv(x,y,t) = \Phi'(x+iy,t), \qquad\quad
  \Phi_t(x+iy,t)
\end{equation}
at each node of the mesh. This gives the velocity components $(u,v)$
directly and is sufficient to compute the pressure via
\begin{equation}\label{eq:press}
  \frac p\rho = C(t) - \phi_t -\frac{1}{2}|\nabla\phi|^2 - gy,
\end{equation}
where $C(t)$ is determined by whatever choice is made in
\eqref{phitileq}. In our code, we choose $C(t)$ so that the mean of
$\tilde\varphi(\alpha,t)$ with respect to $\alpha$ remains zero for
all time. On the free surface, the Laplace-Young condition
$p=p_0-\rho\tau\kappa$ holds, where $\kappa=\theta_\alpha/s_\alpha$
  is the curvature and we have set $p_0=0$. This furnishes boundary
  values for $C(t)-\phi_t$ in \eqref{eq:press}, which is a harmonic
  function in $\Omega$ that we solve for from these boundary values
  using the Cauchy integral framework of Section~\ref{sec:cauchy},
  as explained below.

Numerical evaluation of Cauchy integrals and layer potentials
near boundaries requires care. In Appendix~\ref{sec:helsing}, we adapt
to the spatially periodic setting an idea of Helsing and Ojala
\cite{helsing} for evaluating Cauchy integrals with spectral accuracy
even if the evaluation point is close to (or on) the boundary. Suppose
$f(z)$ is analytic in $\Omega$ and we know its boundary values. Then,
as shown in Appendix~\ref{sec:helsing},
\begin{align}\notag
  &f(z) = \frac1{2\pi i}\int_{\partial\Omega}
  \frac{f(\zeta)}2\cot\frac{\zeta-z}2d\zeta
  \approx \sum_{k=0}^N\sum_{m=0}^{M_k-1}\lambda_{km}(z)
  f(\zeta_k(\alpha_m)), \quad (z\in\Omega), \\
  \label{eq:helsing:quad}
  &\lambda_{km}(z)=\frac{\tilde\lambda_{km}(z)}{
    \sum_{k'm'}\tilde\lambda_{k'm'}(z)}, \quad\;\;
  \tilde\lambda_{km}(z) = \frac1{M_k}\left(\frac12 \cot{\frac{
        \zeta_k(\alpha_m)-z}2}\zeta'(\alpha_m)\right).
\end{align}
The complex numbers $\lambda_{km}(z)$ serve as quadrature weights for
the integral. They express $f(z)$ as a weighted average of the
boundary values $f(\zeta_k(\alpha_m))$. The formula does not break down
as $z$ approaches a boundary point $\zeta_k(\alpha_m)$ since
$\tilde\lambda_{km}(z)\rightarrow\infty$ in that case, causing
$\lambda_{k'm'}(z)$ to approach $\delta_{kk'}\delta_{mm'}$ and $f(z)$
to approach $f(\zeta_k(\alpha_m))$. If $z$ coincides with
$\zeta_k(\alpha_m)$, we set $f(z)=f(\zeta_k(\alpha_m))$.

To evaluate \eqref{uv:phi:phi:t} at the mesh points via
\eqref{eq:helsing:quad}, we just need to compute the values of
$\Phi'(z,t)$ and $\Phi_t(z,t)$ at the boundary points
$z=\zeta_k(\alpha_m,t)$.  These boundary values only have to be
computed once for a given $t$ (which we now suppress in the notation)
to evaluate $\Phi'(z)$ and $\Phi_t(z)$ at all the mesh points of the
fluid. The only values that change with $z$ are the quadrature weights
$\lambda_{km}(z)$, which are easy to compute rapidly in parallel.
Since $\Phi'(z)$ and $\Phi_t(z)$ are single-valued, we include the
contribution of $\Phi_\text{mv}(z)$ in the boundary values.
Equation \eqref{eq:u:on:bdries} gives the needed formulas for
$\Phi'(z)$ on the boundaries.  These formulas are most easily
evaluated via
\begin{equation}\label{eq:phi:prime}
  \Phi'(\zeta_k(\alpha_m)) =
  u-iv= \overline{\der{\phi}{s_k}\mb{\hat{t}}_k +
    \der{\phi}{n_k}\mb{\hat{n}}_k} =
  \left(\der{\phi}{s_k} - i\der{\phi}{n_k}\right)
    \frac{s_{k,\alpha}}{\zeta_k'(\alpha_m)},
\end{equation}
where $\overline{\mb{\hat{t}}_k}=
\overline{\zeta_k'(\alpha)/s_{k,\alpha}}=
s_{k,\alpha}/\zeta_k'(\alpha)$ and
$\overline{\mb{\hat{n}}_k}=-i\overline{\mb{\hat{t}}_k}$.  Formulas for
$\pa\phi/\pa n_k$ were already given in \eqref{eq:dphij:dnk}, where
$\phi=\phi_\text{mv}+\phi_0+\cdots+\phi_N$. A similar calculation
starting from \eqref{eq:u:on:bdries} gives $\pa\phi/\pa s_k$:
\begin{align}
  \notag
  s_{k,\alpha}\der{\phi_\text{mv}}{s_k}
  &= \bigg(V_1 + \frac12\sum_{j=2}^N a_j\bigg)\xi_k'(\alpha)
  + \sum_{j=2}^N a_j\im\left\{\jt
  \frac12\cot\left(\frac{\zeta_k(\alpha)-z_j}2\right)
  \zeta_k'(\alpha)\right\}, \\
  \label{eq:dphij:dsk}
  s_{k,\alpha}\der{\phi_0}{s_k} &= \mp\frac{\delta_{k0}}2\gamma_0(\alpha) +
  \frac1{2\pi}\int_0^{2\pi}
  K_{0k}(\beta,\alpha)\gamma_0(\beta)\,d\beta, 
  \\ \notag
  s_{k,\alpha}\der{\phi_j}{s_k} &= -\frac{\delta_{kj}}2\mbb H \gamma_j(\alpha) -
  \frac1{2\pi}\int_0^{2\pi}
  G_{jk}(\beta,\alpha)\gamma_j(\beta)\,d\beta, \quad
  \left(\begin{aligned} 0\le k\le N \\ 1\le j\le N \end{aligned}\right),
\end{align}
where $0\le k\le N$ in the first two equations. On the solid
boundaries, $\pa\phi/\pa n_k=0$, so only $\pa\phi/\pa s_k$ needs to be
computed in \eqref{eq:phi:prime} when $k\ne0$. In the velocity
potential formulation of Section~\ref{sec:cauchy},
$\{\gamma_j\}_{j=0}^N$ are computed via \eqref{eq:gam:def} first,
before evaluating \eqref{eq:dphij:dnk} and \eqref{eq:dphij:dsk}.

In the velocity potential formulation, we compute $\Phi_t-C(t)$ on the
boundaries, which is needed in \eqref{eq:press}, by solving a system
analogous to \eqref{eq:Awb}, which we denote $\mbb
A\omega^\text{aux}=b^\text{aux}$. The right-hand side is
\begin{equation}\label{eq:pressure:d:prob}
  b^\text{aux}_0(\alpha) = \phi_t\vert_{\Gamma_0}-C(t) =
  \tau\kappa - \frac12|\nabla\phi|^2 - g\eta, \qquad
  b^\text{aux}_k(\alpha) = 0,
\end{equation}
where $k$ ranges from 1 to $N$.  We solve for $\omega^\text{aux}$
using the same code that we use to compute $\omega$ in \eqref{eq:Awb}.
Replacing $\omega$ by $\omega^\text{aux}$ in \eqref{eq:w123} gives
formulas for $\Phi_t-C(t)$ throughout $\Omega$.  Instead of computing
the normal derivative of the real part of the Cauchy integrals on
$\Gamma_0^-$, we now need to evaluate their real and imaginary parts
on $\Gamma_0^-$ and $\Gamma_j^+$ for $1\le j\le N$, using the Plemelj
formula. We regularize the integrand by including the
$\delta_{kj}\frac12\cot\frac{\beta-\alpha}2$ term in
$G_{kj}(\alpha,\beta)$ in Table~\ref{tbl:K:def}, which introduces
Hilbert transforms in the final formulas for $\Phi_t-C(t)$ on the
boundaries. We omit details as they are similar to the calculations of
Section~\ref{sec:cauchy}.

In the vortex sheet formulation, one can
either proceed exactly as above, solving the auxiliary Dirichlet
problem \eqref{eq:pressure:d:prob} by the methods of
Section~\ref{sec:cauchy}, or we can use \eqref{eq:Phi:t}.  The
functions $\omega_{j,t}$ are known from \eqref{eq:gam:def} up to
constants by computing the antiderivatives of $\gamma_{j,t}$ using the
FFT. The constants in $\omega_{j,t}$ for $2\le j\le N$ have no effect
on $\Phi_{j,t}$ in $\Omega$, so we define $\omega_{j,t}$ as the
zero-mean antiderivative of $\gamma_{j,t}$.  We can also do this for
$\omega_{1,t}$ since it only affects the imaginary part of
$\Phi_{1,t}$, due to \eqref{eq:Phij:w:const}, and therefore has no
effect on the pressure.  Varying the constant in $\omega_{0,t}$ by $A$
causes $p/\rho$ to change by $A/2$ throughout $\Omega$, due to
\eqref{eq:Phij:w:const}. We can drop $C(t)$ in \eqref{eq:press} since
the mean of $\omega_{0,t}$ has the same effect. To determine the mean,
we tentatively set it zero, compute the right-hand side of
\eqref{eq:press} at one point on the free surface and compare to the
Laplace-Young condition $p/\rho=-\tau\kappa$. The mean of
$\omega_{0,t}$ is then corrected to be twice the difference of the
results. Once each $\omega_{j,t}$ has been determined, we compute
$\Phi_t$ on the boundaries via the Plemelj formulas applied to
\eqref{eq:Phi:t}, and at interior mesh points using the
quadrature rule \eqref{eq:helsing:quad}.

\subsection{Numerical evaluation of the energy}
\label{sec:energy}

We next derive a formula for the conserved energy in the
multiply-connected setting. A standard calculation for the Euler
equations \cite{chorin:marsden} gives
\begin{equation}\label{eq:chorin:marsden}
  \begin{aligned}
    &\frac{d}{dt} \iint_\Omega \frac\rho2 \mathbf{u} \cdot \mathbf{u}\, dA =
    \iint_\Omega \frac\rho2\frac{D( \mathbf{u} \cdot \mathbf{u})}{Dt}\, dA =
    \iint_\Omega \mathbf{u} \cdot \left( \rho\frac{D\mb u}{Dt}\right)\,dA \\
    &\qquad = - \iint_\Omega  \diver \big(
      \mathbf{u} ( p + \rho g y) \big) dA
    = - \int_{\pa\Omega} (p+\rho g y)\mb u\cdot\mb n_\pm\,ds,
  \end{aligned}
\end{equation}
where $D/Dt$ is the convective derivative and $\mb n_\pm$ is the
outward normal from $\Omega$, which is $\mb n$ on $\Gamma_0$ and $-\mb
n$ on $\Gamma_1,\dots,\Gamma_N$.  On the solid boundaries, $\mb
u\cdot\mb n=0$. On the free surface, $p=-\rho\tau\kappa$, $y=\eta$,
$\mb u\cdot\mb n = U = \zeta_t\cdot\mb n$ and
\begin{equation}
  \begin{aligned}
  &\frac d{dt}\int s_\alpha\,d\alpha = \int
  \frac{\zeta_\alpha}{s_\alpha}\cdot\zeta_{\alpha t}\,d\alpha =
  -\int\re\big(i\theta_\alpha e^{i\theta}\overline{\zeta_t}\big)\,d\alpha =
  -\int\kappa(\zeta_t\cdot\mb n)\,ds, \\
  &\frac d{dt}\int \frac12\eta^2\xi_\alpha\,d\alpha =
  \int \eta(\eta_t\xi_\alpha - \eta_\alpha\xi_t\big)\,d\alpha =
  \int\eta(\zeta_t\cdot\mb n)\,ds.
  \end{aligned}
\end{equation}
Finally, using $\mb u\cdot\mb u=|\nabla\psi|^2$, we have
\begin{equation}\label{eq:KE:psi:phi}
    \iint_\Omega \frac\rho2 \mathbf{u} \cdot \mathbf{u}\, dA =
    \frac12\iint_\Omega \diver(\psi\nabla\psi)\,dA = \frac12\int_{\pa\Omega}
    \psi\der{\psi}{n_\pm}\,ds,
\end{equation}
where $\pa\psi/\pa n_\pm=\nabla\psi\cdot\mb n_\pm=\pm\pa\phi/\pa s$
with the plus sign on the free surface and the minus sign on the solid
boundaries.  On the $j$th solid boundary, $\psi=\psi\vert_j$ is
constant, and we arranged in \eqref{eq:Awb} for $\psi\vert_1=0$.  For
$j\ge2$, $\int_{\Gamma_j}d\phi=-2\pi a_j$. There is no contribution
from the left and right sides of $\Omega$ in \eqref{eq:chorin:marsden}
or \eqref{eq:KE:psi:phi} due to the periodic boundary
conditions. Combining these results shows that
\begin{equation}\label{eq:E:formula}
  E = \frac1{2\pi}\int_0^{2\pi}
    \left(\rho\tau s_\alpha + \frac12\rho g\eta^2\xi_\alpha +
      \frac12(\psi\vert_{\Gamma_0})\varphi_\alpha\right)d\alpha +
    \sum_{j=2}^N \frac12 a_j\,\psi\vert_j
\end{equation}
is a conserved quantity, which we evaluate numerically via the
trapezoidal rule at the $M_0$ points used to discretize the free
surface in Section~\ref{sec:num:discr} above.
We non-dimen\-sionalize $\rho=1$ and $g=1$ and
include the factor of $1/2\pi$ to obtain the average energy per unit
length, which we slightly prefer to the energy per wavelength.  Note
that the stream function is constant on the obstacle boundaries when
time is frozen, but the $\psi\vert_j$ vary in time and have to be
computed to determine $E(t)$. This is easy since we arranged in
\eqref{eq:psi:k} and \eqref{eq:Awb} for $\psi\vert_j=\la\mb
1_j,\omega\ra$ for $2\le j\le N$. Also, $\psi\vert_{\Gamma_0}$ depends
on both $\alpha$ and $t$ since the free surface is generally not a
streamline.

To compute the energy in the vortex sheet formulation, the simplest
approach is to compute $\omega_0=-\int\gamma_0\,d\alpha$ and
$\omega_j=\int\gamma_j\,d\alpha$ as zero-mean antiderivatives and
evaluate the Cauchy integrals \eqref{eq:w123} to obtain $\phi$ and
$\psi$ on the boundaries. The mean of $\omega_j$ for $2\le j\le N$ has
no effect on $\Phi(z)$ in $\Omega$, and the mean of $\omega_0$ and
$\omega_1$ only affect $\phi$ and $\psi$ in $\Omega$ up to a constant,
respectively. This constant in $\phi$ has no effect on the energy $E$
in \eqref{eq:E:formula}, and we replace $\psi\vert_j$ in
\eqref{eq:E:formula} by $(\psi\vert_j-\psi\vert_1)$, which is
equivalent to modifying the mean of $\omega_1$ to achieve
$\psi\vert_{\Gamma_1}=0$. One could alternatively avoid introducing
the stream function in the vortex sheet formulation by replacing
$\psi$ by $\phi$ in \eqref{eq:KE:psi:phi}, which is valid since $\mb
u\cdot\mb u=|\nabla\phi|^2$ as well. But $\pa\Omega$ now has to
include branch cuts to handle the multi-valued nature of $\phi$. This
leads to additional line integrals on paths through the interior of
the fluid that would have to be evaluated using quadrature. So in the
two-dimensional case, it is preferable to take advantage of the
existence of a single-valued stream function when computing the energy
in both the velocity potential and vortex sheet formulations. (In 3D,
  the velocity potential is single-valued, so this complication does
  not arise.)

\section{Solvability of the Integral Equations}
\label{sec:solvability}

In this section we prove invertibility of the operator $\mbb A$ in
(\ref{eq:AB:cor}), the system (\ref{eq:lin:sys2}), and the combined
system (\ref{eq:gam:t:final}) and (\ref{eq:gam:t:aux}). A variant of
\eqref{eq:AB:cor} is treated in Appendix~\ref{sec:psi:specified}.  We
follow the basic framework outlined in Chapter 3 of
\cite{folland1995introduction} to study the integral equations of
potential theory as they arise here.  Many details change due to
imposing different boundary conditions on the free surface versus on
the solid boundaries. The periodic domain also leads to significant
deviation from \cite{folland1995introduction}.  To avoid discussing
special cases, we assume $N\ge2$, though the arguments can be modified
to handle $N=1$ (no obstacles), $N=0$ (no bottom boundary or
  obstacles), or an infinite depth fluid with obstacles.

\subsection{Invertibility of $\mbb A$ in (\ref{eq:AB:cor})}
\label{sec:invert:A1}

After rescaling the rows of $\mbb B$ in (\ref{eq:B:def}) by 2 or $-2$,
it becomes a compact perturbation of the identity in $L^2(\pa\Omega)$.
Thus, its kernel and cokernel have the same finite dimension.  To show
that $\mbb A$ in (\ref{eq:AB:cor}) is invertible in $L^2(\pa\Omega)$,
we need to show that (1): $\mc{V}=\opn{span}\{\mb 1_m\}_{m=2}^N$ is
the entire kernel of $\mbb B$; and (2): $\mc V$ complements the range
of $\mbb B$ in $L^2(\pa\Omega)$. The second condition can be replaced
by (2'): $\mc V\cap\opn{ran}\mbb B=\{0\}$. Indeed, (1) establishes
that the cokernel also has dimension $N-1$, so (2') implies that $\mc
V\oplus\opn{ran}\mbb B=L^2(\pa\Omega)$.  We note that it makes sense
to apply $\mbb A$ and $\mbb B$ to $L^2$ functions, but the $\omega_j$
need to be continuous in order to invoke the Plemelj formulas to
describe the behavior of the layer potentials near the boundary. We
will address this below.

Suppose $\omega=(\omega_0;\dots;\omega_N)\in L^2(\pa\Omega)$ is such
that $\mbb B\omega\in\mc V$, i.e., $\mbb B\omega$ is zero on $\Gamma_0$
and $\Gamma_1$ and takes on constant values $\psi\vert_j$ on
$\Gamma_j$ for $2\le j\le N$. Since $\mbb B\omega$ is continuous, the
$\omega_j$ are also continuous, due to the $\pm(1/2)\mbb I$ terms on
the diagonal of $\mbb B$ in (\ref{eq:B:def}), and since the $\mbb
K_{kj}$ and $\mbb G_{kj}$ operators in (\ref{eq:B:def}) map $L^2$
functions to continuous functions. Let
$\Phi(z)=\Phi_0(z)+\cdots+\Phi_N(z)$ with $\Phi_j(z)$ depending on
$\omega_j$ as in (\ref{eq:w123}).  The real part
$\phi(x,y)=\re\Phi(x+iy)$ satisfies
\begin{equation}\label{eq:phi:unique}
  \Delta\phi=0 \;\; \text{in} \;\; \Omega, \qquad
  \phi=0 \;\; \text{on} \;\; \Gamma_0^-, \qquad
  \der{\phi}{n}=0 \;\; \text{on} \;\; \Gamma_1^+,\dots,\Gamma_N^+.
\end{equation}
Since homogeneous Dirichlet or Neumann conditions are specified on all
the boundaries and one of them is a Dirichlet condition, $\phi\equiv0$
on $\Omega$. This can be proved, e.g., by the maximum principle and the
Hopf boundary point lemma \cite{max:prin:book}. One version of this
lemma states that if $\Omega$ has a $C^1$ boundary and $u$ is harmonic
in $\Omega$, continuous on $\overline\Omega$, and achieves its global
maximum at a point $x_0$ on the boundary where the (outward) normal
derivative $\pa u/\pa n$ exists, then either $\pa u/\pa n>0$ at $x_0$
or $u$ is constant in $\Omega$. Since $\phi$ is a constant function in
$\Omega$, so is its conjugate harmonic function, $\psi=\im\Phi$.  But
$\psi\equiv0$ on $\Gamma_1^+$ since $\mbb B\omega\vert_{\Gamma_1}=0$,
so $\psi\equiv0$ in $\Omega$. We conclude that $\psi\vert_j$, which is the
value of $\psi$ on $\Gamma_j^+$, is zero for $2\le j\le N$. This shows
that $\mc V\cap\opn{ran}\mbb B=\{0\}$.

We have assumed that $\mbb B\omega\in\mc V$ and shown that $\mbb
B\omega=0$.  It remains to show that $\omega\in\mc V$.  Since the
normal derivative of $\phi$ is continuous across the free surface (see
  (\ref{eq:u:on:bdries})), we know that $\pa\phi/\pa n=0$ on
$\Gamma_0^+$.  Next consider a field point $z=x+iy$ with $y$ very
large. From (\ref{eq:w123}), we see that
\begin{equation}\label{eq:lim:y:infty:Phi}
  \lim_{y\rightarrow\infty}\Phi(z)=\frac1{2\pi i}
  \int_0^{2\pi}\frac i2\bigg[\omega_0(\alpha)\zeta_0'(\alpha)+
    i\sum_{j=1}^N \omega_1(\alpha)\zeta_1'(\alpha)\bigg]\,d\alpha = \text{const},
\end{equation}
so $\phi_\infty = \lim_{y\rightarrow\infty}\phi(x,y)$ exists and
does not depend on $x$. For the rest of this section, a
  prime will indicate a component of the complement of the domain
  or a function defined on this complement, rather than a derivative
  as in \eqref{eq:lim:y:infty:Phi}.
Let $\Omega_0'=\{(x,y) : y>\zeta_0(x)\}$. If
there were any point $(x_0,y_0)\in\Omega_0'$ for which
$\phi(x_0,y_0)>\phi_\infty$, then we could choose a $Y>y_0$ large
enough that $\phi(x,Y)<\phi(x_0,y_0)$ for $0\le x\le 2\pi$. Since the
sides of a period cell are not genuine boundaries, the maximum value
of the periodic function $\phi$ over the region $\Omega_0'\cap\{(x,y)
: 0\le x\le 2\pi\,,\, y<Y\}$ must occur on $\Gamma_0^+$. This
contradicts $\pa\phi/\pa n=0$ by the boundary point lemma. Using
the same argument for the minimum value, we conclude that $\phi$ is
constant on $\Omega_0'$. We denote its value by
$\phi\vert_0'=\phi_\infty$.  A similar argument with $\phi$ replaced
by $\psi$ and $y\rightarrow-\infty$ shows that $\psi$ takes on a
constant value $\psi\vert_1'$ in $\Omega_1'=\{(x,y) :
y<\zeta_1(x)\}$. On the interior boundaries $\Gamma_j^-$ of the holes
$\Omega_j'$, we have $\pa\psi/\pa n=0$. Thus $\psi$ satisfies the
homogeneous Neumann problem in each hole, and therefore has a constant
value $\psi\vert_j'$ in each hole.

Since $\omega_0$ gives the jump in $\phi$ across $\Gamma_0$ while
$\omega_j$ for $1\le j\le N$ gives the jump in $\psi$ across
$\Gamma_j$, we conclude that $\omega_j$ is constant on $\Gamma_j$ for
$0\le j\le N$.  Once the $\omega_j$'s are known to be constant, the
integrals (\ref{eq:w123}) can be computed explicitly using
\eqref{eq:d:phi:cyl} to express the antiderivative of the
  integrands in terms of $\Phi_\text{cyl}(\cdot)$. This gives
\begin{equation}\label{eq:Phij:w:const}
  \Phi_j(z) = \frac{\sigma_j\omega_j}{2\pi}\left[
    \Phi_\text{cyl}\big( \zeta_j(\alpha)-z \big) - \frac{\zeta_j(\alpha)-z}2
    \right]_{\alpha=0}^{2\pi}, \qquad
      (0\le j\le N),
\end{equation}
where $\sigma_0=1$ and $\sigma_j=i$ for $1\le j\le N$.  From the
discussion in Section~\ref{sec:phi:cyl:prop}, we conclude that
$\Phi_0(z)=\pm\omega_0/2$ if $z$ is above ($+$) or below ($-$) the
free surface $\zeta_0(\alpha)$; $\Phi_1(z)=\pm i\omega_1/2$ if $z$ is
above ($+$) or below ($-$) the bottom boundary $\zeta_1(\alpha)$; and
$\Phi_j(z)=0$ if $z$ is outside $\Gamma_j$ and $-i\omega_j$ if $z$ is
inside $\Gamma_j$.  For $z\in\Omega$, we conclude that
$\Phi(z)=(-\omega_0+i\omega_1)/2$.  Since we already established that
$\phi$ and $\psi$ are identically zero in $\Omega$, we find that
$\omega_0=0$, $\omega_1=0$, and the other $\omega_j$ are arbitrary
real numbers. Thus, $\ker\mbb B=\mc V$, as claimed.

\subsection{Solvability of the linear systems in the vortex sheet
  strength formulation}
\label{sec:invert:E}

There are two closely related tasks here, the solvability of
(\ref{eq:lin:sys2}) and the solvability of the larger system
consisting of (\ref{eq:gam:t:final}) and (\ref{eq:gam:t:aux}). Noting
that all the operators in these equations involve $\mbb K_{jk}^*$ or
$\mbb G_{jk}^*$, let us generically denote one of these systems by
$\mbb E^*\gamma=b$. In both cases, rescaling the rows of $\mbb E^*$ by
$\pm2$ yields a compact perturbation of the identity, so either $\mbb
E$ and $\mbb E^*$ are invertible or $\dim\ker\mbb E =\dim\ker\mbb E^*
< \infty$.  We will show that $\ker\mbb E=\{ 0\}$ to conclude that
$\mbb E^*$ is invertible.

We begin with (\ref{eq:gam:t:final}) and (\ref{eq:gam:t:aux}). This
system is the adjoint of the exterior version of the problem
considered in Section~\ref{sec:invert:A1} above. In other words, $\mbb
E$ here agrees with $\mbb B$ there, but with all the signs
$\pm(1/2)\mbb I$ reversed. This corresponds to taking limits of the
layer potentials from the opposite side of each boundary via the
Plemelj formulas. Suppose $\mbb E\omega=0$. As argued above, it
follows that each $\omega_j(\alpha)$ is continuous, and that the real
and imaginary parts of the corresponding function
$\Phi(z)=\Phi_0(z)+\cdots+\Phi_N(z)$ satisfy
\begin{equation}
  \phi\vert_{\Gamma_0^+}\equiv0, \qquad \psi\vert_{\Gamma_j^-}\equiv0, \qquad
           (1\le j\le N).
\end{equation}
Since $\psi$ satisfies homogeneous Dirichlet conditions inside each
obstacle, it is zero there.  The $2\pi$-periodic region above the free
surface can be mapped conformally to a finite domain via $w=e^{iz}$,
with $z=i\infty$ mapped to $w=0$. Similarly, the region below the
bottom boundary can be mapped to a finite domain via
$w=e^{-iz}$. Under the former map, $\phi$ becomes a harmonic function
of $w$ and
satisfies homogeneous Dirichlet boundary conditions. Under the latter
map, $\psi$ has these properties. As shown in
Appendix~\ref{sec:bot:hole}, $\phi$ and $\psi$ are also harmonic at
$w=0$ under these maps. Thus, $\phi\equiv0$ above the free
surface and $\psi\equiv0$ below the bottom boundary. Since
$\phi\equiv0$ above the free surface, $\psi$ is constant there, and is
continuous across $\Gamma_0$.  Since $\psi\equiv0$ in $\Omega_j'$ for
$1\le j\le N$ and its normal derivative is continuous across
$\Gamma_j$, we learn that $\psi$ is harmonic in $\Omega$, has a
constant value on $\Gamma_0^-$, and satisfies homogeneous Neumann
conditions on $\Gamma_j^+$ for $1\le j\le N$.  By the maximum
principle and the boundary point lemma, $\psi$ is constant in
$\Omega$, as is its conjugate harmonic function $\phi$. Denote these
constant values by $\psi_0$ and $\phi_0$.  Then $\omega$ is constant
on each boundary, with values $\omega_0=-\phi_0$ and $\omega_j=\psi_0$
for $1\le j\le N$. From (\ref{eq:Phij:w:const}),
$\Phi(z)=(\omega_0+i\omega_1)/2$ for $z$ above the free surface and
$\Phi(z)=-(\omega_0+i\omega_1)/2$ below the bottom boundary. Since
$\phi\equiv0$ above the free surface, $\omega_0=0$.  Since
$\psi\equiv0$ below the bottom boundary, $\omega_1=0$.  Since
$\omega_j=\omega_1$ for $2\le j\le N$, all components of $\omega$ are
zero, and $\ker\mbb E=\{0\}$ as claimed.

The analysis of the solvability of (\ref{eq:lin:sys2}) is nearly
identical, except there is no free surface. Setting $\mbb E\omega=0$
yields $\Phi(z)=\Phi_1(z)+\cdots+\Phi_N(z)$ such that $\psi\equiv0$
inside each cylinder and below the bottom boundary. Continuity of
$\pa_n\psi$ across the boundaries gives a solution of the homogeneous
Neumann problem in $\Omega$ that approaches a constant, $\psi_\infty$,
as $y\rightarrow\infty$. If $\psi(z)$ were to differ from
$\psi_\infty$ somewhere in $\Omega$, the maximum principle and
boundary point lemma would lead to a contradiction. Since $\omega_j$
is the jump in $\psi$ across $\Gamma_j$, it is a constant function
with value $\psi_\infty$.  Below the bottom boundary,
(\ref{eq:Phij:w:const}) gives $\Phi(z)=-i\omega_1/2$, so
$\omega_1=0$. Since $\omega_j=\omega_1$ for $2\le j\le N$, all
components of $\omega$ are zero, and $\ker\mbb E=\{0\}$ as claimed.

\section{Numerical Results}
\label{sec:numerics}

In this section we study the dynamics of a free surface interacting
with multiple obstacles, driven by a background flow of strength
$V_1=1$.

In Section~\ref{sec:flat}, the bottom boundary is flat and we
investigate the effect of varying the circulation parameters $a_j$. In
all three cases considered, the evolution is on track to terminate in
a splash singularity shortly after the final timestep of our numerical
simulation. We evolve the numerical solution on successively finer
grids until proceeding further would cause us to run out of resolution
on the finest grid, based on whether the spatial Fourier modes of
$\theta(\alpha,t)$, $\tilde\varphi(\alpha,t)$ and $\gamma_0(\alpha,t)$
decay below a given tolerance, which we take to be $10^{-12}$ in
double-precision.

In Section~\ref{sec:varbot}, the bottom boundary drops down to form a
basin in which we place a large obstacle. Some of the fluid flows
through the channel boun\-ded by the basin and the obstacle, which pulls
the free surface down around a second smaller obstacle. In this case,
rather than self-intersecting in a splash singularity, the free
surface is on track to collide with the smaller obstacle shortly after
the final timestep computed.

In Section~\ref{sec:connections}, we present numerical evidence to
show that our solutions remain fully resolved with spectral accuracy
at all times shown in the plots of Sections~\ref{sec:flat}
and~\ref{sec:varbot}. We use energy conservation, Fourier mode decay
plots, and quantitative comparison of the solutions computed by the
velocity potential and vortex sheet methods as measures of the
error. We also present the running times of the velocity potential
method and the vortex sheet method for different mesh sizes and study
the effect of floating-point arithmetic on the smoothness of the decay
of the Fourier modes of the solutions. We find that the velocity
potential method is somewhat faster while the vortex sheet method has
somewhat smoother Fourier decay properties.

\subsection{Free-surface flow around three elliptical obstacles}
\label{sec:flat}

In this section we consider a test problem of free-surface flow around
three obstacles in a fluid with a flat bottom boundary at $y=-3$. The
dimensionless gravitational acceleration and surface tension are set
to $g=1$ and $\tau=0.1$, respectively. The obstacles are ellipses
centered at $(x_j,y_j)$ with major semi-axis $q_j$ and minor semi-axis
$b_j$:
\begin{equation}
  \label{eq:ellipse:data}
  \begin{array}{|c|c|c|c|c|r|}
    \hline
    \quad j\quad &\quad x_j \quad &\quad  y_j  \quad &\quad
    q_j \quad &\quad  b_j \quad  & \quad \theta_j \;\quad \\[2pt] \hline
    \quad 2\quad &\quad \pi \quad &\quad -1.00 \quad &\quad
    0.5 \quad &\quad  0.5 \quad & \quad 0.0 \quad \\
    \quad 3\quad &\quad 4.0   \quad &\quad -1.75 \quad &\quad
    0.6 \quad &\quad  0.4 \quad & \quad 1.0 \quad \\
    \quad 4\quad &\quad 2.3 \quad &\quad -1.60 \quad &\quad
    0.7 \quad &\quad  0.3 \quad & \; -0.5 \quad \\ \hline
  \end{array}
\end{equation}
The major axis is tilted at an angle $\theta_j$ (in radians) relative
to the horizontal.  With this geometry, we consider three cases for
the parameters of $\Phi_\text{mv}(z)$ in \eqref{eq:Phi:mv}, namely
\begin{equation}\label{eq:mv:params}
  \begin{array}{|c|c|r|c|c|c|}
    \hline
    & V_1 & a_2 \quad\; & a_3 & a_4 \\ \hline
    \quad\raisebox{-2pt}{problem 1} \quad &
    \quad \raisebox{-2pt}{1.0} \quad &
    \; \raisebox{-2pt}{$-1.0$} \quad &
    \quad  \raisebox{-2pt}{0.0} \quad &
    \quad  \raisebox{-2pt}{0.0} \quad \\[2pt]
    \quad\text{problem 2} \quad & \quad 1.0 \quad &\quad
    0.0 \quad &\quad  0.0 \quad &\quad  0.0 \quad \\
    \quad\text{problem 3} \quad & \quad 1.0 \quad &\quad
    1.0 \quad &\quad  0.0 \quad &\quad  0.0 \quad \\[1pt] \hline
  \end{array}
\end{equation}
The initial wave profile is flat and the initial single-valued part of
the surface velocity potential, $\tilde\varphi(\alpha,0)$, is set to
zero.  The physical evolution governed by the Navier-Stokes equations 
would then develop a boundary layer of vorticity around the bodies that 
would eventually shed in a wake--vortex street regime.  This will leave a 
non-zero circulation around each body.  To see the effects of the circulation 
within the present potential flow framework we consider three cases
with different circulation parameters chosen for the initial conditions.
Since the wave eventually overturns in each case listed in
\eqref{eq:mv:params}, we use the modified HLS representation in which
$P\theta(\alpha,t)$ is evolved and the curve is reconstructed by the
method of Section~\ref{sec:constman}. We solve each problem twice,
once with the velocity potential method of Section~\ref{sec:cauchy}
and once with the vortex sheet method of Section~\ref{sec:laypotalg}.
Identical spatial and temporal discretizations are used for both
methods.

For the spatial discretization, we use $M_1=96$ gridpoints on the
bottom boundary and $M_j=128$ gridpoints on each ellipse boundary,
where $j\in\{2,3,4\}$. The ellipses are discretized uniformly in
$\alpha$ (rather than arc\-length) using the parameterization
$\zeta_j(\alpha)=e^{i\theta_j}\big(q_j\cos(\alpha) + i
  b_j\sin(\alpha)\big)$.  We start with $M_0=256$ gridpoints on the
free surface and add gridpoints as needed to maintain spectral
accuracy as time evolves. This is done by monitoring the solution in
Fourier space and requiring that the Fourier mode amplitudes
$|\hat\theta_k(t)|$ and $|\hat{\tilde\varphi}_k(t)|$ or
$|\hat\gamma_k(t)|$ decay to $10^{-12}$ before $k$ reaches $M_0/2$. We
use the sequence of mesh sizes $M_0$ listed in
  Table~\ref{tbl:M0d}.

\begin{table}
\begin{equation*}
  \begin{array}{r|c|c|c|c|c|c|c|c|c|c|c}
    M_0 \;&\; 256 \;&\; 384 \;&\; 512 \;&\; 768 \;&\; 1152 \;&\;
    1728 \;&\; 2592 \;&\; 3456 \;&\; 4608 \;&\; 6144 \;&\; 7776 \\
    \hline
       d\; & 10  & 15 & 20 & 32 & 56 & 90 & 150 & 200 & 300 & 500 & 700 \\ \hline
    ns_1\, & 17  & 11 &  8 & 16 & 11 &  5 &  6 & 10 &  - &  - & - \\
    ns_2\, & 120 & 30 & 15 & 12 & 13 & 11 &  8 &  4 &  3 &  4 & 3 \\
    ns_3\, & 50  & 36 & 16 & 13 & 10 &  6 &  6 &  3 & 10 & 25 & 4
    \\ \hline
    \text{GEPP}(\tilde\varphi)\,
    & 0.54 & 1.05 & 1.88 & 4.38 & 12.7 & 37.8 & 128 & 300 & 648 & 2010 & 4450  \\
    \text{GEPP}(\gamma_0)\,
    & 0.78 & 1.42 & 2.31 & 5.08 & 14.6 & 43.3 & 178 & 318 & 714 & 2370 & 4970 \\
    \text{GMRES}(\tilde\varphi)\,
    & 0.52 & 0.92 & 1.41 & 3.17 & 8.10 & 22.9 & 76.2 & 181 & 455 & 1349 & 3464 \\
    \text{GMRES}(\gamma_0)\,
    & 0.69 & 1.19 & 1.86 & 3.92 & 10.2 & 26.1 & 107 & 218 & 562 & 1786 & 4232 \\
    \text{GMRES}(\text{GPU},\tilde\varphi)\,
    & 0.26 & 0.44 & 0.64 & 1.27 & 2.82 & 6.88 & 17.9 & 35.8 & 78.6 & 239 & 778
  \end{array}
\end{equation*}
\caption{\label{tbl:M0d}
Parameters used to timestep problems 1--3 from the initial flat
  rest state to a near splash singularity and comparison of running
  times. $M_0$ is the number of spatial gridpoints used to discretize
  the free surface; $d$ is the number of Runge-Kutta steps taken to
  advance time by one macro-step of length $\Delta t=0.025$ separating
  the times at which the output is recorded; $ns_1$, $ns_2$ and $ns_3$
  are the number of macro-steps taken for the given $M_0$ in problems
  1, 2 and 3, respectively; and the last five rows are the wall clock
  running time (in seconds) of one macro-step (i.e.~of $d$ Runge-Kutta
    steps) of the solvers we implemented; see
  Section~\ref{sec:performance} below.}
\end{table}

For time-stepping, we use the 8th order Dormand-Prince Runge-Kutta
scheme described in \cite{hairer:I}. The solution is recorded at equal
time intervals of width $\Delta t=0.025$, which we refer to as
  macro-steps. The timestep of the Runge-Kutta method is set to
$\Delta t/d$, where $d$ increases with $M_0$ as listed in
Table~\ref{tbl:M0d}. These subdivisions are chosen empirically to
maintain stability. We also monitor energy conservation (as explained
  further below) and increase $d$ until there is no further
improvement in the number of digits preserved at the output times
$t\in\mbb N\Delta t$. The number of macro-steps taken for each
  choice of $M_0$ and $d$ in problems 1, 2 and 3 is given in the rows
  labeled $ns_1$, $ns_2$ and $ns_3$, respectively.  In problems
  2 and 3, timesteps are taken until $M_0=7776$ would be
insufficient to resolve the solution through an additional
macro-step $\Delta t$. In problem 1, we stopped at
  $M_0=3456$ as this is sufficient to observe the dynamics we wished
  to resolve.  In all three cases, the solution appears to form a
splash singularity \cite{castro2013finite,castro2012splash} shortly
after the final time reported here. The running times of the
  solver options we tested are given in the last 5 rows of
  Table~\ref{tbl:M0d}; these will be discussed in
  Section~\ref{sec:performance} below.

Figures~\ref{fig:prob1}--\ref{fig:prob3} show the time evolution of
the free surface as it evolves over the cylinders for problems 1--3,
defined in \eqref{eq:mv:params}, along with contour plots of the
magnitude of the velocity. The arrows in the velocity plots are
normalized to have equal length to show the direction of flow. In each
plot, the aspect ratio is 1, i.e., the $x$ and $y$-axes are scaled the
same.  In all three problems, the background flow rate is $V_1=1$ and
there is zero circulation around cylinders 3 and 4.  In panels (a) and
(b) of Figure~\ref{fig:prob1} and panels (a)--(c) of
Figures~\ref{fig:prob2} and~\ref{fig:prob3}, snapshots of the free
surface are shown at equal time intervals over the time ranges
given. The curves are color coded to evolve from green to blue to red,
in the direction of the arrows.  The initial and final times plotted
in each panel are also indicated with black dashed curves.

\begin{figure}[p]
\centering
\includegraphics[width=\linewidth]{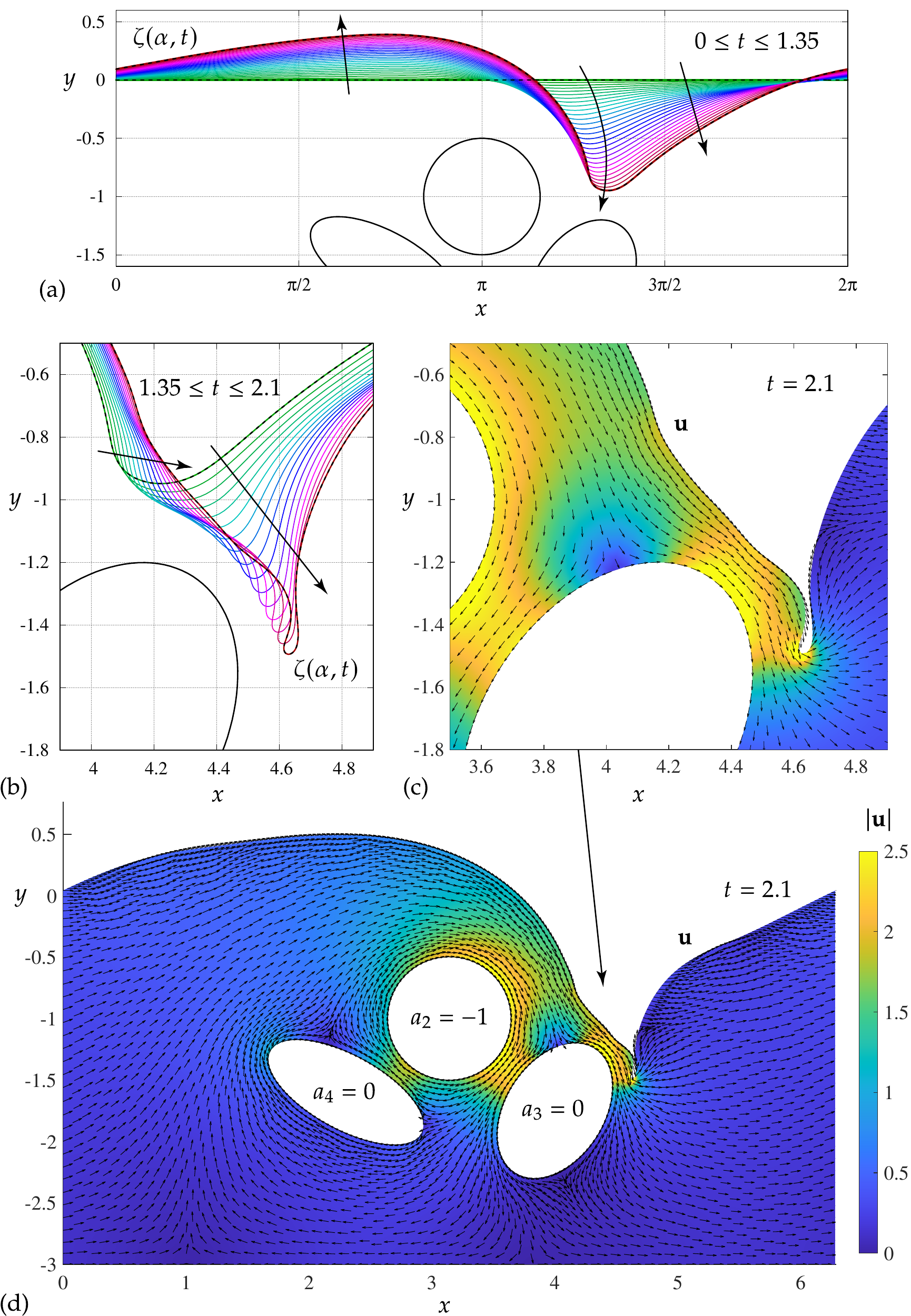}
\caption{\label{fig:prob1} Evolution of the free surface and plots of
  the fluid velocity at the final time computed, $t=2.1$, for problem
  1 of \eqref{eq:mv:params}. The computation breaks down when
    evolved beyond $t=2.1$ with macro-steps of size $\Delta t=0.025$,
    but is on track to self-intersect in a splash singularity. The
  fluid is stagnant below the cylinders and the flow is reversed in
  the channels between cylinder 2 and its neighbors.  As a
  result, the change in velocity potential across the domain is
  0 or $2\pi$ depending on whether the path passes below or above
  cylinder 2.}
\end{figure}

In Figure~\ref{fig:prob1}, the clockwise circulation around cylinder 2
(due to $a_2=-1$) pulls the free surface down to the right of the
cylinder, toward the channel between cylinders 2 and 3. This causes an
upwelling to the left of cylinder 2 in order to conserve mass. At $t=1.35$, we
see in panel (b) that the lowest point on the free surface stops
approaching the channel and begins to drift to the right, around
cylinder 3. The left (upstream) side of the interface (relative to its
  lowest point) accelerates faster than the right side, which causes
the interface to sharpen and fold over itself. Shortly after $t=2.1$,
our numerical solution loses resolution as the left side of the
interface crashes into the right side to form a splash singularity
\cite{castro2013finite,castro2012splash}. The colormap of the contour
plot in panel (c) is the same as in panel (d). We see that the
velocity is largest in magnitude in the region above and to the right
of cylinders 2 and 3, and is relatively small throughout the fluid
otherwise. The change in velocity potential along a path
crossing the domain below all three cylinders is zero in this case
since $V_1+a_2+a_3+a_4=0$, whereas the change along a path crossing
above the cylinders is $2\pi$.

\begin{figure}[p]
\centering
\includegraphics[width=.98\linewidth]{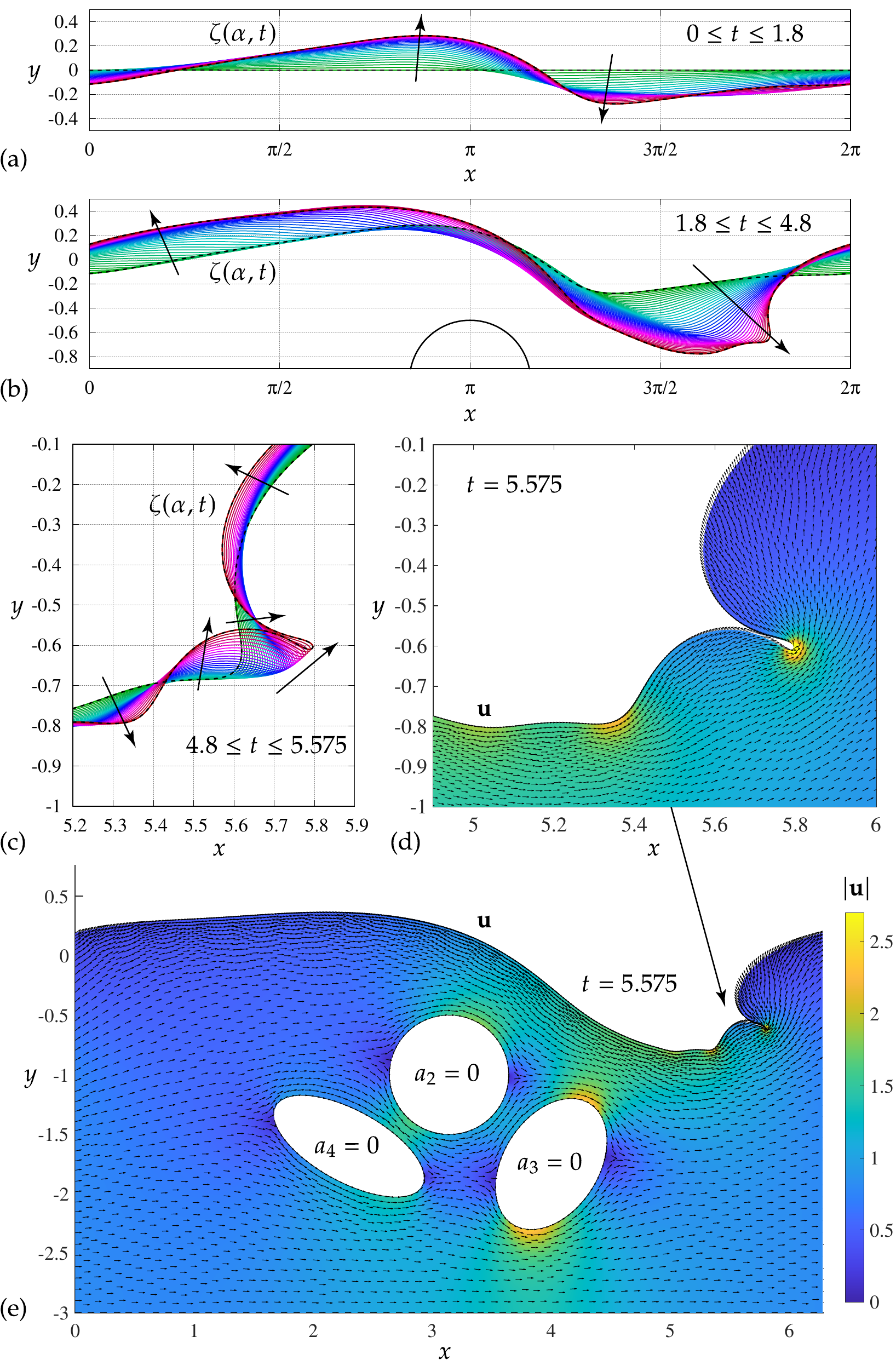}
\caption{\label{fig:prob2} Evolution of the free surface and plots of
  the fluid velocity at the final time computed, $t=5.575$, for
  problem 2 of \eqref{eq:mv:params}. The interface is on track
    to self-intersect in a splash singularity shortly after this, as
    is evident in panel (c). The velocity field in panel (d) on the
    upper surface of the air pocket is nearly tangential, whereas it
    has an appreciable normal component on the lower surface.  This
    provides further evidence that a splash singularity is imminent.}
\end{figure}

In Figure~\ref{fig:prob2}, $a_2$ is set to zero, which causes the
change in velocity potential
along any path across the domain to be $2\pi$, whether it passes
above, below or between the cylinders. As a result, the magnitude of
velocity is more evenly spread throughout the fluid. This magnitude is
largest below cylinder 3 and above cylinders 2 and 3, where the width
of the fluid domain is smallest. Similar to problem 1, the free
surface initially drops to the right of the cylinders and rises to the
left, but it does not get pulled toward the channel between
  cylinders 2 and 3 since the flow is not reversed there this
  time. Nevertheless, the free surface eventually folds over itself,
  but farther downstream and at a later time than in problem 1.
Panel (b) shows the development of an air pocket expanding into
  the fluid as it travels down and to the right. This air pocket
sharpens in panel (c) to form a splash singularity shortly after
$t=5.575$.

\begin{figure}[p]
\centering
\includegraphics[width=\linewidth]{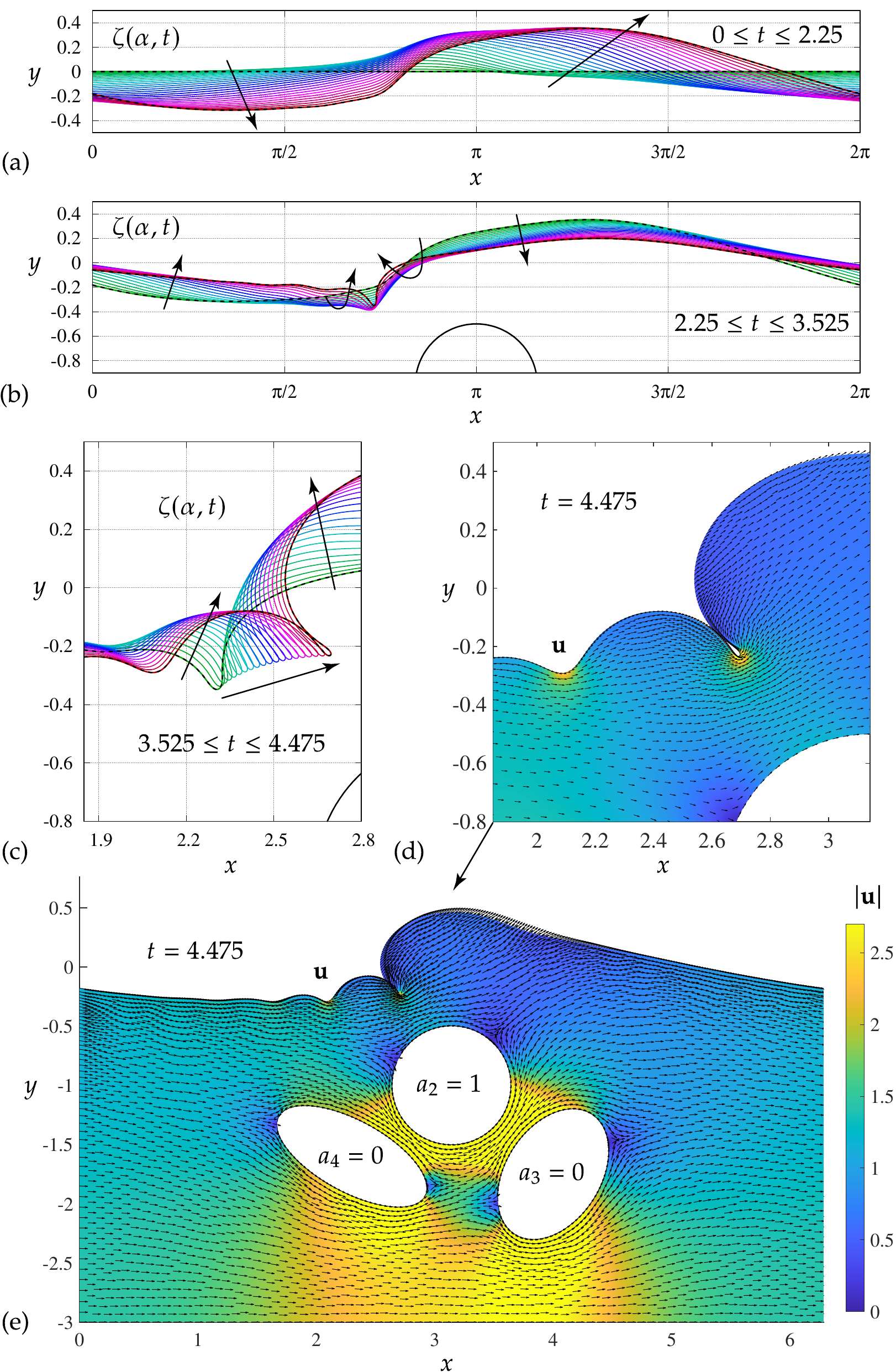}
\caption{\label{fig:prob3} Evolution of the free surface and plots of
  the fluid velocity for problem 3 at $t=4.475$, when the stream
  function on the rigid boundaries is $\psi\vert_1=0$,
  $\psi\vert_3=2.243$, $\psi\vert_4=2.580$ and $\psi\vert_2=3.387$,
  and the fluid flux in the channels is
  $(\psi\vert_2-\psi\vert_4)=0.807$ and
  $(\psi\vert_2-\psi\vert_3)=1.144$.}
\end{figure}

In Figure~\ref{fig:prob3}, $a_2$ is set to 1. The counter-clockwise
circulation around cylinder 2 causes the change in velocity potential
to be $2\pi$ along a path
crossing the domain above all three cylinders and to be $4\pi$ along a path
passing below any subset of the cylinders that includes cylinder 2.
The magnitude of velocity is largest in the channels between cylinder
2 and its neighbors, and below all three cylinders. The net flux
below cylinder 3 is still larger than that passing between cylinders 2
and 3, as noted in the caption.  There is an upwelling of the free
surface above and to the right of cylinder 2 with a drop in fluid
height to the left of the cylinders, which is the opposite of what
happens in problems 1 and 2. Capillary waves form at the free surface
ahead of the cylinders, with the largest oscillation eventually
folding over to form a splash singularity. In the final stages of this
process, shown in panel (c), a structure resembling a Crapper wave
\cite{crapper,akers2014gravity} forms, which travels slowly to the
right as the fluid flows faster around and below it (left to
  right). As it evolves, the sides of this structure slowly approach
each other while also slowly rotating counter-clockwise.

\begin{figure}[p]
\centering
\includegraphics[width=.9\linewidth]{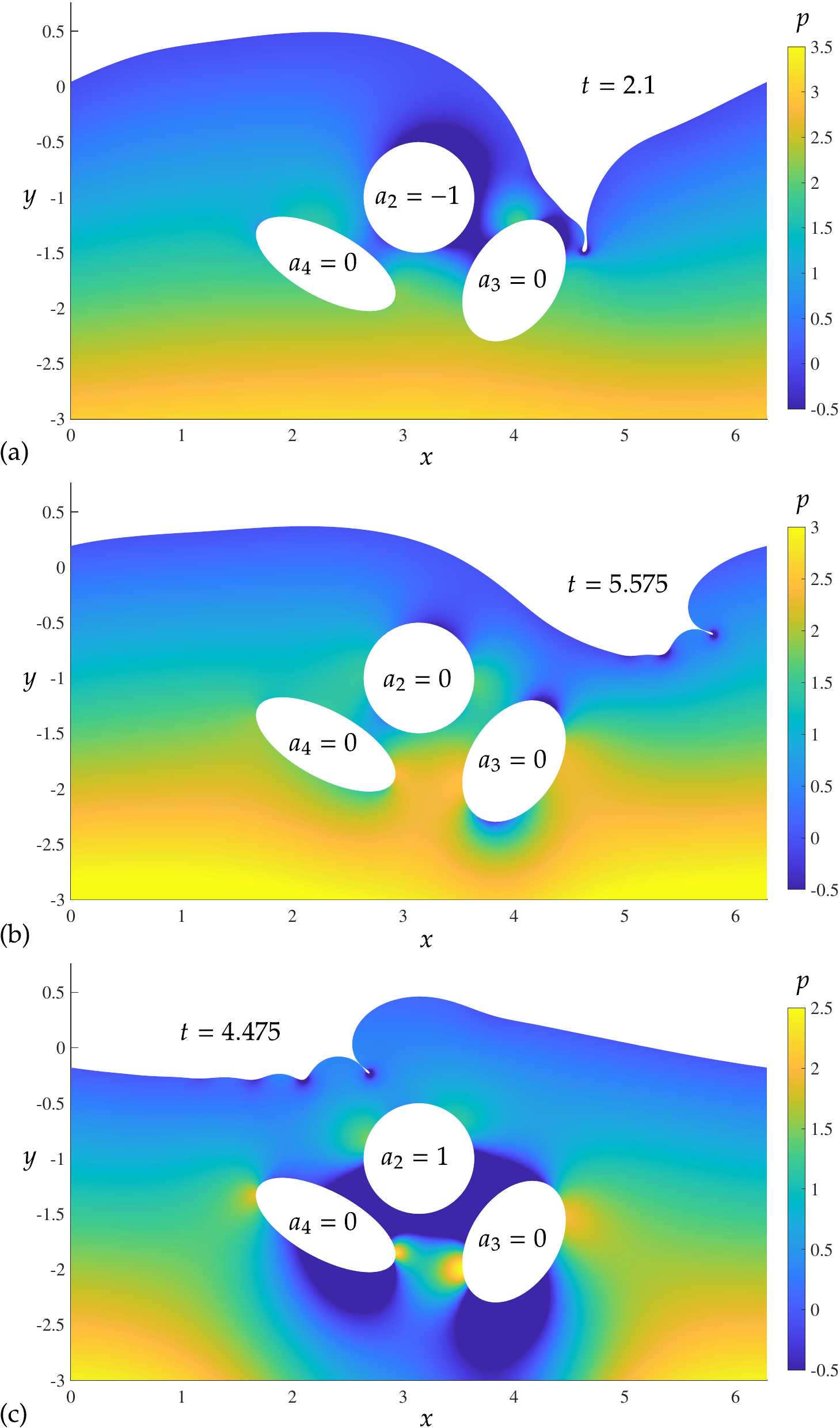}
\caption{\label{fig:pressure} Pressure in the fluid at the
final times computed for problems 1--3 of \eqref{eq:mv:params}.}
\end{figure}

Figure~\ref{fig:pressure} shows the pressure in the fluid at the final
times shown in Figures~\ref{fig:prob1}--\ref{fig:prob3}.  On the free
surface, the pressure is given by the Laplace-Young condition,
$p=p_0-\rho\tau\kappa$, where we take $p_0=0$, $\rho=1$ and
$\tau=0.1$. Setting $p_0=0$ means pressure is measured relative to the
ambient pressure, so negative pressure is allowed. The curvature
$\kappa=\theta_\alpha/s_\alpha$ is positive when the interface curves
to the left (away from the fluid domain) as $\alpha$ increases,
i.e., moving along the interface from left to right. Inside the fluid,
we compute $p$ using \eqref{eq:press} and
  \eqref{eq:pressure:d:prob}, as explained in Section~\ref{sec:fvp}
  above. From \eqref{eq:press}, we see that pressure increases with
fluid depth and decreases in regions of high velocity, up to the
correction $\phi_t$ in \eqref{eq:press}, which is a
harmonic function satisfying homogeneous Neumann boundary conditions
on the solid boundaries.

In all three panels of Figure~\ref{fig:pressure}, the pressure is
visibly lower near the capillary wave troughs, especially the largest
trough that folds over into a structure similar to a Crapper wave
before the splash singularity forms. As the circulation parameter
$a_2$ increases from $-1$ in panel (a) to 0 in panel (b) and to 1 in
panel (c), the pressure decreases near the bottom boundary. Problem 3
has higher velocities than problems 1 and 2 below the cylinders and in
the channels between cylinders, which leads to smaller pressures in
these regions in panel (c) than in panels (a) and (b). This effect
would have been even more evident if we had used the same colorbar
scaling in all three plots, but this would have washed out some of the
features of the plots.

The contour plots of Figures~\ref{fig:prob1}-\ref{fig:prob3} confirm
that the fluid velocity increases in the neighborhood of the capillary
wave troughs (where the pressure is lower), and is quite large
below the Crapper wave structure. We also see in
Figure~\ref{fig:pressure} that in problem 1, which involves negative
circulation around the top-most body, the pressure above
  this obstacle is significantly lower than elsewhere.
This sheds light on an effect
observed experimentally \cite{wu1972cavity} whereby an air bubble
can be permanently trapped between the top of an airfoil and the free
surface.  We will investigate this phenomenon in more detail in future
work.

\subsection{Free-surface flow in a geometry with variable bottom topography}
\label{sec:varbot}

Next we consider an example (problem 4) in which the bottom boundary
drops off rapidly and later rises again, forming a basin in
between. We define $\zeta_1(\alpha)=\alpha+i\eta_1(\alpha)$ with
$\eta_1(\alpha)$ satisfying
\begin{equation}
  \eta_1(0)=-\frac12, \qquad
  \eta_1'(\alpha) = -5\sin^{63}\alpha, \qquad 0\le\alpha\le2\pi.
\end{equation}
The 63rd power of $\sin(\alpha)$ is close to zero except near
$\alpha=\frac\pi2$ and $\alpha=\frac{3\pi}2$, causing $\eta_1(\alpha)$
to be quite flat in both the shallow and deep regions. The fluid depth
ranges from $\eta_1(0)=-0.5$ to $\eta_1(\pi)=-2.072774$, and is
$2\pi$-periodic since $\sin^{63}\alpha$ has zero mean. We place a
large ellipse in the center of the basin to create a channel between
the ellipse and the bottom boundary boundary.  We force some of the
fluid to flow through the channel by setting $a_2=1/2$. We place a
smaller ellipse near the entrance of the channel and set the
circulation parameter of this ellipse to be $a_3=0$.  In the notation of
\eqref{eq:ellipse:data}, the ellipse positions, sizes and circulation
parameters are given by
\begin{equation}
  \label{eq:ellipse:data4}
  \begin{array}{|c|c|c|c|c|c|r|}
    \hline
    \multicolumn{7}{|c|}{\text{Obstacle data in
        problem 4}} \\ \hline
    \quad j \quad & \quad x_j \quad & \quad  y_j  \quad & \quad
    q_j \quad & \quad  b_j \quad  & \quad \theta_j & \quad a_j \quad \\[2pt] \hline
    \quad 2\quad &\quad \pi \quad &\quad -9/8 \quad &\quad
    4/3 \quad &\quad  3/4 \quad & \quad 0.0 \quad & \quad 0.5 \quad \\
    \quad 3\quad &\quad 1.8   \quad &\quad -0.4 \quad & \quad
    0.2 \quad &\quad  0.1 \quad & \quad 0.0 \quad & \quad 0.0 \quad \\ \hline
  \end{array}
\end{equation}
As in Section~\ref{sec:flat}, we also set $V_1=1$, $\tau=0.1$ and $g=1$.

\begin{figure}[p]
\centering
\includegraphics[width=.96\linewidth]{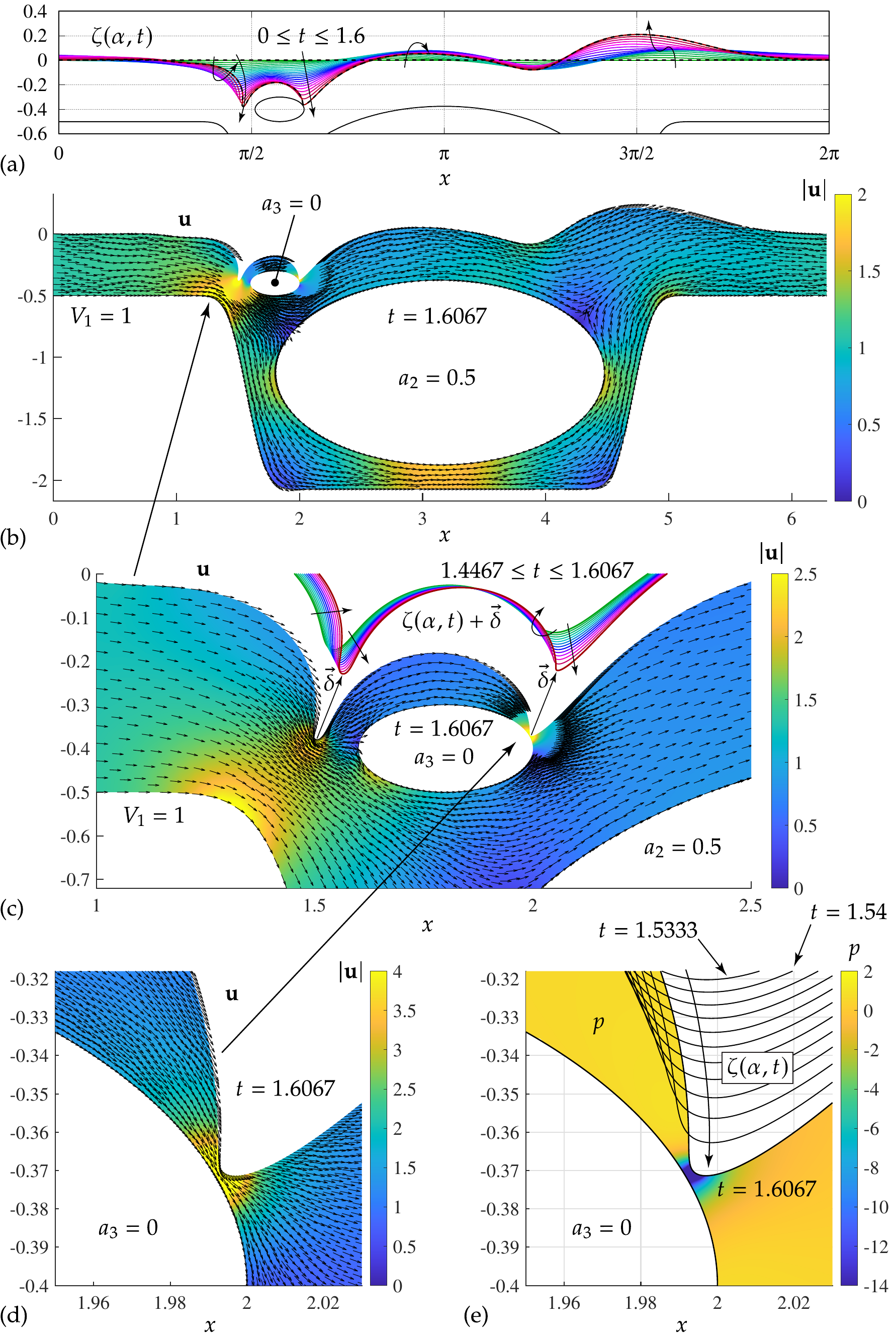}
\caption{\label{fig:prob4u} Time evolution of the free surface and
  plots of velocity and pressure at $t=1.6067$. The time increments
  shown in panels (a), (c) and (e) are $0.08$, $0.01333$ and
  $0.00667$, respectively. In panel (c), an offset
  $\vec\delta=(0.06,0.15)$ was added to the free surface plots to
  avoid obscuring the contour plot. In panel (e), the free surface
  approaches the obstacle at an accelerating rate as the gap shrinks
  and the curvature of the free surface grows. The numerical solution
  is still resolved with spectral accuracy at $t=1.6067$, but we cannot
  not proceed without further mesh refinement.}
\end{figure}

In Figure~\ref{fig:prob4u}, panel (a) shows the time evolution of the
free surface for $0\le t\le1.6$, computed using the velocity potential
method.  The curves are shown at equal time intervals of size $\Delta
t=0.08$. They are color coded to evolve from green to blue to red in
the direction of the arrows. The interface is initially flat and the
periodic part of the velocity potential is initially set to zero,
$\tilde\varphi(\alpha,0)=0$.  Setting $V_1=1$, $a_2=1/2$ and $a_3=0$
causes the fluid entering the domain from the left to split into two
parts, one flowing down through the channel between the basin and the
large ellipse and the other flowing above the large ellipse from left
to right. This leads to a stagnation point in the upper left quadrant
of the large ellipse, as shown in panel (b) at $t=1.6067$. A similar
stagnation point is present in the upper right quadrant, where the two
streams recombine to flow over the right edge of the basin. This leads
to an upwelling of fluid above the right edge of the basin, as shown
in panels (a) and (b). The opposite occurs at the left edge of the
basin, where the free surface is pulled down by the fluid entering the
channel below. The presence of the small obstacle causes the free
surface to form two air pockets moving downward on either side of the
obstacle.  We exclude arrows in the velocity direction field near
these air pockets in panel (b) to avoid obscuring the contour plots.

Panel (c) of Figure~\ref{fig:prob4u} provides a magnified view of the
flow near the small obstacle, including the direction field near the
upstream air pocket (to the left of the obstacle).  The contour plot
and direction field correspond to $t=1.6067$. We also show the time
evolution of the free surface leading up to this state, with $t$
ranging from $1.44667$ to $1.60667$, plotted at time increments of
$0.01333$. We plot $\zeta(\alpha,t)+\vec\delta$, where we have
introduced the offset $\vec\delta=(0.06,0.15)$ to separate the
evolving curves from the contour plot. As time evolves, each of the
air pockets sharpens as it is pulled further into the fluid. The
upstream air pocket begins to form a Crapper-wave structure similar to
those seen in Figures~\ref{fig:prob1}--\ref{fig:pressure} in
Section~\ref{sec:flat}, and would likely form a splash
singularity at a later time. But before this happens, the right air
pocket approaches the obstacle and appears on track to collide with it
shortly after $t=1.6067$.

Panel (d) of Figure~\ref{fig:prob4u} shows the magnitude and direction
of the velocity in the neighborhood of the point of closest approach
at $t=1.6067$. The fluid moves fastest in the small gap between the
free surface and the obstacle. The increased speed is caused by a
large pressure drop in the gap, shown in panel (e), which causes the
fluid to accelerate as it approaches the gap and decelerate
afterwards. The pressure drop is caused by surface tension and the
high curvature of the interface near the gap. Also shown in panel (e)
are snapshots of the interface from $t=1.53333$ to $t=1.60667$ in
increments of $\Delta t=0.00667$. These curves are plotted in black
since the direction of motion is clear. The free surface appears to
approach the obstacle at an accelerating rate as it sharpens. We
believe an impact will occur with a simultaneous blow-up of the
curvature there, though it is possible that the interface will slide
around the obstacle before crashing into it or crash into it before
forming a curvature singularity. Further investigation
would likely require using a non-uniform grid (rather than arclength
  parameterization) and adapting the ideas of
Appendix~\ref{sec:helsing} to handle the close approach of the
interface to the obstacle without excessive mesh refinement; however,
these are topics for future research.

\begin{figure}[t]
\centering
\includegraphics[width=.95\linewidth]{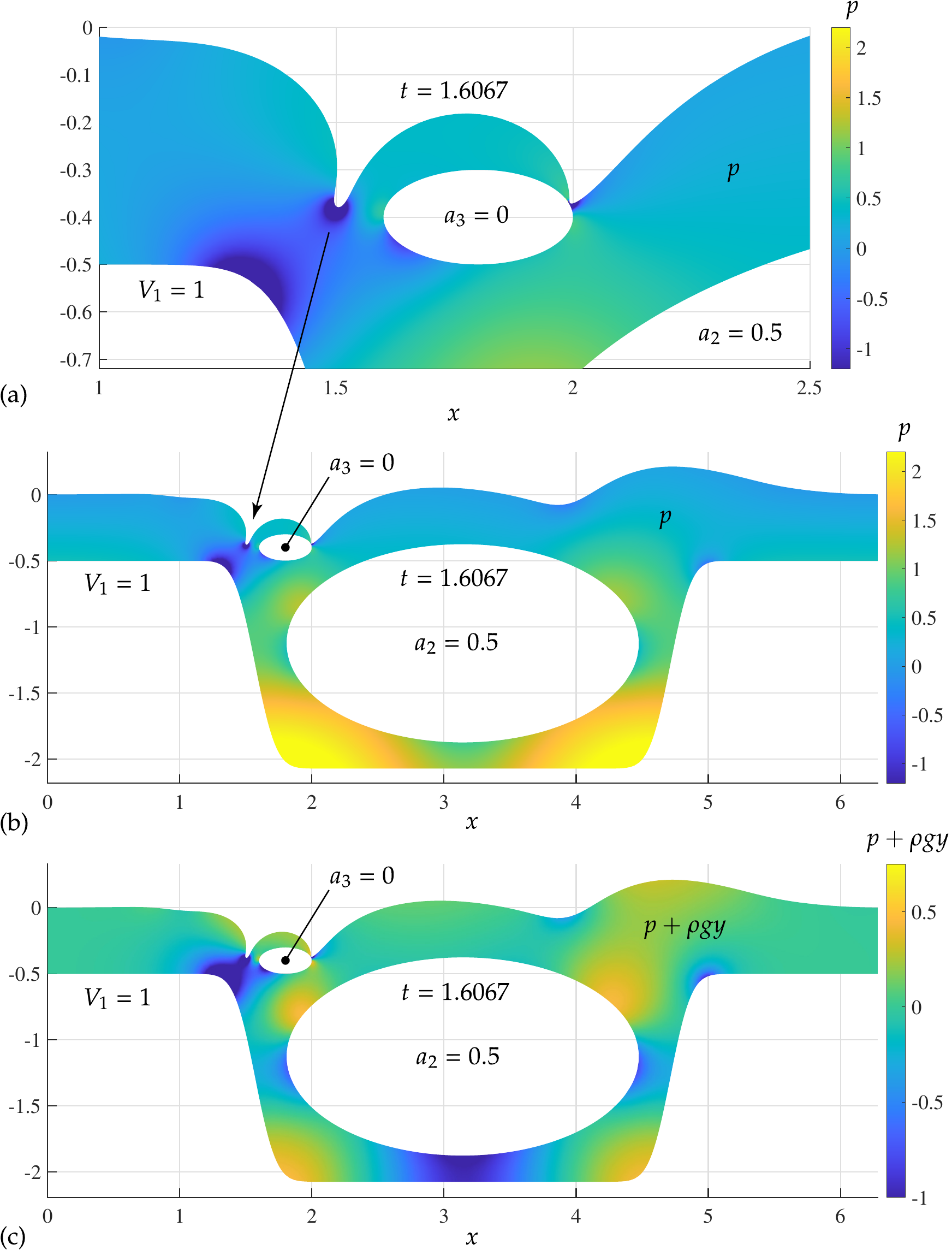}
\caption{\label{fig:prob4p} The pressure is lowest where the free
  surface has high curvature and where the fluid velocity is highest,
  e.g., due to flowing through a constriction. The hydrostatic term
  $-\rho gy$ in \eqref{eq:press} is responsible for the high-pressure
  regions in the bottom corners of the basin in panel (b). This term
  has been eliminated by adding $\rho gy$ to $p$ in panel (c).}
\end{figure}

Zooming out from panel (e) of Figure~\ref{fig:prob4u} and rescaling
the colorbar yields the pressure plot (at $t=1.6067$) shown in panel
(a) of Figure~\ref{fig:prob4p}. Comparison with panel (c) of
Figure~\ref{fig:prob4u} shows that the pressure is lowest where the
velocity is highest, with local minima occurring under the two air
pockets of the free surface, at the left edge of the basin where the
bottom boundary drops off, and along part of the lower-left boundary
of the small obstacle.  Zooming out further to the entire domain gives
the pressure plot in panel (b). The scaling of the colorbar is the
same as in panel (a). Here the effects of the hydrostatic term $-\rho
gy$ in the formula \eqref{eq:press} for $p$ are clearly seen, with the
largest values of pressure occurring in the bottom corners of the
basin. In panel (c) we plot the deviation from this hydrostatic state,
which would be the equilibrium pressure if the fluid were at rest with
a flat free surface. We see a large negative deviation where the fluid
flows fastest and where the free surface curves upward most rapidly,
and positive deviations near the stagnation points on the large
obstacle, in the bottom corners of the basin, and in the upwelling
above the right edge of the basin.

\begin{table}
  \begin{equation*}
    \begin{array}{r|c|c|c|c|c|c|c}
      M_0 \;&\; 768 \;&\;  1536 \;&\;  3072 \;&\; 5184 \;&\;
      9216 \;&\; 12288 \;&\; 16384 \\
      M_3 \;&\; 256 \;&\;  256 \;&\;  256 \;&\; 288 \;&\;
      768 \;&\; 1296 \;&\; 1536 \\ \hline
      d\;  &   8 & 20 & 50 & 90 & 210 & 300 & 450 \\ \hline
      ns\; & 138 & 48 & 30 & 18 & 6 &  2/3 &  1/3 \\ \hline
      \text{GEPP}(\tilde\varphi)\,
      & 2.80 & 13.7 & 77.2 & 370 & 3103 & 9605 & 29184 \\
      \text{GEPP}(\gamma_0)\,
      & 4.14 & 17.1 & 97.7 & 419 & 3734 & 10779 & 35437 \\
      \text{GMRES}(\tilde\varphi)\,
      & 1.98 & 9.07 & 59.6 & 260 & 2019 & 6278 & 15925 \\
      \text{GMRES}(\gamma_0)\,
      & 3.25 & 13.8 & 93.1 & 334 & 2831 & 9897 & 27921 \\
      \text{GMRES}(\text{GPU},\tilde\varphi)\,
      & 0.62 & 2.37 & 12.6 & 48.7 & 364 & 937  & 2492
    \end{array}
  \end{equation*}
  \caption{\label{tbl:M0M3} Parameters used to timestep problem
      4 from the initial flat state to a near collision of the free
      surface with an obstacle. We set $M_1=768$ and $M_2=M_3$ in each
      case. Here $d$ and $ns$ have the same meanings as in
      Table~\ref{tbl:M0d}, except that the macro-step size is $\Delta
      t=0.00667$ instead of $0.025$. The fractional steps with
      $ns=2/3$ and $1/3$ indicate a change in mesh size part-way
      through the final macro-step from $t=1.6$ to $1.60667$, shown in
      panel (e) of Figure~\ref{fig:prob4u}. The last 5 rows report the
      wall-clock running time (in seconds) of one macro-step for the
      solvers we implemented.}
\end{table}

We used the sequence of meshes and stepsizes listed in
Table~\ref{tbl:M0M3} to evolve the solution with spectral accuracy
from $t=0$ to $t=1.6067$, the time at which the velocity and pressure
are plotted in Figures~\ref{fig:prob4u} and~\ref{fig:prob4p}.  In all
cases, we discretize the bottom boundary with $M_1=768$ points and set
$M_2=M_3$ for the two obstacles. $M_3$ must increase as the interface
approaches the small obstacle in order to maintain spectral accuracy
in the Fourier representation of $\omega_3(\alpha,t)$. It would have
been sufficient to use $M_2=256$ throughout the computation since the
free surface does not approach the large obstacle; however, for
simplicity, our code assumes each obstacle has the same number of
gridpoints.  In the terminology of Table~\ref{tbl:M0d}, we have set
the macro-step size to $\Delta t=0.00667$. This is the temporal
spacing of the curves plotted in panel (e) of
Figure~\ref{fig:prob4u}. The curves in panels (a) and (c) were plotted
with time increments of $0.08$ and $0.01333$, which are every 12th and
every 2nd macro-step, respectively.

In each column of Table~\ref{tbl:M0M3}, $ns$ macro-steps were taken
with the listed values of $M_0$, $M_3$ and $d$, where $d$ is the
number of Runge-Kutta steps per macro-step. The last 5 rows report the
wall-clock running time (in seconds) of evolving the solution through
one macro-step using the velocity potential or vortex sheet method
with Gaussian elimination or GMRES to solve the linear systems that
arise. We also implemented a GPU-accelerated version of the GMRES
solver in the velocity potential framework.
The final macro-step to evolve the solution from
$t=1.60000$ to $t=1.60667$ was done in two stages with the parameters
listed in the last two columns of the table. Both $d$ and the running
times are scaled in the table to correspond to one full macro-step.
Multiplication by $ns$ gives the total number of Runge-Kutta steps and
the total computational time of that phase of the numerical solution.
The running times of the solvers we tested will be discussed further
in Section~\ref{sec:performance} below.

\subsection{Fourier mode decay, energy conservation and comparison
  of results}
\label{sec:connections}

In this section we compare the numerical results of the velocity
potential and vortex sheet formulations for the test problems
\eqref{eq:mv:params}.  Since we have taken the single-valued part of
the surface velocity potential to be zero initially, i.e.,
$\tilde\varphi(\alpha,0)=0$, we have to compute the corresponding
initial vortex sheet strength $\gamma_0(\alpha,0)$ to solve an
equivalent problem using the vortex sheet formulation. This is easily
done within the velocity potential code by first computing
$\omega_j(\alpha,0)$ by solving \eqref{eq:B:def} and then evaluating
$\gamma_0(\alpha,0)=-\omega_0'(\alpha,0)$ in
\eqref{eq:gam:def}. Because $V_1$ and possibly $a_2$ are nonzero, this
initial condition $\gamma_0(\alpha,0)$ is nonzero for each of the
three problems \eqref{eq:mv:params}.

\begin{figure}[t]
\centering
\includegraphics[width=\linewidth]{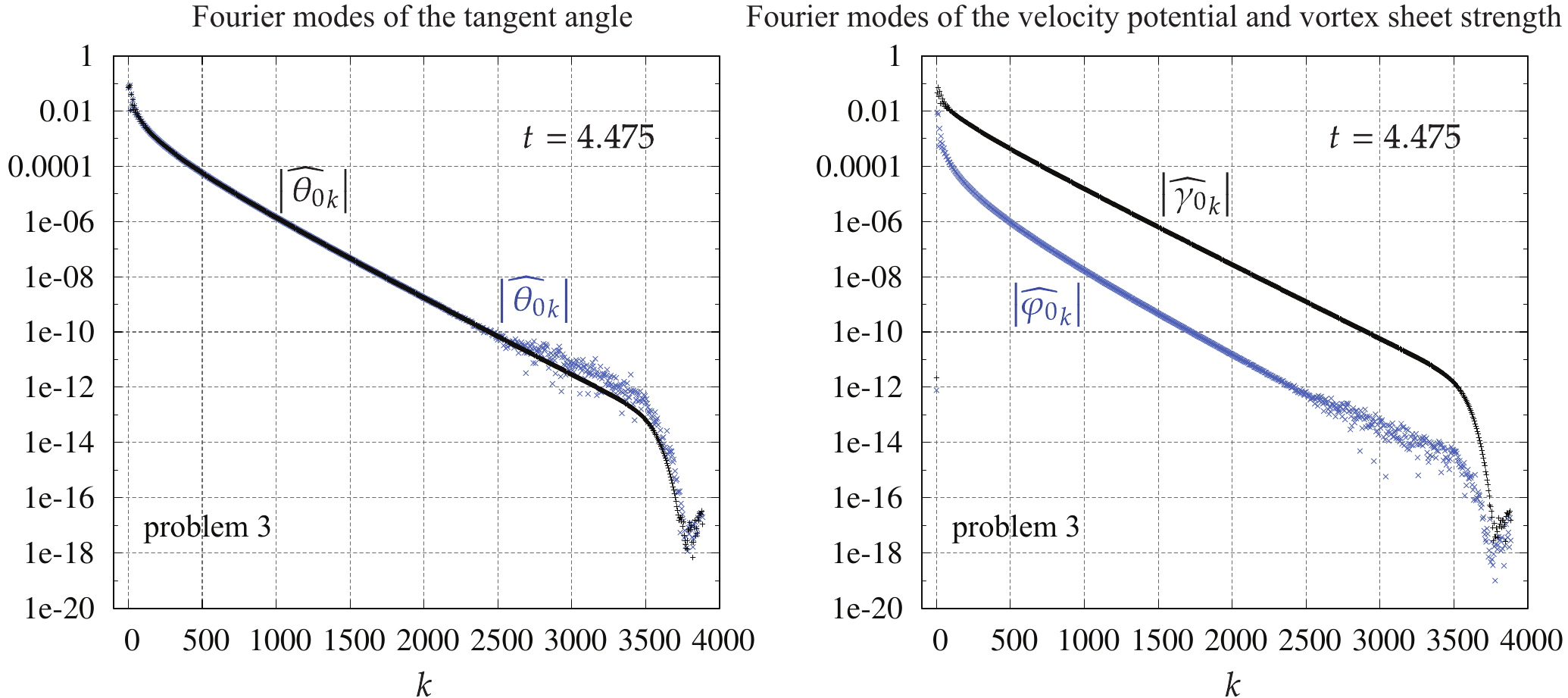}
\caption{\label{fig:fourier} Fourier mode amplitudes of
  $\theta_0(\alpha,t)$ and $\varphi_0(\alpha,t)$ or
  $\gamma_0(\alpha,t)$ for problem 3 of \eqref{eq:mv:params} at
  $t=4.475$, computed using the velocity potential (blue) or vortex
  sheet (black) formulations.}
\end{figure}

Figure~\ref{fig:fourier} shows the Fourier mode
amplitudes of $\theta(\alpha,t)$ and $\varphi_0(\alpha,t)$ or
$\gamma_0(\alpha,t)$ for problem 3 at the final time computed,
$t=4.475$. The results are similar for problems 1 and 2 at $t=2.1$ and
$t=5.575$, respectively, so we omit them.  At $t=4.475$ in problem 3,
there are $M_0=7776$ gridpoints on the free surface, so the Fourier
mode index ranges from $k=0$ to $3888$. We only plotted every fifth
data point (with $k$ divisible by 5) so that individual markers can be
distinguished from one another. The blue and black markers show the
results of the velocity potential and vortex sheet formulations,
respectively.
In both formulations, the Fourier modes decay to $10^{-12}$ before a
rapid drop-off due to the Fourier filter occurs.  Beyond $k=2500$, the
Fourier modes of the velocity potential formulation begin to look
noisy and scattered, which suggests that roundoff errors are having an
effect. This is not seen in the vortex sheet formulation. A possible
explanation is that because $\widehat{\varphi_0}_k$ decays faster than
$\widehat{\gamma_0}_k$, there is some loss of information in storing
$\varphi_0(\alpha,t)$ in double-precision to represent the state of the
system relative to storing $\gamma_0(\alpha,t)$.  Indeed, combining
\eqref{eq:d:phi:mv}, \eqref{eq:d:phi:j} and \eqref{eq:d:phi:0} in the
velocity potential formulation gives the same formula
\eqref{eq:U:from:gam:j} for the normal velocity $U$ in the vortex
sheet formulation, but we have to solve for the $\omega_j$ and then
differentiate these to obtain the $\gamma_j$ before computing $U$ in
the velocity potential formulation.

This is not a complete explanation for the smoother decay of
$\widehat{\gamma_0}_k$ as the right-hand sides of
\eqref{eq:gam:t:final} and \eqref{eq:gam:t:aux}, which govern
$\gamma_{j,t}(\alpha,t)$, contain an extra $\alpha$-derivative
relative to the right-hand side of \eqref{phitileq} for
$\tilde\varphi_t(\alpha,t)$. But it appears that the dispersive nature
of the evolution equations and the Fourier filter suppress roundoff
noise caused by this $\alpha$-derivative. We emphasize that the
smoother decay of Fourier modes in the vortex sheet formulation does
not necessarily mean that these results are more accurate than the
velocity potential approach. The $\alpha$-derivatives in the
right-hand sides of \eqref{eq:gam:t:final} and \eqref{eq:gam:t:aux}
may cause just as much error as arises in computing
$\gamma_j(\alpha,t)$ from $\tilde\varphi_0$, but it is smoothed out
more effectively in Fourier space for the vortex sheet formulation. A
higher-precision numerical implementation would be needed to
investigate the accuracy of each method independently, which is beyond
the scope of the present work.

\begin{figure}[t]
\centering
\includegraphics[width=\linewidth]{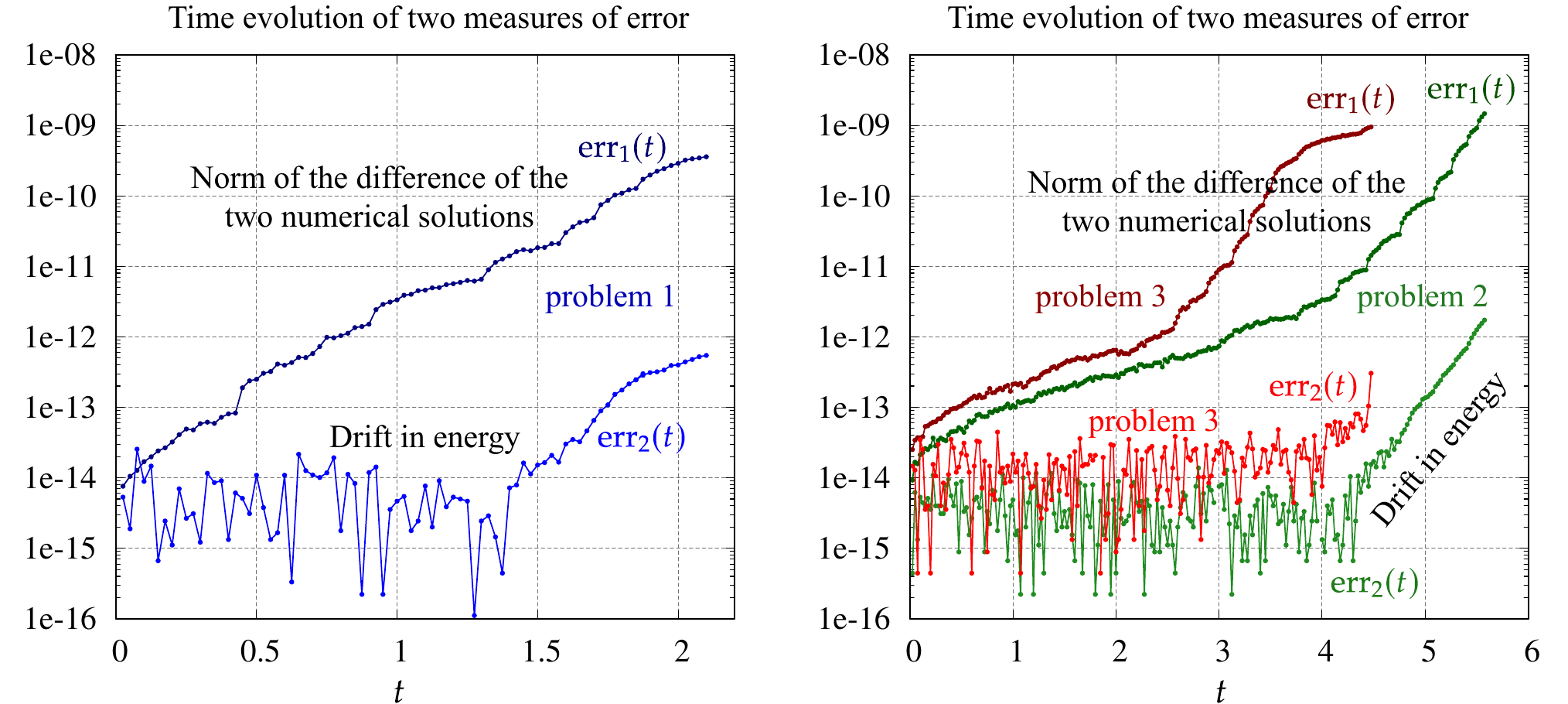}
\caption{\label{fig:error} Time evolution of $\text{err}_1(t)$ and
  $\text{err}_2(t)$ from \eqref{eq:err1} and \eqref{eq:err2} for
  problems 1, 2 and 3 from the initial flat state to the final time
  computed for each problem, just before the splash singularity.}
\end{figure}

In Figure~\ref{fig:error}, we plot the norm of the difference of the
numerical solutions obtained from the velocity potential and vortex
sheet formulations for problem 1 (left) and problems 2 and 3 (right).
Since the tangent angle $\theta(\alpha,t)$ is computed directly in
both formulations, we use
\begin{equation}\label{eq:err1}
  \text{err}_1(t)=\sqrt{\frac1{2\pi}\int_0^{2\pi}|\theta_\text{vp}
    (\alpha,t) -\theta_\text{vs}(\alpha,t)|^2\,d\alpha}
\end{equation}
as a measure of the discrepancy between the two calculations, where vp
and vs refer to `velocity potential' and `vortex sheet.' The results
are plotted in blue, green and red for problems 1, 2 and 3,
respectively. In all three cases, $\text{err}_1(t)$ grows
exponentially in time from the initial flat state to the final time
computed, just before the splash singularity would occur. This
exponential growth is likely due to nearby solutions of the water wave
equations diverging from one another with an exponential growth rate
with this configuration of obstacles, background flow, and circulation
parameters $a_j$. We do not believe the exponential growth is due to
numerical instabilities in the method beyond those associated with
dynamically increasing the number of mesh points, $M_0$, and
timesteps, $d$, per time increment plotted, $\Delta t=0.025$, as
listed in Table~\ref{tbl:M0d}.  At the final time, $\text{err}_1(t)$ has
only grown to around $10^{-9}$ in spite of the rapid change in
$\theta(\alpha,t)$ by $2\pi$ radians over a short range of
$\alpha$ values when traversing the structures resembling Crapper
waves in Figures~\ref{fig:prob1}--\ref{fig:pressure}.

As a second measure of error, we also plot in Figure~\ref{fig:error}
the change in energy from the initial value,
\begin{equation}\label{eq:err2}
  \text{err}_2(t)=E(t)-E(0),
\end{equation}
for all three problems, shown in lighter shades of blue, green and
red. In each numerical calculation, this change in energy remains in
the range $10^{-16}$--$10^{-14}$ for early and intermediate times.
For comparison, the values of $E(0)$ are
\begin{equation}
  \begin{array}{r|ccc}
    \text{problem}\; & 1 & 2 & 3 \\ \hline
    \raisebox{-1pt}{$E(0)$}\; &
    \; \raisebox{-1pt}{$0.79004$} \; &
    \; \raisebox{-1pt}{$1.29626$} \; &
    \; \raisebox{-1pt}{$3.71426$}
  \end{array}
\end{equation}
At later times, $\text{err}_2(t)$ begins to grow exponentially at a
rate similar to that of $\text{err}_1(t)$ but remains 3--4 orders of
magnitude smaller.  Thus, while energy conservation is a necessary
condition for maintaining accuracy, it tends to under-predict the
error of a numerical simulation.

\subsection{Running time and performance}
\label{sec:performance}

The wall-clock times listed in Tables~\ref{tbl:M0d} and~\ref{tbl:M0M3}
above were obtained by running our C++ implementation of the velocity
potential and vortex sheet methods with 24 OpenMP threads on a server
with two 12-core 3.0 GHz Intel Xeon Gold 6136 processors.  The rows
labeled GEPP$(\tilde\varphi)$ or GEPP$(\gamma_0)$ in the tables
correspond to using Gaussian elimination with partial pivoting to
solve \eqref{eq:Awb} in the velocity potential method
or \eqref{eq:lin:sys2}, \eqref{eq:gam:t:final} and
\eqref{eq:gam:t:aux} in the vortex sheet method. The rows
labeled GMRES$(\tilde\varphi)$ or GMRES$(\gamma_0)$ correspond to
using the generalized minimal residual method \cite{golub:book} to
solve these linear systems.

We also implemented a version of the GMRES code in which the matrix
entries of these linear systems are computed and stored on a graphics
processing unit (GPU) and the matrix-vector multiplications of the
GMRES algorithm are performed on the GPU. This approach is effective
as $O(M^2i_\text{max})$ work is done on the GPU for each linear system
solved, with only $O(M\,i_\text{max})$ communication cost between the
CPU and GPU, where
\begin{equation}\label{eq:M:tot}
  M=M_0+M_1+\cdots M_N
\end{equation}
is the size of the linear systems \eqref{eq:Awb},
\eqref{eq:gam:t:final} and \eqref{eq:gam:t:aux} and $i_\text{max}$ is
the number of GMRES iterations required for convergence (typically
  30--120, as discussed below).  This part of the code was written in
Cuda and run on the same server, which has an Nvidia Tesla
P100-PCIE-16GB GPU with 3584 cores running at 1.2 GHz. Less
expensive operations such reconstructing $\zeta(\alpha,t)$
from $P\theta(\alpha,t)$, computing $\alpha$-derivatives,
and applying the Fourier filter \eqref{eq:filter} at the end
of each Runge-Kutta step are still done on the CPU.


\begin{figure}[t]
\centering
\includegraphics[width=\linewidth]{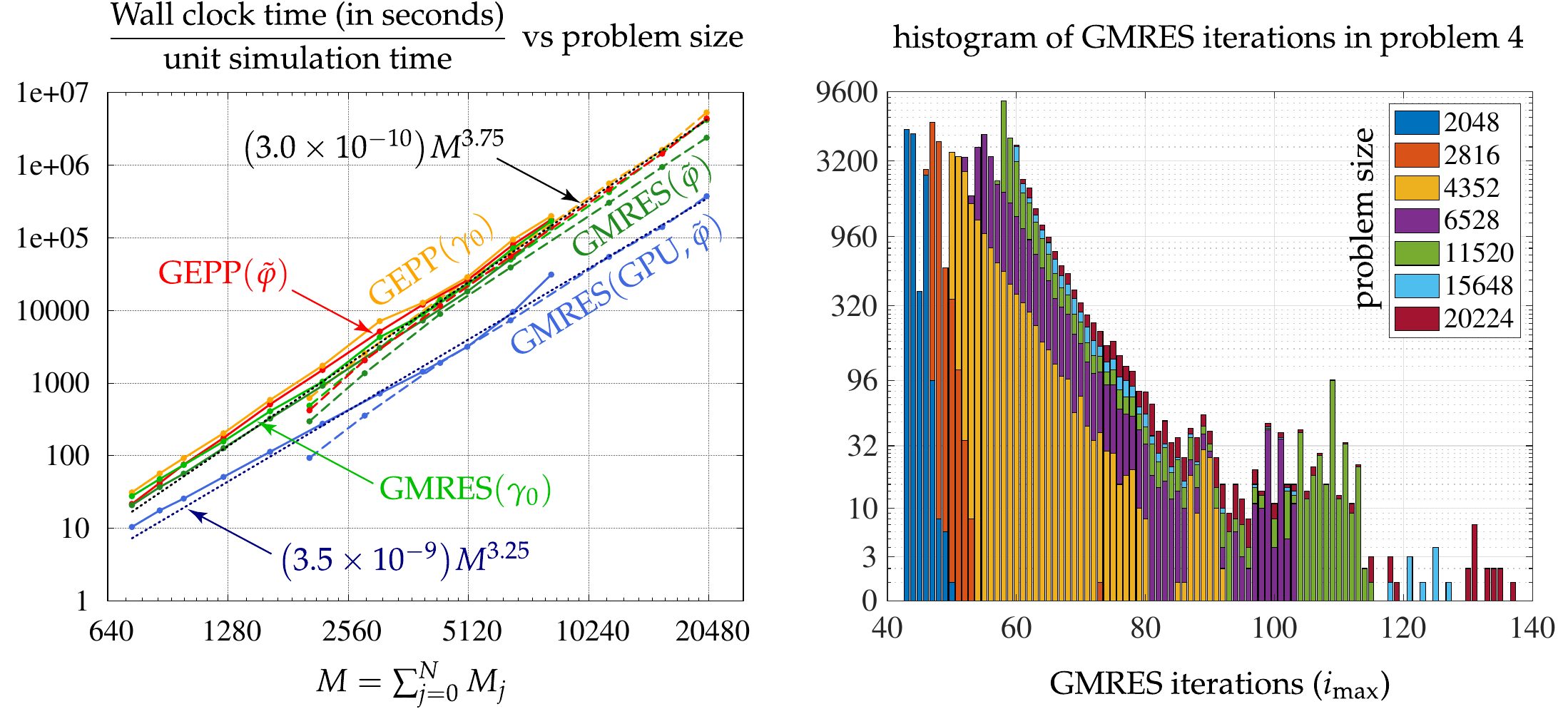}
\caption{\label{fig:performance} Wall-clock running time per unit
  simulation time versus problem size for the velocity potential
  $(\tilde\varphi)$ and vortex sheet $(\gamma_0)$ formulations using
  Gaussian elimination (GEPP) or GMRES to solve the linear systems
  that arise in problems 1--3 and 4, and histogram of the number of
  GMRES iterations required to achieve convergence in problem 4.}
\end{figure}

The performance results are plotted on a log-log plot in the left
panel of Figure~\ref{fig:performance}. Since problem 4 has a different
macro-step size $\Delta t$ from problems 1--3, we divided the
wall-clock running times by $\Delta t$ to obtain the wall-clock time
per unit simulation time as a function of problem size, $M$.  The
bottom 5 rows of Table~\ref{tbl:M0d} are plotted with solid lines that
extend from $M=736$ to $8256$ while the bottom 5 rows of
Table~\ref{tbl:M0M3} are plotted with dashed lines that extend from
$M=2048$ to $20224$.  The orange and light green curves show the GEPP
and GMRES results for the vortex sheet method, respectively, while the
red and dark green curves show the GEPP and GMRES results for the
velocity potential method. The blue curves show the results of the
GPU-accelerated GMRES code for the velocity potential method. We did
not implement the GPU variant of the GMRES algorithm in the vortex
sheet framework.

We find that the wall-clock running time of the GEPP algorithm
scales like $O(M^{3.75})$ at the larger grid sizes used in problem~4.
This is demonstrated with the dotted black line in the left panel
of Figure~\ref{fig:performance}.
%
%
One expects $O(M^{4.5})$ scaling, with a factor of $O(M^3)$ from the
cost of Gaussian elimination and a factor of $d$ from
Tables~\ref{tbl:M0d} and~\ref{tbl:M0M3}, the number of Runge-Kutta
steps taken per macro-step of fixed size $\Delta t$.  A small-scale
decomposition analysis \cite{HLS1,HLS2} shows that the surface tension
terms in \eqref{phitileq} and \eqref{eq:gam:t:final} make the systems
mildly stiff, and $d$ should grow like $O(M_0^{3/2})$ to maintain
stability using an explicit Runge-Kutta method. Since $M_0\ge (3/8)M$
in all cases considered here, this gives an $O(M^{4.5})$ growth rate.
But the linear algebra runs more efficiently on a multi-core CPU for
larger problem sizes, which reduces the wall-clock running time to
$O(M^{3.75})$ in this test problem. This modest reduction does not
change the fact that the method becomes expensive far large grid
sizes. For example, the total running time of the GEPP solver for
problem 4 was 12.4 hours in the velocity potential framework and
14.8 hours in the vortex sheet framework, and 36\% of this time was
devoted in each case to evolving the solution through the final
macro-step via the last two columns of Table~\ref{tbl:M0M3}.

Switching to GMRES eliminates the $O(M^3)$ operations of the
LU-factor\-ization but requires several matrix-vector multiplications
to be performed iteratively, each costing $O(M^2)$ operations. We
denote the number of iterations required to reduce the norm of the
residual to $10^{-15}$ times the norm of the right-hand side by
$i_\text{max}$.  There is still a factor of $d$, which scales like
$O(M^{3/2})$, so the expected running time is
$O(M^{3.5}i_\text{max})$.  As with GEPP on the CPU, the GPU makes more
efficient use of its many cores for larger problem sizes. Thus, even
though $i_\text{max}$ grows somewhat with $M$, as discussed below, we
find that the wall-clock running time scales like $O(M^{3.25})$, shown
by the dotted navy line in the left panel of
Figure~\ref{fig:performance}.  Instead of 12.4 hours, the running time
of problem 4 drops to 1.4 hours using the GPU in the velocity
potential framework, with 28.6\% of the time devoted to the final
macro-step.

Next we investigate how the number of GMRES iterations needed for
convergence, $i_\text{max}$, depends on $M$.  A benefit of
discretizing second-kind integral equations is that the condition
number remains $O(1)$ as the mesh size approaches zero. But in the
current case, the mesh is refined in response to the domain evolving
toward increasingly complicated geometries, which affects the
condition number. The right panel of Figure~\ref{fig:performance}
shows a histogram of the resulting $i_\text{max}$ for each of the
linear systems solved during the course of evolving problem 4 in the
velocity potential framework.  Each column of Table~\ref{tbl:M0M3}
corresponds to a batch of Runge-Kutta steps with a fixed value of
$M$. Each of these Runge-Kutta steps contributes 12 values of
$i_\text{max}$ to the histogram (as it is a 12-stage, 8th order
  scheme), and there are a handful of additional $i_\text{max}$ values
in the histogram from computing the energy at the output times.

We color-code the histogram results by problem size. We see that
colors corresponding to larger problem sizes occupy bins further to
the right in the histogram, which means that as the free surface
evolves to a more complicated state and requires more gridpoints, the
number of GMRES iterations also increases. But the change is not
drastic. Tables~\ref{tbl:imax:3} and~\ref{tbl:imax:4} give the average
value of $i_\text{max}$ for each linear system with a given problem
size $M$ that arises in problems 3 and 4, where problem 3 is
representative of problems 1--3, which have the same grid
parameters. We see in Table~\ref{tbl:imax:4}, for example, in the
velocity potential framework, that increasing $M$ from 2048 to 20224
in problem 4 causes the average value of $i_\text{max}$ to increase
from $43.0$ to only $70.0$.

\begin{table}
\begin{equation*}
  \begin{array}{r|c|c|c|c|c|c|c|c|c|c|c}
    M \;&\; 736 \;&\; 864 \;&\; 992 \;&\; 1248 \;&\; 1632 \;&\;
    2208 \;&\; 3072 \;&\; 3936 \;&\; 5088 \;&\; 6624 \;&\; 8256 \\
    \hline
    \text{restart parameter}\;
    & 50 & 50 & 50 & 50 & 50 & 50 & 50 & 60 & 60 & 70 & 90 \\ \hline
    \text{average } i_\text{max}(\tilde\varphi) \,
    & 31.6 & 34.0 & 34.7 & 35.1 & 36.2 & 37.2 & 39.0 & 40.0  & 41.0
    & 47.2 & 66.5 \\
    \text{average } i_\text{max}(\gamma_0) \,
    & 33.2 & 35.6 & 36.2 & 36.6 & 37.5 & 38.2 & 39.1 & 40.1 & 42.2 &
    47.1 & 67.9 \\ \hline
    \text{\% restarted}(\tilde\varphi) \, & 0 & 0 & 0 & 0 & 0 & 0 & 0 &
    0 & 0 & 0.10 & 0.50 \\
    \text{\% restarted}(\gamma_0)\, &\, 0.017\, &\, 0.092\, & 0 & 0.020 & 0 & 0 &
    0 & 0 & 0.749 & 0 & 3.97 \\ \hline
    \text{avg. } i_\text{max}(\gamma_0;\text{aux}) \,
    & 20 & 20 & 20 & 20 & 20 & 20 & 20 & 20 & 20 & 20 & 20.6 \\
    \text{\% restarted}(\gamma_0;\text{aux}) \, & 0 & 0 & 0 & 0 & 0 & 0 & 0 &
    0 & 0 & 0 & 0
  \end{array}
\end{equation*}
\caption{\label{tbl:imax:3}
  GMRES performance in problem 3.
    The rows labeled $i_\text{max}$ give the average number of GMRES
    iterations required for convergence. Also shown are the
    percentage of cases in which GMRES was restarted due to the
    iteration count reaching the restart parameter. The rows labeled
    ``aux'' refer to the auxiliary linear system \eqref{eq:lin:sys2}
    of the vortex sheet formulation.}
\end{table}


\begin{table}
\begin{equation*}
  \begin{array}{r|c|c|c|c|c|c|c}
    M \, &\; 2048 \;&\; 2816 \;&\; 4352 \;&\; 6528 \;&\;
    11520 \;&\; 15648 \;&\; 20224 \\ \hline
    \text{restart parameter} \,
    & 60 & 70 & 80 & 90 & 100 & 110 & 120 \\ \hline
    \text{average } i_\text{max}(\tilde\varphi) \,
    & 43.0 & 46.9 & 53.2 & 59.1 & 62.7 & 63.6 & 70.0 \\
    \text{average } i_\text{max}(\gamma_0) \, & 63.0 & 52.5 &
    56.4 & 57.2 & 59.7 & 105 & 116 \\ \hline
    \text{\% restarted}(\tilde\varphi) \, & 0 &\, 0.0087\, & 0.54 & 0.68 &
    2.0 & 0.42 & 0.88 \\
    \text{\% restarted}(\gamma_0) \, & 72 & 2.4 & 6.0 & 0.18 & 0.59 &
    63 & 66 \\ \hline
    \text{avg. } i_\text{max}(\gamma_0;\text{aux}) \, &
    26.4 & 26.4 & 26.3 & 26.4 & 27.0 & 27.2 & 27.4 \\
    \text{\% restarted}(\gamma_0;\text{aux}) \, & 0 & 0 & 0 & 0 & 0 & 0 & 0
  \end{array}
\end{equation*}
\caption{\label{tbl:imax:4}
GMRES performance in problem 4. Each column corresponds to a batch
  of Runge-Kutta steps with a fixed spatial discretization. In the first
  column and last two columns, a large fraction of the linear systems
  in the vortex sheet formulation stall on the first GMRES cycle, but
  then converge rapidly after a single restart. This causes the average
  $i_\text{max}$ to be 45--65\% larger in the vortex sheet framework than
  the velocity potential framework in these columns.}
\end{table}

Also listed in Tables~\ref{tbl:imax:3} and~\ref{tbl:imax:4} are the
GMRES restart parameters used, as well as the percentage of linear
systems of each problem size in which the iteration count reached the
restart parameter before the convergence criterion was reached,
triggering a restart. In problem 3, this was rare, occurring less that
1\% of the time in all cases except $M=8256$ with the vortex sheet
method, when it occurred just under 4\% of the time.  But in problem
4, while still rare for the velocity potential method, it was common
for GMRES to stall in the first cycle of iterations and then converge
rapidly after a single restart. As seen in Table~\ref{tbl:imax:4},
this occurred 72\% of the time with $M=2048$, 63\% of the time with
$M=15648$, and 66\% of the time with $M=20224$. In these cases, we
find that increasing the restart parameter has a negative impact on
performance as the extra iterations of the first GMRES cycle are not
as effective at reducing the residual as they would have been had a
restart already occurred. But we also find that reducing the restart
parameter to a small value like 10 or 12 does not work well as the
number of restart cycles can then increase significantly. With the
restart parameters listed in Table~\ref{tbl:imax:3}
and~\ref{tbl:imax:4}, convergence was always reached with at most one
restart in all cases encountered. For background on selecting
optimal restart parameters in GMRES, see \cite{baker:gmres:restart}.

In the vortex sheet approach, one also has to solve the auxiliary
linear system \eqref{eq:lin:sys2} for $\gamma_1$, \dots,
$\gamma_N$. This system only involves discretization points on the
solid boundaries, which do not become more geometrically complicated
as the free surface evolves. We find that the value of $i_\text{max}$
for this auxiliary system is extremely stable, taking on the value of
20 or 21 for all 58671 linear systems that arose in timestepping
problem 3, independent of $M$ in \eqref{eq:M:tot}, and ranging between
26 and 32 in all cases that arose in timestepping problem 4, with the
average value increasing from 26.4 to 27.4 as $M$ changes from 2048 to
20224. There were no instances of a restart occurring for the
auxiliary problem in problems 3 or 4.

%

\begin{table}
  \begin{equation*}
    \begin{array}{r|c|c|c|c||c|c|c|c||c|c|c|c}
      \text{solver}\, & \multicolumn{4}{|c||}{GEPP} &
      \multicolumn{4}{|c||}{GMRES} & \multicolumn{4}{|c}{GMRES(GPU)} \\ \hline
\text{problem}\, & 1 & 2 & 3 & 4 & 1 & 2 & 3 & 4 & 1 & 2 & 3 & 4 \\ \hline
\text{total time}(\tilde\varphi)\, &\,
1.17 \,&\, 7.3 \,&\, 21.3 \,&\, 12.4 \,&\,
0.71 \,&\, 5.3 \,&\, 14.9 \,&\, 8.0 \,&\,
0.157 \,&\, 1.11 \,&\, 2.84 \,&\, 1.41 \\ \hline
\text{\% increase}(\gamma_0)\, &\,
13.1 \,&\, 14.5 \,&\, 15.9 \,&\, 19.0 \,&\,
24 \,&\, 26 \,&\, 29 \,&\, 49 \,&\,
- \,&\, - \,&\, - \,&\, -
\end{array}
  \end{equation*}
  \caption{\label{tbl:tot:run} Total running times (in hours) of
      the velocity potential method on problems 1--4 with the solvers
      implemented and the percentage increase in running time of the
      vortex sheet method.}
\end{table}

We find that the vortex sheet method takes 13--50\% longer to solve
problems 1--4 when implemented using the same solver.
Table~\ref{tbl:tot:run} gives the total running times (in hours) of
the velocity potential method and the percentage increase in running
time of the vortex sheet method.  With Gaussian elimination, the
discrepancy is due to having to solve the auxiliary problem
\eqref{eq:lin:sys2} in addition to the system \eqref{eq:gam:t:final}
and \eqref{eq:gam:t:aux}. The latter system is computationally
equivalent to the system \eqref{eq:Awb} in the velocity potential
method. However, the auxiliary problem is smaller (of size
  $M-M_0=M_1+\cdots+M_N$), adding 13--19\% rather than doubling the
running time in problems 1--4. With GMRES, the performance is somewhat
worse, ranging from 24--29\% slower in problems 1--3 and 49\% slower
in problem 4.  This is partly due to the additional cost of the
auxiliary problem, and also due to a larger average number of GMRES
iterations being needed in the vortex sheet method, especially in
problem 4, as noted in Table~\ref{tbl:imax:4}.

We remark that $d$ can be reduced to $O(M)$ via the HLS small-scale
decomposition \cite{HLS1,HLS2} using an implicit-explicit Runge-Kutta
scheme \cite{carpenter} or exponential time-differencing scheme
\cite{cox:matthews}. The latter has been implemented in
\cite{quasi:ivp}, for example. But these are 4th or 5th order schemes,
so there is a trade-off between the stability constraints of an 8th
order explicit method and the accuracy constraints of an IMEX or ETD
scheme. We did not explore this here, but it would impact the scaling
of running time versus problem size in both the GEPP and GMRES
approaches discussed above.

We also note that GMRES is faster than Gaussian elimination, though
not by as much as one might predict from an operation count.  While
factoring the matrix requires $O(M^3)$ operations, they run at level 3
BLAS speed.  By contrast, GMRES requires $i_\text{max}$ matrix-vector
multiplications to be computed sequentially, each involving $O(M^2)$
operations that run at level 2 BLAS speed. For example, when $M=20224$
in problem 4, the average value of $i_\text{tot}$ was $70$, and the
ratio of flops between GEPP and GMRES is roughly $(2M^3/3)/(70\times
  2M^2)\approx 96$, but the ratio of running times (from the last
  column of Table~\ref{tbl:M0M3}) is only $29184/15925=1.83$. Using
the GPU improves this significantly to $29184/2492=11.7$. In future
work, one could try switching to a block variant of GMRES
\cite{baker:gmres:block}, which might further improve the performance
of the iterative approach by reducing the number of iterations and
enabling much of the work to be done using level 3 BLAS routines. One
could also explore the use of fast algorithms
\cite{rokhlin85,nishimura02} to reduce the $O(M^2)$ cost of forming
the matrices and performing GMRES iterations, but the current approach
is likely to remain competitive since it is easy to parallelize and
could be run on a supercomputer for larger matrix sizes.


\section{Conclusion}
\label{sec:conclude}

We presented two spectrally accurate numerical methods for computing
the evolution of gravity-capillary water waves over obstacles and
variable bottom topography. The methods are closely related, differing
in whether the surface velocity potential or the vortex sheet strength
is evolved on the free surface, along with its position. The kinematic
variable governing the free surface position can be the graph-based
wave height $\eta(x,t)$ or the tangent angle $\theta(\alpha,t)$
introduced by Hou, Lowengrub and Shelley.  In the latter case, we
showed how to modify the curve reconstruction by evolving only the
projection $P\theta$ and using algebraic formulas to determine the
mean value $P_0\theta$ and curve length $L=2\pi s_\alpha$ from
$P\theta$. This prevents $O(\Delta t^2)$ errors in internal
Runge-Kutta stages from causing errors in high-frequency modes that do
not cancel when the stages are combined into a full timestep. The
bottom boundary and obstacles can be parameterized arbitrarily; we do
not assume equal arclength parameterizations.

We derived an energy formula that avoids line integrals over branch
cuts through the fluid by taking advantage of the existence of a
single-valued stream function. This formula does not generalize to 3D,
but also is not necessary in 3D since the velocity potential is
single-valued in that case. We overcame a technical challenge in the
velocity potential method by correcting a nontrivial kernel by
modifying the equations to solve for the stream function values on the
solid boundaries. This issue does not arise in the vortex sheet method
unless the energy is being computed using the velocity potential
approach. We also derived formulas for velocity and pressure in the
fluid that retain spectral accuracy near the boundaries using a
generalization of Helsing and Ojala's method \cite{helsing} to the
periodic case. This method also is limited to 2D as it makes use of
complex analysis through the Cauchy integral formula or the residue
theorem. A different approach will need to be developed in 3D in
future work.

The angle-arclength formulation is convenient for studying overturning
waves, which we demonstrate in a geometry with three elliptical
obstacles and another with two obstacles and a basin-shaped
  bottom boundary. In all cases, a flat initial interface develops
one or more localized indentations that sharpen into
  overhanging wave structures. Often these structures resemble Crapper
  waves with walls that become narrower as time evolves and appear on
  track to terminate with a splash singularity where the curve
self-intersects. In one case, the wave structure appears on
  track to collide with one of the obstacles, with the fluid velocity
  in the gap between the free surface and the obstacle increasing as
  the gap shrinks and as the curvature of the free surface above the gap
  grows. Both methods are demonstrated to be spectrally accurate,
with spatial Fourier modes exhibiting exponential decay. By monitoring
this decay rate, it is easy to adaptively refine the mesh by
increasing the number of gridpoints on the free surface or obstacles
as necessary.

In the test problems in which the free surface eventually
  self-intersects in a splash singularity, we
increased the number of gridpoints $M_0$ on the free surface through the
sequence given in Table~\ref{tbl:M0d}, which ranges from $256$
initially to $7776$ just before the splash singularity. In the
  test problem in which a collision with an obstacle occurs, we
  increased $M_0$ to 16384 at the end. Although we cannot evolve
  all the way to the singularity,
  the solutions remain fully resolved in Fourier space  at
  all times reported, and energy is conserved to 12--15
  digits. In Section~\ref{sec:connections}, we computed several of
  these solutions using both the velocity potential and vortex sheet
  methods and compared them to each other to corroborate the accuracy
  predicted by monitoring energy conservation and the decay of the
  Fourier modes of $\theta(\alpha,t)$, $\tilde\varphi(\alpha,t)$,
  $\gamma_0(\alpha,t)$ and the $\omega_j(\alpha,t)$ or
  $\gamma_j(\alpha,t)$ for $1\le j\le N$ at the output times. While
  energy conservation under-predicts the error, the independent
  calculations agree with each other to at
  least 9 digits of accuracy at the final times computed.

Our assessment is that the velocity potential method is simpler to
derive and somewhat easier to implement since there is only one
``solve'' step required to obtain the $\omega_j$ from $\tilde\varphi$
versus having to solve \eqref{eq:lin:sys2}, \eqref{eq:gam:t:final} and
\eqref{eq:gam:t:aux} for the $\gamma_j$ and $\gamma_{j,t}$ in the
vortex sheet formulation. The vortex sheet formulation was also
  found to require more GMRES iterations to reduce the norm of the
  residual to $10^{-15}$ times that of the right-hand side.  This
  leads to longer running times for the vortex sheet method
  method (by 13--50\%), as seen in Table~\ref{tbl:tot:run}.
The biggest difference we observe in the
numerical results is that high-frequency Fourier modes continue to
decay smoothly in the vortex sheet formulation but are visibly
corrupted by roundoff-error noise in the velocity potential method. We
speculate that there is some loss of information in storing
$\tilde\varphi(\alpha,t)$ in double-precision to represent the state of
the system relative to storing $\gamma_0(\alpha,t)$.
This could also explain why it takes fewer iterations for
  GMRES to drive the residual in $\tilde\varphi$ to zero in comparison
  to $\gamma_0$. To the
extent that one can neglect the compact perturbations of the identity
in \eqref{eq:gam:t:final} and \eqref{eq:gam:t:aux}, the equations are
in conservation form with $\gamma_{j,t}$ equal to the
$\alpha$-derivative of a flux function, which seems to suppress
roundoff error noise in the high-frequency Fourier modes. Additional
work employing higher-precision numerical calculations would be needed
to determine if the smoother decay of Fourier modes in the vortex
sheet approach leads to greater accuracy over the velocity potential
method.

A natural avenue of future research would be to compute steady-state
gravity-capillary waves with background flow over obstacles and study
their stability.  For prediction and design purposes, we are also
interested in comparing our numerical results to laboratory
experiments such as towed obstacles or bottom topographies in a
wavetank.  Of particular interest is the application of this new
numerical technique to experimental observations of air bubbles
permanently trapped between the free surface and the top of a tilted,
submerged airfoil \cite{wu1972cavity}.  Further, we hope to consider
time dependent motion of the obstacles and identify a mechanism for
physically selecting the circulation parameters.

On the numerics side, future goals include
  the development of a non-uniform grid spacing algorithm that can be
  dynamically adjusted to resolve emerging singularities without
  losing spectral accuracy; adapting the technique of evaluating
  Cauchy integrals near boundaries outlined in
  Appendix~\ref{sec:helsing} to model the final stages of a splash
  singularity of the free surface with itself or an obstacle; and
  developing fast-multipole methods \cite{nishimura02} to improve the
  scaling of running time versus problem size observed in
  Section~\ref{sec:performance}.  Generalizing the methods to three
dimensions would also be a useful extension with many
applications.  We include some remarks on the 3D problem in
Appendix~\ref{sec:3d}.

\vspace*{5pt}

{
  \footnotesize
\noindent
\textbf{Funding:} This work was supported in part by the National Science
  Foundation under award numbers DMS-1907684 (DMA), NSF DMS-1352353 \& DMS-1909035 (JLM), DMS-1716560 (JW) and
  DMS-1910824 (RC \& RM); by the Office of Naval Research under award number ONR N00014-18-1-2490 (RC \& RM); and by the Department of Energy, Office of Science,
  Applied Scientific Computing Research, under award number
  DE-AC02-05CH11231 (JW).  JLM wishes to thank the Mathematical Sciences Research Institute for hosting him while a portion of this work was completed.
}

\vspace*{5pt}
{
\footnotesize
\noindent
\textbf{Conflict of interest:} The authors declare no competing interests.
}

\appendix

\section{Verification of the HLS equations}
\label{appendix:HLS}

In Section~\ref{sec:constman}, we proposed evolving only $P\theta$ via
(\ref{eq:Pth:t}) and constructing $P_0\theta$, $s_\alpha$ and
$\zeta(\alpha)$ from $P\theta$ via (\ref{eq:CS:th}) and
(\ref{eq:xi:eta:recon}). Here we show that both equations of
(\ref{eq:theta:t}) hold even though $P_0\theta$ and $s_\alpha$
are computed algebraically rather than by solving ODEs,
and that these equations, in turn, imply that the curve
kinematics are correct, i.e., $(\xi_t,\eta_t)=U\mb n + V\mb t$.

From (\ref{eq:CS:th}), we have $S_t = P_0\left[(\cos
    P\theta)(P\theta)_t\right]$, $C_t = -P_0\left[(\sin
    P\theta)(P\theta)_t\right]$, and
\begin{equation}\label{eq:P0th:t}
  \begin{aligned}
    (P_0\theta)_t &= \frac{-CS_t + SC_t}{C^2+S^2} = -s_\alpha\left[
      (\cos P_0\theta)S_t + (\sin P_0\theta)C_t\right] \\
    &= -\frac{s_\alpha}{2\pi}\int_0^{2\pi}
    \Big[ (\cos P_0\theta)(\cos P\theta) -
      (\sin P_0\theta)(\sin P\theta)\Big](P\theta)_t\,d\alpha \\
    &= -\frac{s_\alpha}{2\pi}\int_0^{2\pi} (\cos\theta)P
    \left(\frac{U_\alpha + V\theta_\alpha}{s_\alpha}\right)d\alpha
    = \frac{1}{2\pi}\int_0^{2\pi}(s_\alpha^{-1}-\cos\theta)(
      U_\alpha + V\theta_\alpha)\,d\alpha.
  \end{aligned}
\end{equation}
In the last step, we used (\ref{eq:P:cos:sin}) and the fact that $P$
is self-adjoint.  Combining (\ref{eq:Pth:t}) and (\ref{eq:P0th:t}), we
obtain
\begin{equation}
  \theta_t = \frac{U_\alpha + V\theta_\alpha}{s_\alpha} - \frac{1}{2\pi}
  \int_0^{2\pi}(\cos\theta)(U_\alpha + V\theta_\alpha)\,d\alpha.
\end{equation}
We must show that the second term is zero.  This follows from
$V_\alpha = P(\theta_\alpha U)$ in (\ref{definitionOfV}). Indeed,
\begin{align*}
  \int (\cos\theta)(V\theta_\alpha)\,d\alpha &=
  \int V\partial_\alpha[\sin\theta]\,d\alpha
  = -\int (\sin\theta) P(\theta_\alpha U)\,d\alpha \\
  &= -\int (\sin\theta)(\theta_\alpha U)\,d\alpha =
  \int U\partial_\alpha[\cos \theta]\,d\alpha
  = -\int (\cos\theta) U_\alpha\,d\alpha,
\end{align*}
where the integrals are from 0 to $2\pi$ and we used
(\ref{eq:P:cos:sin}).  Similarly, we have
\begin{equation}\label{eqn:sig:t:tmp}
  \begin{aligned}
    s_{\alpha t} &= -s_\alpha^3[CC_t + SS_t] =
    -s_\alpha^2\left[(\cos P_0\theta)C_t - (\sin P_0\theta)S_t\right] \\
    &= \frac{s_\alpha^2}{2\pi}\int (\sin\theta)
    P\left(\frac{U_\alpha + V\theta_\alpha}{s_\alpha}\right)\,d\alpha
    = \frac{s_\alpha}{2\pi}\int
    (\sin\theta)(U_\alpha + V\theta_\alpha)\,d\alpha.
\end{aligned}
\end{equation}
Using $V_\alpha = P(\theta_\alpha U)$ again, we find that
\begin{equation*}
  \begin{aligned}
    \int (\sin\theta)(V\theta_\alpha)\,d\alpha &= \int
    -V\partial_\alpha[\cos\theta]\,d\alpha
    = \int (\cos\theta)P[\theta_\alpha U]\,d\alpha \\
    &= \int (\cos\theta-s_\alpha^{-1})(\theta_\alpha U)\,d\alpha =
    \int U\pa_\alpha[\sin\theta]\,d\alpha -
    \frac1{s_\alpha}\int\theta_\alpha U\,d\alpha.
  \end{aligned}
\end{equation*}
Combining this with (\ref{eqn:sig:t:tmp}), we obtain $s_{\alpha t} =
-P_0[\theta_\alpha U]$, as claimed.

As for the second assertion that $(\xi_t,\eta_t)=U\mb n + V\mb t$,
note that the equations of (\ref{eq:theta:t}) are equivalent to
\begin{equation}
  \pa_t \Big[s_\alpha e^{i\theta}\Big] = \pa_\alpha\Big[ (V+iU)e^{i\theta} \Big].
\end{equation}
By equality of mixed partials, the left-hand side equals
$\pa_\alpha\big[\zeta_t\big]$, so we have $\zeta_t =
(V+iU)e^{i\theta}$ up to a constant that could depend on $t$ but not
$\alpha$. However, to enforce $\xi_t(0)=\pa_t0=0$ in (\ref{eq:xi:eta:recon}),
  we choose $V(0)$ in (\ref{eq:V:from:Va}) so that the real part of
$(V+iU)e^{i\theta}$ is zero at $\alpha=0$. We conclude that
$\zeta_t-(V+iU)e^{i\theta}=ia$, where $a$ is real and could depend on
time but not~$\alpha$. We need to show that $a=0$. Note that
\begin{equation}
  2\pi a = \int_0^{2\pi}a\xi_\alpha\,d\alpha =
  \int_0^{2\pi} (0,a)\cdot\mb{\hat n}s_\alpha\,d\alpha =
  \int_\Gamma \big[(\xi_t,\eta_t)\cdot\mb{\hat n} - U \big]ds.
\end{equation}
The divergence theorem implies that $\int_\Gamma U\,ds=0$. This is
because $\nabla\phi$ is single-valued and divergence free in $\Omega$;
$U=\nabla\phi\cdot\mb{\hat n}$ on $\Gamma$; 
$\nabla\phi\cdot\mb{\hat n}=0$ on the solid boundaries; and
  $(\nabla\phi\cdot\mb{\hat n})\vert_{x=2\pi}=-(\nabla\phi\cdot\mb{\hat n})\vert_{x=0}$
  since $\nabla\phi$ is periodic while $\mb{\hat n}$ changes sign.
From (\ref{eq:xi:eta:recon}), $\int_0^{2\pi}\eta\xi_\alpha\,d\alpha=0$
for all time. Differentiating, we obtain
\begin{equation}
  0 = \int_0^{2\pi} [\eta_t\xi_\alpha - \xi_t\eta_\alpha]\,d\alpha
  = \int_\Gamma (\xi_t,\eta_t)\cdot\mb{\hat n}\,ds.
\end{equation}
Thus $a=0$ and $\zeta_t=(V+iU)e^{i\theta}$, as claimed.

\section{Variant Specifying the Stream Function on the Solid Boundaries}
\label{sec:psi:specified}

The integral equations of Section~\ref{sec:int:eq:densities} are
tailored to the case where $V_1$, $a_2$, \dots, $a_N$ in the
representation (\ref{eq:Phi:decomp}) for $\Phi$ are given and the
constant values $\psi\vert_k$ are unknown. If instead $\psi$ is
completely specified on $\Gamma_k$ for $1\le k\le N$, then we would
have to solve for $a_2$, \dots, $a_N$ along with the $\omega_j$. In
this scenario, $\varphi=\phi\vert_{\Gamma_0^-}$ is given on the free
surface, from which we can extract $V_1$ as the change in $\varphi$
over a period divided by $2\pi$.  So we can write
\begin{equation}\label{eq:phi:alt}
  \Phi(z) = \check\Phi(z) + V_1z = \left(\tilde\Phi(z)+
  \sum_{j=2}^N a_j[\omega]\Phi_\text{cyl}(z-z_j)\right) + V_1z,
\end{equation}
where $a_j[\omega] = \la\mb 1_j,\omega\ra=
\frac1{2\pi}\int_0^{2\pi}\omega_j\,d\alpha$ are now functionals that
extract the mean from $\omega_2$, \dots, $\omega_N$. Instead of
(\ref{eq:AB:cor}), we would define
\begin{equation}\label{eq:A:alt}
  \mbb A\omega = \mbb B\omega +\sum_{m=2}^N \begin{pmatrix}
    \phi_\text{cyl}(\zeta_0(\alpha)-z_m) \\
    \psi_\text{cyl}(\zeta_1(\alpha)-z_m) \\
    \vdots \\
    \psi_\text{cyl}(\zeta_N(\alpha)-z_m)\end{pmatrix}
  \la \mb 1_m,\omega\ra.
\end{equation}
The right-hand side $b$ in (\ref{eq:Awb}) would become
$b_0(\alpha)=\big[\varphi(\alpha)-V_1\xi(\alpha)\big]$ and
$b_k(\alpha)=[\psi(\zeta_k(\alpha))-V_1\eta_k(\alpha)]$, where
$\varphi$ and $\psi\vert_{\Gamma_k}$ are given. The latter would
usually be constant functions, though a nonzero flux through the
cylinder boundaries can be specified by allowing
$\psi\vert_{\Gamma_k}$ to depend on~$\alpha$. However, we still
  require $\psi\vert_{\Gamma_k}$ to be
periodic (since the stream function is single-valued in our
  formulation), so the net flux out of each cylinder must be zero.

We now prove invertibility of this version of $\mbb A$, which maps
$\omega$ to the restriction of the real or imaginary parts of
$\check\Phi(z)$ to the boundary.  We refer to these real or imaginary
parts as the ``boundary values'' of $\check\Phi$. In the same way,
$\mbb B$ maps $\omega$ to the boundary values of $\tilde\Phi$ in
(\ref{eq:B:def}).  Note that $\mbb A$ differs from $\mbb B$ by a rank
$N-1$ correction in which a basis for $\mc V=\ker\mbb B$ is mapped to
a basis for the space $\mc R_\text{cyl}$ of boundary values of
$\opn{span}\{\Phi_\text{cyl}(z-z_j)\}_{j=2}^N$. From
Section~\ref{sec:invert:A1}, we know that
$\dim\big(\opn{coker}(\mbb B)\big)=N-1$, so we just have to show that
$\mc R_\text{cyl}\cap\opn{ran}(\mbb B)=\{0\}$. Suppose the boundary
values of $\Phi_c(z)=\sum_{j=2}^N a_j\Phi_\text{cyl}(z-z_j)$ belong to
$\opn{ran}(\mbb B)$. Then there are dipole densities $\omega_j$ such
that the corresponding sum of Cauchy integrals
$\tilde\Phi(z)=\sum_{j=0}^N\Phi_j(z)$ has these same boundary
values. The imaginary part, $\tilde\psi$, satisfies the Laplace
equation in $\Omega$, has the same Dirichlet data as $\psi_c$ on
$\Gamma_1,$ \dots, $\Gamma_N$, and the same Neumann data as $\psi_c$
on $\Gamma_0$ (due to $\pa_n\psi=\pa_s\phi$).  Since solutions are
unique, $\tilde\psi=\psi_c$.  But the conjugate harmonic function to
$\tilde\psi$ is single-valued while that of $\psi_c$ is
multiple-valued unless all the $a_j=0$. We conclude that $\mc
R_\text{cyl}\cap\opn{ran}(\mbb B)=\{0\}$, as claimed.

\section{Cauchy Integrals, Layer Potentials and Sums Over Periodic Images}
\label{sec:cauchy:laypot}

In this section we consider the connection between Cauchy integrals
and layer potentials and the effect of summing over periodic images
and renormalization. As is well-known
\cite{singularIntegralEquations}, Cauchy integrals are closely related
to single and double layer potentials through the identity
\begin{equation}
  \frac{d\zeta}{\zeta-z} = d\,\log(\zeta-z) =
  d\,\log r + i\,d\theta = \frac{dr}r + i\,d\theta,
\end{equation}
where $\zeta-z=re^{i\theta}$.  We adopt the sign convention of
electrostatics \cite{kress,jackson} and define the Newtonian potential
as $N(\zeta,z)=-(2\pi)^{-1}\log|\zeta-z|$.  The double-layer potential
(with normal $\mb{n}_\zeta$ pointing left from the curve $\zeta$, as
  in Section~\ref{sec:cauchy} above) has the geometric interpretation
\begin{equation}\label{eq:dNdz:dthds}
  \der{N}{n_\zeta} = \nabla_\zeta N(\zeta,z)\cdot\mb{n}_\zeta =
  \frac1{2\pi}\frac{(x-\xi,y-\eta)}{(x-\xi)^2+(y-\eta)^2}\cdot
  \frac{(-\eta_\alpha,\xi_\alpha)}{(\xi_\alpha^2+\eta_\alpha^2)^{1/2}} =
  \frac1{2\pi}\frac{d\theta}{ds}.
\end{equation}
For a closed contour in the complex plane, we have
\begin{equation}
  \frac1{2\pi i}\int_\Gamma \frac{\omega(\zeta)}{\zeta-z}\,d\zeta =
  \int_\Gamma \der{N}{n_\zeta}\omega(\zeta)\,ds +
  i\int_\Gamma N(\zeta,z)\Big(-\frac{d\omega}{ds}\Big)ds,
\end{equation}
so, if $\omega$ is real-valued, the real part of a Cauchy integral is
a double-layer potential with dipole density $\omega$ while the
imaginary part is a single-layer potential with charge density
$-d\omega/ds$.  In the spatially periodic setting, the real
part of the two formulas in (\ref{eq:w123}) may be written
\begin{equation}\label{eq:double:layer}
  \begin{aligned}
  \phi_0(z) &= \frac1{2\pi}\int_0^{2\pi} \im\left\{
  \frac{\zeta'(\alpha)}2\cot\frac{\zeta(\alpha) - z}2\right\}
  \omega_0(\alpha)\,d\alpha \\
  &= \lim_{M\rightarrow\infty}\sum_{m=-M}^M \frac1{2\pi}\int_0^{2\pi}
  \im\left\{\frac{\zeta'(\alpha)}{\zeta(\alpha)+2\pi m - z}\right\}
  \omega_0(\alpha)\,d\alpha \\
  &= \lim_{M\rightarrow\infty}\sum_{m=-M}^M \int_0^{2\pi}
  \der{N}{n_\zeta}\big(\zeta(\alpha)+2\pi m,z\big)\,
  \omega_0(\alpha)s_\alpha\,d\alpha  \\
  &= PV\int_{-\infty}^\infty \der{N}{n_\zeta}(\zeta_j(\alpha),z)\,
  \omega_0(\alpha)s_\alpha\,d\alpha
  \end{aligned}
\end{equation}
and
\begin{alignat}{2}
  \notag
  \phi_j(z) &= \frac1{2\pi} \int_0^{2\pi} \re\left\{
  \frac{\zeta_j'(\alpha)}2\cot\frac{\zeta_j(\alpha) - z}2\right\}
  \omega_j(\alpha)\,d\alpha \\
  \notag
  &= \frac1{2\pi}\int_0^{2\pi} -\log\left|\sin\frac{
    \zeta_j(\alpha)-z}2\right|
  \omega_j'(\alpha)\,d\alpha \\
  \label{eq:single:layer1}
  &= \lim_{M\rightarrow\infty}\sum_{m=-M}^M
  \int_0^{2\pi} N(\zeta_j(\alpha)+2\pi m,z)\,\omega_j'(\alpha)\,d\alpha
  &\quad &(1\le j\le N) \\
  \label{eq:single:layer2}
  &= \lim_{M\rightarrow\infty}\int_{-2\pi M}^{2\pi(M+1)}
  N(\zeta_j(\alpha),z)\,\omega_j'(\alpha)\,d\alpha, &
  \quad &(j=1 \text{ only}).
\end{alignat}
Equation (\ref{eq:single:layer1}) follows from
Euler's product formula $\sin w =
w\prod_{m=1}^\infty\big(1-(w/m\pi)^2\big)$, which gives
\begin{equation}\label{eq:log:sin:w2}
  -\frac1{2\pi}\log\left|\sin\frac{\zeta_1(\alpha)-z}2\right|=
  \lim_{M\to\infty}\sum_{m=-M}^M \bigg( N\big(
      \zeta_1(\alpha)+2\pi m,z\big) - c_m \bigg),
\end{equation}
where $c_0=-\frac1{2\pi}\log 2$ and $c_m=-\frac1{2\pi}\log|2\pi m|$ if
$m\ne0$. It was possible to drop the terms $c_m$ in (\ref{eq:single:layer1})
and (\ref{eq:single:layer2}) since $\omega_j'(\alpha)$ is integrated
over a period of $\omega_j(\alpha)$. However, these terms have to be
retained to express (\ref{eq:single:layer2}) as a principal value
integral,
\begin{equation}
  \phi_1(z) = PV\int_{-\infty}^\infty N_1(\alpha,z)\,\omega_1'(\alpha)
  \,d\alpha, \quad
  \left(\begin{gathered}
    N_1(\alpha,z) = N\big(\zeta_1(\alpha),z\big) - c_m \\
    2\pi m\le \alpha<2\pi(m+1)
    \end{gathered}\right)\!.
\end{equation}
Through (\ref{eq:log:sin:w2}), we can regard $\log|\sin(w/2)|$ as a
renormalization of the divergent sum of the Newtonian potential over
periodic images in 2D.  Setting aside these technical issues, it is
conceptually helpful to be able to interpret $\phi_0(z)$ and
$\phi_j(z)$ from (\ref{eq:w123}) as double and single layer potentials
with dipole and charge densities $\omega_0(\alpha)$ and
$\omega_j'(\alpha)/s_\alpha$, respectively, over the real line or over
the periodic array of obstacles.  Of course, it is more practical in
2D to work directly with the formulas involving complex cotangents
over a single period, but (\ref{eq:double:layer}) and
(\ref{eq:single:layer1}) are a useful starting point for
generalization to 3D.

\section{Alternative Derivation of the Vortex Sheet Strength Equation}
\label{appendix:gam:t}

In this appendix, we present an alternative derivation of
(\ref{eq:gam:t:final}) that makes contact with results reported
elsewhere \cite{ambroseMasmoudi1,BMO} in the absence of solid
boundaries.  As in Section~\ref{sec:cauchy}, the velocity potential is
decomposed into $\phi(z)=\tilde\phi(z) + \phi_\text{mv}(z)$ where
$\tilde\phi(z)$ is the sum of layer potentials and $\phi_\text{mv}(z)$
is the multi-valued part. We also define $\mb W$ as in
(\ref{eq:W:tot}), where the component Birkhoff-Rott integrals $\mb
W_{0j}$ are given in complex form by
\begin{equation}\label{eq:BR:def}
  \begin{aligned}
    W_{k0}^*(\alpha) &=
    \frac{1}{2\pi i}PV\!\!\int_0^{2\pi}
    \frac12\cot\frac{\zeta_k(\alpha)-\zeta_0(\beta)}2
    \gamma_0(\beta)\,d\beta,\\
    W_{kj}^*(\alpha) &=
    -\frac1{2\pi i}PV\!\!\int_0^{2\pi}
    \frac12\cot\frac{\zeta_k(\alpha)-\zeta_j(\beta)}2
    i\gamma_j(\beta)\,d\beta.
  \end{aligned}
\end{equation}
The Plemelj formulas (\ref{eq:u:on:bdries}) imply that when the
interface is approached from the fluid region,
\begin{equation}\label{fromPlemelj}
  \nabla\phi = \mb{W} +\frac{\gamma_0}{2s_{\alpha}}\mathbf{\hat{t}}.
\end{equation}
Recall that $\varphi(\alpha,t)=\phi(\zeta(\alpha,t),t)$ is the
restriction of the velocity potential to the free surface as it
evolves in time, and note that
$\varphi_{\alpha}=s_{\alpha}\nabla\phi\cdot\mathbf{\hat{t}}.$ Solving
for $\gamma_0,$ then, we have
\begin{equation}\label{eq:gam:from:phi:W}
  \gamma_0=2\varphi_{\alpha}-2s_{\alpha}\mb{W} \cdot\mathbf{\hat{t}}.
\end{equation}
Differentiating with respect to time, we get
\begin{equation}\nonumber
  \gamma_{0,t}=2\varphi_{\alpha t}-2s_{\alpha t}\mb{W} \cdot\mathbf{\hat{t}}
  -2s_{\alpha}\mb{W} _{t}\cdot\mathbf{\hat{t}}
  -2s_{\alpha}\mb{W} \cdot\mathbf{\hat{t}}_{t}.
\end{equation}
In Section~\ref{sec:evol:gam}, we avoided directly taking time
derivatives of $\gamma_0(\alpha,t)$, $\mb W(\alpha,t)$ and
$\varphi(\alpha,t)$, which lead to more involved calculations here due
to the moving boundary.  We know that
$\mathbf{\hat{t}}_{t}=\theta_{t}\mathbf{\hat{n}},$ and that
$\theta_{t}=(U_{\alpha}+V\theta_{\alpha})/s_{\alpha}.$ We substitute
these to obtain
\begin{equation}\label{gammaT1}
  \gamma_{0,t}=2\varphi_{\alpha t}-2s_{\alpha t}\mb{W} \cdot\mathbf{\hat{t}}
  -2s_{\alpha}\mb{W} _{t}\cdot\mathbf{\hat{t}}
  -2U(U_{\alpha}+V\theta_{\alpha}).
\end{equation}
We now work on the equation for $\varphi_{\alpha t}.$  As was done in
\cite{ambroseMasmoudi1}, the convective derivative (\ref{eq:convect:phi})
together with the Bernoulli equation gives
\begin{equation}\label{bernoulli1}
  \varphi_{t}=\nabla\phi\cdot(U\mathbf{\hat{n}}+V\mathbf{\hat{t}})
  -\frac{1}{2}|\nabla\phi|^{2}-\frac p\rho - g \eta_0.
\end{equation}
We write $\mb{W} =U\mathbf{\hat{n}}+(\mb{W}
  \cdot\mathbf{\hat{t}})\mathbf{\hat{t}}$, substitute
\eqref{fromPlemelj} into \eqref{bernoulli1}, and use $\mb{W}
\cdot\mb{W} =U^{2}+(\mb{W} \cdot\mathbf{\hat{t}})^{2}$:
\begin{equation}\nonumber
  \varphi_{t}=U^{2}+V(\mb{W} \cdot\mathbf{\hat{t}})+
  \frac{\gamma_0 V}{2s_{\alpha}}
  -\frac{1}{2}\big(U^2 + (\mb{W} \cdot\mathbf{\hat{t}})^{2}\big)
  -\frac{\gamma_0}{2s_{\alpha}}(\mb{W} \cdot\mathbf{\hat{t}})
  -\frac{\gamma_0^{2}}{8s_{\alpha}^{2}} - \frac p\rho - g \eta_0.
\end{equation}
We differentiate with respect to $\alpha$:
\begin{equation}\label{phiAlphaT}
\begin{aligned}
  \varphi_{\alpha t}&=UU_{\alpha}
  +V_{\alpha}(\mb{W} \cdot\mathbf{\hat{t}})
   +V(\mb{W} \cdot\mathbf{\hat{t}})_{\alpha}
   +\left(\frac{\gamma_0 V}{2s_{\alpha}}\right)_\alpha \\
  &\qquad -(\mb{W} \cdot\mathbf{\hat{t}})
  (\mb{W} \cdot\mathbf{\hat{t}})_{\alpha}
   -\left(\frac{(\mb W\cdot\mb{\hat t})\gamma_0}{2s_{\alpha}}\right)_{\alpha}
   -\left(\frac{\gamma_0^2}{8s_{\alpha}^{2}}\right)_\alpha -
   \frac{p_{\alpha}}\rho - g \eta_{0,\alpha}.
\end{aligned}
\end{equation}
We substitute \eqref{phiAlphaT} into \eqref{gammaT1}, noticing that
the $UU_{\alpha}$ terms cancel:
\begin{multline}\nonumber
  \gamma_{0,t}=2V_{\alpha}(\mb{W} \cdot\mathbf{\hat{t}})
  +2V(\mb{W} \cdot\mathbf{\hat{t}})_{\alpha}+
  \left(\frac{\gamma_0 V}{s_{\alpha}}\right)_\alpha
  -2(\mb{W} \cdot\mathbf{\hat{t}})(\mb{W} \cdot\mathbf{\hat{t}})_{\alpha}
  -\left(\frac{(\mb{W} \cdot\mathbf{\hat{t}})\gamma_0}{s_{\alpha}}
    \right)_{\alpha}\\
  -\left(\frac{\gamma_0^2}{4s_{\alpha}^{2}}\right)_\alpha
  -2\frac{p_{\alpha}}\rho - 2g \eta_{0,\alpha}
  -2s_{\alpha t}\mb{W} \cdot\mathbf{\hat{t}}
  -2s_{\alpha}\mb{W} _{t}\cdot\mathbf{\hat{t}}
  -2UV\theta_{\alpha}.
\end{multline}
We group this as follows:
\begin{multline}\nonumber
  \gamma_{0,t}=-2\frac{p_{\alpha}}\rho +
  \left(\frac{(V-\mb{W} \cdot\mathbf{\hat{t}})\gamma_0}{s_\alpha}
    \right)_{\alpha}
  -2s_{\alpha}\mb{W} _{t}\cdot\mathbf{\hat{t}}-
  \left(\frac{\gamma_0^2}{4s_{\alpha}^{2}}\right)_\alpha
  -2g \eta_{0,\alpha}
  \\
  +\left[2V_{\alpha}(\mb{W} \cdot\mathbf{\hat{t}})+2V(\mb{W}
      \cdot\mathbf{\hat{t}})_{\alpha}
    -2(\mb{W} \cdot\mathbf{\hat{t}})(\mb{W} \cdot\mathbf{\hat{t}})_{\alpha}
    -2s_{\alpha t}\mb{W} \cdot\mathbf{\hat{t}}-2UV\theta_{\alpha}
    \right].
\end{multline}
The quantity in square brackets simplifies considerably using the
equations $V_{\alpha}=s_{\alpha t}+\theta_{\alpha}U,$
$U=\mb{W} \cdot\mathbf{\hat{n}},$ and
$\mathbf{\hat{t}}_{\alpha}=\theta_{\alpha}\mathbf{\hat{n}}.$ Together
with the boundary condition for the pressure (the
  Laplace-Young condition), we obtain
\begin{equation}\label{eq:gam:t:ver1}
  \gamma_{0,t} =
  \left(2 \tau\frac{\theta_\alpha}{s_{\alpha}}
    + \frac{(V-\mb{W} \cdot\mathbf{\hat{t}})\gamma_0}{s_{\alpha}}
    - \frac{\gamma_0^2}{4s_{\alpha}^{2}} - 2g\eta_0\right)_\alpha
  - 2s_{\alpha}\mb{W} _{t}\cdot\mathbf{\hat{t}} 
  +2(V-\mb{W} \cdot\mathbf{\hat{t}})(\mb{W} _{\alpha}\cdot\mathbf{\hat{t}}).
\end{equation}
This agrees with the equation for $\gamma_{0,t}$ as found in
\cite{ambroseMasmoudi1} if one assumes $(s_\alpha)_\alpha=0$.
The calculation of \cite{ambroseMasmoudi1} has no solid
boundaries and a second fluid above the first, which we
take to have zero density when comparing to (\ref{eq:gam:t:ver1}).

Our final task is to compute $s_\alpha\mb W_t\cdot\mb{\hat t} =
(\mb W_{00,t} + \cdots + \mb W_{0N,t})\cdot(s_\alpha \mb{\hat t})$
in the right-hand side of (\ref{eq:gam:t:ver1}). Differentiating
(\ref{eq:BR:def}) with respect to time for $1\le j\le N$ gives
\begin{equation*}
  W_{0j,t}^*(\alpha,t) = -\frac1{2\pi}\int_0^{2\pi}\frac12\cot\frac{
    \zeta(\alpha,t)-\zeta_j(\beta)}2\gamma_{j,t}(\beta,t)\,d\beta
  + \frac{\zeta_t(\alpha,t)}{\zeta'(\alpha,t)} W_{0j,\alpha}^*(\alpha,t).
\end{equation*}
Here, as above, a prime denotes $\pa_\alpha$ and we note that the
solid boundaries remain stationary in time. Suppressing $t$ in the
arguments of functions again, we conclude that for $1\le j\le N$,
\begin{equation}\label{eq:saW0jt}
  s_\alpha \mb W_{0j,t}\cdot\mb{\hat t} = \re\{\zeta'(\alpha)
  W_{0j,t}^*(\alpha)\} =
  -\frac1{2\pi}\int_0^{2\pi} G_{j0}(\beta,\alpha)
  \gamma_{j,t}(\beta)\,d\beta +
  \zeta_t\cdot \mb W_{0j,\alpha},
\end{equation}
where $\zeta_t$ is treated as the vector $(\xi_t,\eta_t)$ in the dot
product. When $j=0$, we regularize the integral
\begin{equation*}
  \zeta'(\alpha)W_{00}^*(\alpha) = -\frac i2\mbb H\gamma_0(\alpha)
  + \frac1{2\pi i}
  \int_0^{2\pi} \left[\frac{\zeta'(\alpha)}2\cot\frac{\zeta(\alpha)-
      \zeta(\beta)}2 - \frac12\cot\frac{\alpha-\beta}2\right]
  \gamma_0(\beta)\,d\beta
\end{equation*}
and then differentiate both sides with respect to time
\begin{equation*}
  \begin{aligned}
    \zeta_t'(\alpha)W_{00}^*(\alpha) &+ \zeta'(\alpha)W_{00,t}^*(\alpha) = \\
    & -\frac i2\mbb H\gamma_{0,t}(\alpha) + \frac1{2\pi i}\int_0^{2\pi}
    \left[\frac{\zeta'(\alpha)}2\cot\frac{\zeta(\alpha)-
        \zeta(\beta)}2 - \frac12\cot\frac{\alpha-\beta}2\right]
    \gamma_{0,t}(\beta)\,d\beta \\
    & \qquad + \frac1{2\pi i}\int_0^{2\pi}\left(\pa_t\left[
        \frac{\zeta'(\alpha)}2\cot\frac{\zeta(\alpha)-
        \zeta(\beta)}2\right]\right)\gamma_0(\beta)\,d\beta.
  \end{aligned}
\end{equation*}
Observing that
$\zeta_t'W_{00}^*=\big([\zeta_tW_{00}^*]_\alpha-\zeta_tW_{00,\alpha}^*\big)$,
we find that
\begin{equation}\label{eq:saW00t}
  \begin{aligned}
    s_\alpha\mb W_{00,t}\cdot\mb{\hat t} &=
    \re\{\zeta'(\alpha)W_{00,t}^*(\alpha)\} \\
    &= -\big(\zeta_t\cdot\mb W_{00}\big)_\alpha +
    \zeta_t \cdot \mb W_{00,\alpha} +
    \mbb K_{00}^*\gamma_{0,t}(\alpha) \\
    & \qquad + \re\left\{\frac1{2\pi i}\int_0^{2\pi}
    \left(\pa_\alpha\left[
        \frac{\zeta_t(\alpha)-\zeta_t(\beta)}2
        \cot\frac{\zeta(\alpha)-\zeta(\beta)}2\right]\right)
    \gamma_0(\beta)\,d\beta\right\}.
  \end{aligned}
\end{equation}
Finally, setting $\mb
W_\text{mv}=\nabla\phi_\text{mv}(\zeta(\alpha,t))$, we compute
\begin{equation}\label{eq:saWmv}
  \begin{aligned}
    s_\alpha\mb W_{\text{mv},t}\cdot\mb{\hat t} &=
    \re\{\zeta'(\alpha)W_{\text{mv},t}^*\} =
    \re\{\zeta'(\alpha)\pa_t\Phi'_\text{mv}(\zeta(\alpha,t))\}
    = \re\{\zeta'\Phi''_\text{mv}\zeta_t\} \\
    &= \re\{\zeta_t\pa_\alpha \Phi'_\text{mv}(\zeta(\alpha,t))\}
    = \re\{\zeta_t W_{\text{mv},\alpha}^*\}
    = \zeta_t\cdot\mb W_{\text{mv},\alpha}.
    \end{aligned}
\end{equation}
When (\ref{eq:saW0jt}), (\ref{eq:saW00t}) and (\ref{eq:saWmv}) are
combined and substituted into (\ref{eq:gam:t:ver1}), several of the
terms cancel:
\begin{align*}
  -2\sum_{j=0}^N \zeta_t\cdot\mb W_{0j,\alpha} &
  -2\zeta_t\cdot\mb W_{\text{mv},\alpha} +
  2(V-\mb W\cdot\mb{\hat t})(\mb W_\alpha\cdot\mb{\hat t}) \\[-10pt]
  &= -2(U\mb{\hat n} + V\mb{\hat t})\cdot\mb W_\alpha +
  2(V-\mb W\cdot\mb{\hat t})(\mb W_\alpha\cdot\mb{\hat t}) \\
  &= -2(U\mb{\hat n} + (W\cdot\mb{\hat t})\mb{\hat t})\cdot \mb W_\alpha
  = -2\mb W\cdot\mb W_\alpha = -(\mb W\cdot\mb W)_\alpha.
\end{align*}
Also, in (\ref{eq:saW00t}), $\zeta_t\cdot\mb W_{00}$ cancels the
$\zeta_t(\alpha)$ term in the integrand, leaving behind a principal
value integral.  Including the other terms of (\ref{eq:gam:t:ver1}),
moving the unknowns to the left-hand side, and dividing by 2, we
obtain (\ref{eq:gam:t:final}).

\section{Treating the Bottom Boundary as an Obstacle}
\label{sec:bot:hole}

The conformal map $w=e^{-iz}$ maps the infinite, $2\pi$-periodic
region $\Omega_1'$ below the bottom boundary to a finite domain, with
$-i\infty$ mapped to zero. Let $w_j=e^{-iz_j}$ denote the images of
the points $z_j$ in \eqref{eq:Phi:mv}, which are used to represent
flow around the obstacles via multi-valued velocity potentials.  We
also define the curves
\begin{equation}\label{eq:upsilon:def}
  \Upsilon_j(\alpha)=e^{-i\zeta_j(\alpha)}, \qquad\quad (0\le j\le N),
\end{equation}
which traverse closed loops in the $w$-plane, parameterized
clockwise. The image of the fluid region lies to the right of
$\Upsilon_0(\alpha)$ and to the left of $\Upsilon_j(\alpha)$ for $1\le
j\le N$.  The terms $V_1z$ and $a_j\Phi_\text{cyl}(z-z_j)$ appearing
in \eqref{eq:Phi:mv} all have a similar form in the new variables,
\begin{equation}
  V_1z(w) = V_1i\log w, \qquad
  a_j\Phi_\text{cyl}(z(w)-z_j) = a_j \big(i\log w - i\log(w-w_j)\big).
\end{equation}
We can think of $V_1z$ as a multiple-valued complex potential on the
$2\pi$-periodic domain of logarithmic type with center at
$z_1=-i\infty$. It maps to $V_1i\log(w-w_1)$ in the $w$-plane, where
$w_1=0$. From \eqref{eq:phi:up}, we see that the $n$th sheet of the
Riemann surface for $\Phi_\text{cyl}(z(w)-z_j)$ is given by
$-i\opn{Log}(1-w_j/w)+2\pi n$, which has a branch cut from the origin
to $w_j$. When traversing the curve $w=\Upsilon_k(\alpha)$ with
$\alpha$ increasing, the function $\Phi_\text{cyl}(z(w)-z_j)$
decreases by $2\pi$ if $k=j$, increases by $2\pi$ if $k=1$, and
returns to its starting value for the other boundaries, including the
image of the free surface ($k=0$). This is done so that only the
$V_1z$ term has a multiple-valued real part on $\Gamma_0$, which
simplifies the linear systems analyzed in
Section~\ref{sec:invert:A1}--\ref{sec:invert:E} above.

The cotangent-based Cauchy integrals $\Phi_j(z)$ in (\ref{eq:w123})
transform to $(1/w)$-based Cauchy integrals in the new variables,
aside from an additive constant in the kernels \cite{kress}. In more
detail,
\begin{equation}
  \frac{d\Upsilon_j}{\Upsilon_j-w} =
  \frac{-ie^{-i\zeta_j}\,d\zeta_j}{e^{-i\zeta_j}-e^{-iz}} =
  \frac{ie^{-i(\zeta_j-z)/2}\,d\zeta_j}{e^{i(\zeta_j-z)/2}-e^{-i(\zeta_j-z)/2}} =
  \left(\frac12\cot\frac{\zeta_j-z}2-\frac i2\right)d\zeta_j.
\end{equation}
For $1\le j\le N$, we then have
\begin{equation}
  \Phi_j(z(w))
  = \frac1{2\pi i}\int_0^{2\pi}\frac{i\omega_j(\alpha)}{
    \Upsilon_j(\alpha)-w}
  \Upsilon_j'(\alpha)\,d\alpha + \frac1{2\pi i}\int_0^{2\pi}
  \left(\frac i2\right)
  i\omega_j(\alpha)\,\zeta_j'(\alpha)\,d\alpha,
\end{equation}
with a similar formula for $\Phi_0(z(w))$, replacing
$i\omega_j(\alpha)$ by $\omega_0(\alpha)$. The second term is a
constant function of $w$ that prevents $\mb 1_1$ from being
annihilated by $\mbb B$ in Section~\ref{sec:int:eq:densities}. This is
the primary way in which the bottom boundary differs from the other
obstacles in the analysis of
Sections~\ref{sec:invert:A1}--\ref{sec:invert:E}.

We note that $\tilde\Phi(z(w))=\sum_{j=0}^N\Phi_j(z(w))$ is analytic
at $w=0$, which allows us to conclude that if its real or imaginary
part satisfies Dirichlet conditions on $\Gamma_1^-$, it is zero in
$\Omega_1'$.  A similar argument using $w=e^{iz}$ works for the region
$\Omega_0'$ above the free surface, which was needed in
Section~\ref{sec:invert:E} above.

\section{Evaluation of Cauchy Integrals Near Boundaries}
\label{sec:helsing}

In this section we describe an idea of Helsing and Ojala
\cite{helsing} to evaluate Cauchy integrals with spectral accuracy
even if the evaluation point is close to the boundary. We modify the
derivation to the case of a $2\pi$-periodic domain, which means the
$\frac1z$ Cauchy kernels in \cite{helsing} are replaced by
$\frac12\cot\frac z2$ kernels here. The key idea is to first compute
the boundary values of the desired Cauchy integral $f(z)$. The
interior values are expressed in terms of these boundary values. From
the residue theorem, we have
\begin{equation}\label{eq:res:thm}
  f(z) = \frac1{2\pi i}\int_{\partial\Omega}
  \frac{f(\zeta)}2\cot\frac{\zeta-z}2d\zeta, \qquad
  1 = \frac1{2\pi i}\int_{\partial\Omega}
  \frac12\cot\frac{\zeta-z}2d\zeta,
\end{equation}
where $\pa\Omega=\cup_{k=0}^N \Gamma_k$.  Multiplying the second
equation by $f(z)$ and subtracting from the first, we obtain
\begin{equation}
  \frac1{2\pi i}\int_{\partial\Omega}
  \frac{f(\zeta)-f(z)}2\cot\frac{\zeta-z}2d\zeta = 0,
  \qquad (z\in\Omega).
\end{equation}
The integrand is a product of two analytic functions of $z$ and
$\zeta$, namely $\frac{\zeta-z}2\cot\frac{\zeta-z}2$ and the divided
difference $f[\zeta,z]=\big(f(\zeta)-f(z)\big)/(\zeta-z)= \int_0^1
f'(z+(\zeta-z)\alpha)\,d\alpha$. In particular,
$f[\zeta,\zeta]=f'(\zeta)$ is finite, and the $k$th partial derivative
of $f[\zeta,z]$ with respect to $\zeta$ is bounded, uniformly in $z$,
by $\max_{w\in\Omega}|f^\e{k+1}(w)|/(n+1)$. Thus, the integrand is
smooth and the integral can be approximated with spectral accuracy
using the trapezoidal rule,
\begin{equation}
  \sum_{k=0}^N \frac1{M_k}\sum_{m=0}^{M_k-1}
  \frac{f(\zeta_k(\alpha_m)) - f(z)}2\cot\frac{\zeta_k(\alpha_m)-z}{2}
  \zeta'(\alpha_m) \approx 0.
\end{equation}
Solving for $f(z)$ gives
\begin{equation}
  f(z) \approx \frac{
    \sum_{k=0}^N \frac1{M_k}\sum_{m=0}^{M_k-1}
  \frac{f(\zeta_k(\alpha_m))}2\cot\frac{\zeta_k(\alpha_m)-z}2
  \zeta'(\alpha_m)}{
    \sum_{k=0}^N \frac1{M_k}\sum_{m=0}^{M_k-1}
  \frac12\cot\frac{\zeta_k(\alpha_m)-z}2
  \zeta'(\alpha_m)}, \qquad (z\in\Omega).
\end{equation}
In \eqref{eq:helsing:quad}, we interpret this as a quadrature
rule for evaluating the first integral of \eqref{eq:res:thm}
that maintains spectral accuracy even if $z$ approaches
or coincides with a boundary point $\zeta_k(\alpha_m)$.

\section{Remarks on Generalization to Three Dimensions}
\label{sec:3d}

We anticipate that both methods of this paper generalize to 3D with
some modifications.  One aspect of the problem becomes easier in 3D,
namely that the velocity potential is single-valued. However, one
loses complex analysis tools such as summing over periodic images in
closed form with the cotangent kernel and making use of the residue
theorem to accurately evaluate layer potentials near the boundary.

The velocity potential method can be adapted to 3D by replacing
constant boundary conditions for the stream function on the solid
boundaries with homogeneous Neumann conditions for the velocity
potential. This entails using a double layer potential on the free
surface and single layer potentials on the remaining boundaries. In
her recent PhD thesis \cite{huang_thesis}, Huang shows how to do this
in an axisymmetric HLS framework. She implemented the method to study
the dynamics of an axisymmetric bubble rising in an infinite
cylindrical tube. One of the biggest challenges was finding an analog
of the Hilbert transform to regularize the hypersingular integral that
arises for the normal velocity. Huang introduces a three-parameter
family of harmonic functions involving spherical harmonics for this
purpose. This method can handle background flow along the axis of
symmetry, but many technical challenges remain for the
non-axisymmetric case, e.g., for doubly-periodic boundary conditions
in the horizontal directions.

Analogues of the vortex sheet method in three dimensions have been
developed previously in various contexts.  Caflisch and Li
\cite{caflisch:li} work out the evolution equations in a Lagrangian
formulation of a density-matched vortex sheet with surface tension in
an axisymmetric setting.  Nie \cite{nieAxisymmetric} shows how to
incorporate the HLS method to study axisymmetric, density-matched
vortex sheets. In his recent PhD thesis, Koga \cite{koga_thesis}
studies the dynamics of axisymmetric vortex sheets separating a
``droplet'' from a density-matched ambient fluid. He develops a
mesh-refinement scheme based on signal processing and shows how to
regularize singular axisymmetric Biot-Savart integrals with new
quadrature rules. Koga implements these ideas using graphics
processing units (GPUs) to accelerate the computations.

The non-axisymmetric problem with doubly-periodic boundary conditions
has been undertaken by Ambrose \emph{et al.}
\cite{ambroseSiegelTlupova}. They propose a generalized isothermal
parameterization of the free surface, building on work of Ambrose and
Masmoudi \cite{ambrose:masmoudi}, which possesses several of the
advantages of the HLS angle-arclength parameterization in 2D.  The
context of \cite{ambroseSiegelTlupova} is interfacial Darcy flow in
porous media, which also involves Birkhoff-Rott integrals in 3D:
\begin{equation}\label{3DBR}
  \mb W(\vec{\alpha})=\frac{1}{4\pi}\mathrm{PV}\!\!\iint 
  \left(\omega_{\alpha}\mathbf{X}_{\beta}-\omega_{\beta}
    \mathbf{X}_{\alpha}\right)\times
  \frac{\mathbf{X}-\mathbf{X}'}{
    |\mathbf{X}-\mathbf{X}'|^{3}}\ d\vec{\alpha}'.
\end{equation}
Here $\vec{\alpha}=(\alpha,\beta),$ and the surface is given by
$\mathbf{X}(\vec{\alpha})=(\xi(\alpha), \eta (\alpha), \zeta
  (\alpha))$ with $\zeta$ now the $z$-coordinate instead of the
complexified surface.  In the integrand, the subscripts $\alpha$ and
$\beta$ represent derivatives with respect to these variables, and
quantities without a prime are evaluated at $\vec{\alpha}$ while
quantities with a prime are evaluated at $\vec{\alpha}'.$ The domain
of integration is $\mathbb{R}^{2}.$ The quantity $\omega$ is, as in
the 2D problem, the source strength in the double layer potential.

The lack of a closed formula for the sum over periodic images in
\eqref{3DBR} contributes to the computational challenge of
implementing the method in 3D. In \cite{ambroseSiegelTlupova}, a fast
method for calculation of this integral is introduced, based on Ewald
summation.  This involves splitting the calculation of the integral
into a local component in physical coordinates and a complementary
calculation in Fourier space; the method is optimized so that the two
sums take similar amounts of work.  We expect that the single layer
potentials that occur at solid boundaries in the multiply-connected
case of the present paper could be computed similarly in 3D.

\bibliographystyle{abbrv}
 

\end{document}